\documentclass[aps,11pt,nofootinbib]{revtex4-1}
\pdfoutput=1
\usepackage{graphicx}
\usepackage[margin=1in]{geometry}
\usepackage{amsmath}
\usepackage{amssymb}
\usepackage{color}
\usepackage[utf8]{inputenc}
\usepackage{hyperref}
\usepackage[normalem]{ulem} % needed for \sout command to be removed later
\usepackage{cancel} %needed for \cancel command in math mode
\usepackage{comment}

\makeatletter
\renewcommand\@make@capt@title[2]{%
  \@ifx@empty\float@link{\@firstofone}{\expandafter\href\expandafter{\float@link}}%
   {\textbf{#1}}\@caption@fignum@sep#2\quad
}%
\makeatother

\newcommand{\be}{\begin{equation}}
\newcommand{\ee}{\end{equation}}

\newcommand{\eper}{E}
\newcommand{\mper}{M}
\newcommand{\qper}{Q}
\newcommand{\Rsec}{\mathcal{R}}
\newcommand{\fPBH}{f_\mathrm{PBH}}

\newcommand{\msun}{\mathrm{M}_\odot}

\newcommand{\au}{\textsc{au}}
\newcommand{\pc}{\mathrm{pc}}
\newcommand{\kms}{\mathrm{km\,s^{-1}}}
\newcommand{\renc}{R_\mathrm{enc}}
\newcommand{\gyr}{\mathrm{Gyr}}

\newcommand{\epssa}{\epsilon_\mathrm{SA}}
\newcommand{\epsoct}{\epsilon_\mathrm{oct}}

\usepackage{bib_macros}

\begin{document}

\author{Sam Young$^1$}
\email{syoung@mpa-garching.mpg.de}
\author{Adrian S. Hamers$^{1}$}
\email{hamers@mpa-garching.mpg.de}

\affiliation{1) Max Planck Institute for Astrophysics, Karl-Schwarzschild-Strasse 1, 85748 Garching bei Muenchen, Germany,}

\date{\today}

\title{The impact of distant fly-bys on the rate of binary primordial black hole mergers}

\begin{abstract}
By performing Monte Carlo simulations of the evolution of binary primordial black hole (PBH) systems, we estimate the effect of distant encounters with single PBHs upon the coalescence time and merger rate of binary PBHs. We find that, for models where PBHs compose a large fraction of dark matter, $f_\mathrm{PBH}\sim 1$, the expected fractional change in coalescence time is negligible, of order $10^{-6}$ for most binaries. For models with significantly lower PBH abundances, $f_\mathrm{PBH}\ll 1$, we find that the average change in binary lifetime due to encounters can be as large as $\mathcal{O}(10^{-2})$, with a small number of binaries experiencing an order unity change in lifetime. 
In the absence of encounters, we also compare the use of an analytic approximation for the coalescence time to numerically evolving the binary system, finding that the analytic approximation results in an order $10\%$ error in the coalescence time. However, when these effects are taken into consideration, there is a negligible change to the calculated merger rate, placing previous constraints on the PBH abundance arising from observed gravitational wave signals from merging binary black holes on a more secure footing.

\end{abstract}

\maketitle

\tableofcontents

\section{Introduction}

Since the first detection of gravitational waves by the LIGO scientific collaboration on September 14, 2015 from the merger of 2 black holes (BHs) \cite{Abbott:2016blz}, as well as subsequent detections \cite{Abbott:2016nmj,Abbott:2017vtc,Abbott:2017gyy,Abbott:2017oio,LIGOScientific:2018mvr, Abbott:2020uma,LIGOScientific:2020stg} there has been speculation on the origin of these BHs. Did they form through an astrophysical channel, or did the BHs have a primordial origin? 

Astrophysical channels can broadly be characterized into two types: mergers arising from the evolution of isolated binary stars (e.g., \citep{1973NInfo..27...70T,1997MNRAS.288..245L,1993MNRAS.260..675T,2002ApJ...572..407B,2003MNRAS.342.1169V,2007PhR...442...75K,2012ApJ...759...52D,2013ApJ...779...72D,2014ApJ...789..120B,2016Natur.534..512B,2018MNRAS.473.4174Z,2018PhRvD..98h4036G,2019ApJ...870L..18Q,2020A&A...635A..97B,2020A&A...636A.104B}), 
or dynamical evolution in dense stellar systems or multiple-star systems (e.g., \citep{1993Natur.364..423S,2000ApJ...528L..17P,2004Natur.428..724P,2006ApJ...637..937O,2012ApJ...757...27A,2011ApJ...741...82T,2014ApJ...781...45A,2014MNRAS.441.3703Z,2015ApJ...799..118P,2015PhRvL.115e1101R,2016MNRAS.459.3432M,2016MNRAS.460.3494S,2016MNRAS.463.2443K,2016ApJ...831..187A,2016ApJ...832L...2R,2016PhRvD..93h4029R,2017ApJ...836...39S,2017ApJ...836L..26C,2017ApJ...840L..14S,2017ApJ...841...77A,2017ApJ...846..146P,2018MNRAS.480L..58A,2018PhRvD..98l3005R,2018ApJ...855..124S,2018ApJ...853..140S,2018ApJ...865....2H,2018ApJ...856..140H,2018PhRvD..97j3014S,2018MNRAS.477.4423A,2018PhRvL.120o1101R,2018ApJ...853...93R,2018ApJ...856..140H,2018ApJ...860....5G,2018ApJ...864..134R,2019MNRAS.483..152A,2019MNRAS.486.4443F,2019MNRAS.487.5630H,2019PhRvD.100d3010S,2020MNRAS.493.3920F}). The former typically involves binary interactions such as mass transfer and common envelope evolution ultimately producing a close pair of BHs, whereas the latter involves multi-body interactions between BHs and other objects. Typically, astrophysical channels predict a rate which peaks at low redshift, since cosmic star formation peaked at around $z=2$ (e.g., \citep{2014ARA&A..52..415M}).

Primordial black holes (PBHs) could form through a variety of mechanisms in the early universe, including the collapse of large density perturbations \cite{1967SvA....10..602Z,Hawking:1971ei,Carr:1974nx}, from cosmic strings \cite{Hawking:1987bn}, or from bubble collisions \cite{Hawking:1982ga} (see \citet{2020arXiv200602838C} for a recent review). Part of the motivation for considering PBHs as a candidate for the LIGO BHs is the observed spins of the merging BHs --- which can be difficult to explain from astrophysical BHs (e.g. \citep{2019MNRAS.483.3288P}), but is a natural prediction for PBHs \cite{Clesse:2017bsw,Postnov:2019tmw,Fernandez:2019kyb,DeLuca:2019buf,DeLuca:2020bjf}.

The question of whether LIGO had detected PBHs was quickly investigated by \citet{Bird:2016dcv}, \citet{Clesse:2016vqa} and \citet{2016JCAP...11..036B}, and the initial findings were that the observed merger rate matched closely with the merger rate predicted if dark matter (DM) was composed entirely of PBHs. Since then, however, the calculation has been refined, notably including the formation of binary systems in the early universe \cite{Bringmann:2018mxj,Raidal:2018bbj,Ballesteros:2018swv,Raidal:2017mfl,Chen:2018czv,Ali-Haimoud:2017rtz,Sasaki:2016jop}, and the current consensus is that the observed merger rate is too low for PBHs to make up the entirety of DM, implying that, at most, PBHs could compose $\mathcal{O}(0.1\%)$ of DM, $\fPBH\lesssim0.001$. See \citet{DeLuca:2020qqa} for a recent discussion of PBHs and the LIGO/Virgo observations.

However, such calculations of the merger rate today typically ignore the effect of other nearby objects on the evolution of binary primordial black holes (BPBHs), although attempts have been made to account for this. \citet{Vaskonen:2019jpv} considered the disruption of binaries located in haloes undergoing core collapse, which slightly weakened constraints on the PBH abundance coming from the observed merger rate, whilst \citet{Raidal:2018bbj} used an $N$-body approach to study binary PBHs in the early universe, finding that initial binaries are likely to be disrupted if the abundance of PBHs is large, $\fPBH\gtrsim0.1$. 

In this paper, we will study the effect of ``fly-bys'' (other PBHs passing near the binary system) in the late universe. A simple calculation was performed in \citep{Ali-Haimoud:2017rtz}, which found that the chance of single PBHs passing by closely enough to have a significant effect on the coalescence time of BPBH was unlikely, although the approximate calculation employed therein may not be accurate given the highly eccentric nature of BPBHs \citep{2019MNRAS.487.5630H}. Here, we will investigate the cumulative effect of many such fly-bys using more accurate Monte Carlo methods that employ analytic equations to predict the effects of each fly-by.

In particular, we will use the secular approximation, in which the BPBH orbital period is much shorter than the passage time-scale of the perturbing body (e.g., \citep{1975MNRAS.173..729H,1996MNRAS.282.1064H,2009ApJ...697..458S,2018MNRAS.476.4139H,2019MNRAS.487.5630H,2019MNRAS.488.5192H}). This approximation works well for the overwhelming majority of perturbers in our scenario, as we will show in section \ref{sect:sec}. We combine Monte Carlo sampling of perturbers with the decay of the BPBH orbit due to the emission of GWs, similar in approach to \citep{2019PhRvD.100d3010S} who considered BH binaries of astrophysical origin in globular clusters.

The organisation of this paper is as follows:
\begin{itemize}
    \item Section \ref{sect:formation} will discuss the formation of binaries and initial conditions of BPBHs in the early universe;
    \item Section \ref{sect:sec} discusses the secular regime relevant for distant fly-bys;
    \item Section \ref{sect:estimate} presents an analytic estimate of the change in coalescence time due to the effect of fly-bys;
    \item Section \ref{sect:mc} deals with how the evolution of binary systems is calculated using a Monte Carlo approach;
    \item Section \ref{sect:results} presents the results of the Monte Carlo simulations, and finally;
    \item Section \ref{sect:discuss} discusses the conclusions and implications from our investigation.
    
\end{itemize}

\section{Binary formation and initial conditions}
\label{sect:formation}

 We will be performing a Monte Carlo procedure to model the evolution of binary systems as other PBHs pass nearby, and to this end, we require that the initial distribution of binary systems and their orbital parameters closely match those expected in the early universe. Therefore, in this section we will discuss the formation of binary PBH systems in the early universe, and the calculation of their initial conditions. In order to do this with, we will follow the derivation of initial conditions given in \citet{Raidal:2018bbj}.
 
 We will assume throughout that PBHs form with a Poissonian spatial distribution, which is consistent with a Gaussian distribution of primordial fluctuations. The presence of primordial non-Gaussianity is expected to have a significant impact not only on the PBH abundance \cite{Bullock:1996at,Ivanov:1997ia,Byrnes:2012yx,Shandera:2012ke,Young:2013oia,Young:2014oea,Young:2015cyn,Franciolini:2018vbk,Yoo:2019pma,Atal:2019cdz,Atal:2018neu}, but also the initial clustering \cite{Tada:2015noa,Young:2015kda,Suyama:2019cst}, mass function, and the merger rate observed today \cite{Young:2019gfc}. Note that, whilst the density contrast $\delta$ is expected to be significantly non-Gaussian even if the curvature perturbation $\zeta$ is Gaussian \cite{Young:2019yug,Yoo:2018kvb,Kawasaki:2019mbl,Kalaja:2019uju,DeLuca:2019qsy}, this will not affect the initial Poissonian spatial distribution.

\subsection{Primordial black hole mass}
We will consider the case that PBHs form from the collapse of large-amplitude overdensities in the early universe. Assuming PBHs form with a relatively narrow mass function (for example, arising from a narrow peak in the primordial power spectrum leading to enhanced PBH formation at those scales), the mass function can be well approximated with a log-normal distribution \cite{Gow:2020bzo,Young:2019gfc}
\be
\psi(m) = \frac{1}{\sqrt{2\pi}\sigma_m m}\exp\left(-\frac{1}{2}\frac{\ln^2(m/m_\mathrm{c})}{\sigma_m^2} \right),
\label{eqn:logNormalMassFn}
\ee
where $m$ is the PBH mass, $m_\mathrm{c}$ is the mass at which the distribution peaks, and $\sigma_m$ is the width of the distribution. In this paper, we will treat $m_\mathrm{c}$ and $\sigma_m$ as free parameters, although a best-fit model to the black hole coalescence events observed by LIGO suggests $m_\mathrm{c} \approx 20\,\msun$ and $\sigma_m \approx 0.5$ \cite{Raidal:2018bbj}. Note that this definition of the mass function is normalised to integrate to unity, $\int \mathrm{d} m \, \psi(m)=1$ (and can therefore be interpreted as the probability distribution function (PDF) for PBH mass). To describe the total abundance of PBHs, we will use the parameter $\fPBH$, which is the fraction of DM composed of PBHs,
\be
f_{\mathrm{PBH}} = \frac{\Omega_\mathrm{PBH}}{\Omega_{\mathrm{CMD}}},
\ee
where $\Omega$ is the density parameter for PBHs or cold dark matter (CDM). The number density $n$ of PBHs can therefore be expressed as
\be
n = \frac{\fPBH\rho_{\mathrm{CDM}}}{\bar{m}},
\label{eqn:numberDensity}
\ee
where $\bar{m}=m_\mathrm{c}\exp\left(\sigma_m^2/2\right)$ is the mean PBH mass. We note that, since the PBH density evolves as matter, the (mean) number density expressed in comoving coordinates is constant with respect to time. %We also define the differential number density 
%\be
%n=\frac{\mathrm{d}N}{\mathrm{d}m}.
%\ee

\subsection{Initial semi-major axis}
In the absence of primordial non-Gaussianity, PBHs are expected to follow a Poissonian spatial distribution. The PDF for the radial distance $r_0$ from a given PBH to its nearest neighbor at formation is then given by 
\be
P_{\mathrm{nn}}(r_0) = 4\pi r_0^2 n\exp\left( -\frac{4}{3}\pi r_0^3 n \right),
\ee
where $n$ is the number density of PBHs at the time of formation. In order for a pair of PBHs to form a binary system, we require that the PBHs are allowed to decouple from the Hubble flow and fall towards each other without being disrupted by other nearby perturbations/PBHs. To achieve this, we include an exclusion zone around the pair, in which no other PBHs are found --- which is a factor $A$ larger than the initial separation of the PBH pair.
The probability of finding the nearest neighbor in the range $r\rightarrow r+\mathrm{d}r$, and no other PBHs within the exclusion zone (a sphere of radius $Ar$) is then
\be
P_{\mathrm{nn,ex}}(r_0) = 4\pi r_0^2 n\exp\left( -\frac{4}{3}\pi (A r_0)^3 n \right).
\label{eqn:NearestNeighbour}
\ee
We will later want the distribution for the initial separation of binary systems (rather than PBH pairs which may not form binaries), and so the distribution is correctly normalised as
\be
P_{\mathrm{bin}}(r_0) = 4\pi r_0^2 n A^3 \exp\left( -\frac{4}{3}\pi (Ar_0)^3 n \right),
\label{eqn:initialSeparation}
\ee
where the factor $A^3$ can be considered a normalisation factor to ensure that $\int \mathrm{d}r_0 \, P_{\mathrm{bin}}(r_0)=1$, since we want the distribution of $r_0$ for PBH pairs which form binaries. 

The number density of initial pairs forming binary systems, with initial separation in the range $r_0\rightarrow r_0+\mathrm{d}r_0$ and masses in the range $m_{1,2}\rightarrow m_{1,2}+\mathrm{d}m$, can be calculated by multiplying the relevant probabilities:
\be
\mathrm{d}n_{\mathrm{bin}} = 2\pi r_0^2 n \exp\left( -\frac{4}{3}\pi (Ar_0)^3 n \right)\mathrm{d}r_0 \mathrm{d}n(m_1)\mathrm{d}n(m_2),
\ee
where a factor of $1/2$ is included to avoid overcounting, and $\int \mathrm{d}n(m)=n$. This equation is equivalent to equation (2.18) in \citet{Raidal:2018bbj}, with $4\pi r_0^2 \mathrm{d}r_0 = \mathrm{d}V(x_0)$ and $\frac{4}{3}\pi (Ar_0)^3 n = \tilde{N}(y)$.

When PBHs form significantly close to each other relative to the average, the pair can be considered a matter overdensity, which eventually decouples from the Hubble flow and ``collapses'' to form a binary system. We follow \citet{Raidal:2018bbj} and define the quantity,
\be
\delta_b \equiv \frac{3(m_1+m_2)}{2\pi r_0^3 \rho_\mathrm{M}}.
\ee
where $\rho_\mathrm{M}$ is the matter energy density. During radiation domination, such matter perturbations collapse when $\rho_\mathrm{r} a^{-4} = \delta_b \rho_\mathrm{M} a^{-3}$, where $\rho_\mathrm{r}$ is the radiation energy density. An approximate estimate for the decoupling scale factor is then given by
\be
a_\mathrm{dc} \equiv \frac{a_\mathrm{eq}}{\delta_b},
\ee
where the $\mathrm{eq}$ represents matter-radiation equality. The approximate value for the initial semi-major axis $r_a$ of the binary when it decouples from the Hubble flow is given by \citet{Raidal:2018bbj}
\be
r_a \approx 0.1 a_\mathrm{dc} r_0 = \frac{0.1 a_\mathrm{eq}r_0}{\delta_b}.
\ee
An estimate for the initial semi-major axis can therefore be calculated from the PBHs masses and initial separation, for which the PDFs are known. As will be seen, our conclusion is not sensitive to small errors in the initial conditions, and so the approximations made here are considered acceptable.

\subsection{Initial angular momentum}

After a PBH pair decouples from the Hubble flow, in the absence of external perturbations/objects modifying the gravitational field around the PBH pair, the pair would fall straight back towards each other and immediately coalesce. However, nearby density perturbations and PBHs can provide a torque to the system, imparting sufficient angular momentum to the pair to prevent a head-on collision and instead form a stable binary system.

When describing the orbits of binary systems, it will be helpful for use the eccentricity $e$ to describe the ellipticity, defined as
\be
e = \frac{r_\mathrm{a}-r_\mathrm{p}}{r_\mathrm{a}+r_\mathrm{p}},
\ee
where $r_\mathrm{a}$ is the apoapsis, and $r_\mathrm{p}$ is the periapsis. However, when calculating the distribution of initial conditions, it will be more helpful to describe the dimensionless angular momentum $j$, related to the eccentricity as
\be
j = \sqrt{1-e^2}.
\ee

\citet{Raidal:2018bbj} gives an order of magnitude estimate for the initial angular momentum
\be
j_0 \approx 0.4\frac{\fPBH}{\delta_b},
\ee
under the assumption that most of the torque is generated by the nearest PBH to the pair. Depending on the exact configuration of nearby density perturbations and PBHs, the actual angular momentum $j$ will vary relative to this value, and the PDF itself varies depending on the expected number of PBHs in the exclusion zone, $\tilde{N}$. \citet{Raidal:2018bbj} provides several forms for the PDF depending, depending on the value of $\tilde{N}(y)$. In the limit $\tilde{N}\rightarrow 0$, the PDF of the initial angular momentum $j$ is given by a power-law:
\be
P(j) = \frac{j/j_0^2}{\left( 1+j^2/j_0^2 \right)^{3/2}}.
\label{eqn:powerLaw}
\ee
In the limit $\tilde{N}\rightarrow \infty$, a Gaussian-like PDF is found instead:
\be
P(j) = \frac{2j}{\sigma_j^2}\exp\left( -\frac{j^2}{\sigma_j^2} \right),
\label{eqn:gaussLike}
\ee
with $\sigma_j$ given by
\be
\sigma_j^2 = \frac{6}{5}j_0^2\left( \frac{\langle m^2 \rangle}{\langle m \rangle^2 \tilde{N}(y)} + \frac{\sigma_\mathrm{M}^2}{\fPBH^2} \right).
\ee
Here, $\sigma^2_\mathrm{M} \equiv \frac{\Omega_\mathrm{M}}{\Omega_\mathrm{DM}}\langle \delta_\mathrm{M}^2\rangle$ is the re-scaled variance of the matter density perturbation, $\Omega_\mathrm{M}$ and $\Omega_\mathrm{DM}$ are the density parameter for matter and DM respectively. We will follow \citet{Ali-Haimoud:2017rtz} and take $\langle \delta_\mathrm{M}^2\rangle = 0.005^2$ when a numerical estimate is required.

If the variance of matter perturbations is dominated by the Poisson noise generated by PBHs, the power-law PDF is expected to hold --- that is, if $\fPBH\lesssim \sigma_\mathrm{M}$. As can be seen in figures \ref{fig:a_dist} and \ref{fig:j_dist}, for PBHs expected to merge around today, whilst the tails of the distributions can be quite different, the peaks of the distributions for $j$ are quite similar, and so we will proceed by using the power-law distribution to generate initial conditions --- which has a negligible impact on our results.

The required orbital parameters can then be sampled by generating random masses, $m_1$ and $m_2$, and initial separation $r_0$ for the binary PBHs. These values are used to infer an initial semi-major axis $r_a$, and characteristic angular momentum $j_0$, which are then used to generate a distribution for the angular momentum, from which a random angular momentum $j$ is drawn, and the eccentricity calculated.

\subsection{Distribution of initial conditions for PBHs merging today}

Here we will briefly discuss the typical values for the initial orbital parameters for the binary PBH systems. A full analysis of the entire parameter range will not be considered here (the interested reader can find a more detailed and thorough analysis in \citet{Ali-Haimoud:2017rtz} and \citet{Raidal:2018bbj}) --- but a brief summary is included here, because it is useful to consider the typical values which we might expect to find for binary PBHs expected to be merging today, and how these depend on several key parameters.

The first thing to consider is the abundance of PBHs. The higher the number density of PBHs, the closer PBHs will initially be. This means that for large $\fPBH$, we expect to find a smaller initial separation --- and this smaller separation means a stronger gravitational binding, meaning an earlier decoupling from the Hubble flow, and an even smaller initial semi-major axis. Therefore, $\fPBH$ will have a strong impact on the distribution of the initial semi-major axis. 

The mass function will have a similar effect. If the average PBH mass is smaller, this will imply a larger number density of PBHs (assuming the same $\fPBH$) --- and so a higher average PBH mass will typically imply a larger initial semi-major axis. 

Here, we make the simple assumption that two PBHs which form close to each other will eventually decouple from the Hubble flow and form a binary, if there are no other PBHs nearby to disrupt this process. This is parameterised by an exclusion radius: we require that there are no other PBHs within a radius $A r_0$ (recall that $r_0$ is the initial separation of the binary). Choosing a higher value for $A$ implies that PBH pairs which form binaries would have a smaller initial separation, and lower initial semi-major axis. A sensible choice for $A$ is likely to be $2 \lesssim A\lesssim \mathcal{O}(5)$, and we find that this has a small effect on the initial separation, of order unity.

Now, let us turn our attention to the initial angular momentum. After the PBHs form, the PBH pair decouples from the Hubble flow and the PBHs begin to fall back towards each other --- before beginning to oscillate around each other. Most of the torque, which provides the initial angular momentum, is expected to originate from the nearest PBH to the fledgling binary --- the higher $\fPBH$ is, the closer the nearest neigbour is likely to be, and the larger the angular momentum is likely to be.

In addition, the torque is strongest when the nearest neighbour is close relative to the binary separation, before the binary decouples from the Hubble flow --- therefore, the longer a system takes to decouple (due to a larger initial separation for example), the higher the total effect of the torque is likely to be. We note that, interestingly, the power-law distribution of $j$ is actually independent of the PBH mass function (although the specific masses of the PBHs in the binary do enter indirectly through $j_0$).

When considering binary systems which are expected to be merging today, a larger initial semi-major axis $r_a$ would require a smaller initial angular momentum $j$ and vice-versa (see equation \eqref{eqn:approxTime} later in the paper). Taken at face value, the simple arguments presented above present a somewhat contradictory picture --- a larger $f_{\mathrm{PBH}}$ implies $r_a$ should be smaller for binaries merging today, whilst also implying $j$ should be smaller. In the end, it is the smaller semi-major axis argument which is more important (owing partly to the fact that a large decrease in $r_a$ can be cancelled by a small increase in $j$ to give the same coalescence time).

\begin{figure*}[t!]
%\vspace{-1.5cm}
 \centering
  \includegraphics[width=0.49\textwidth]{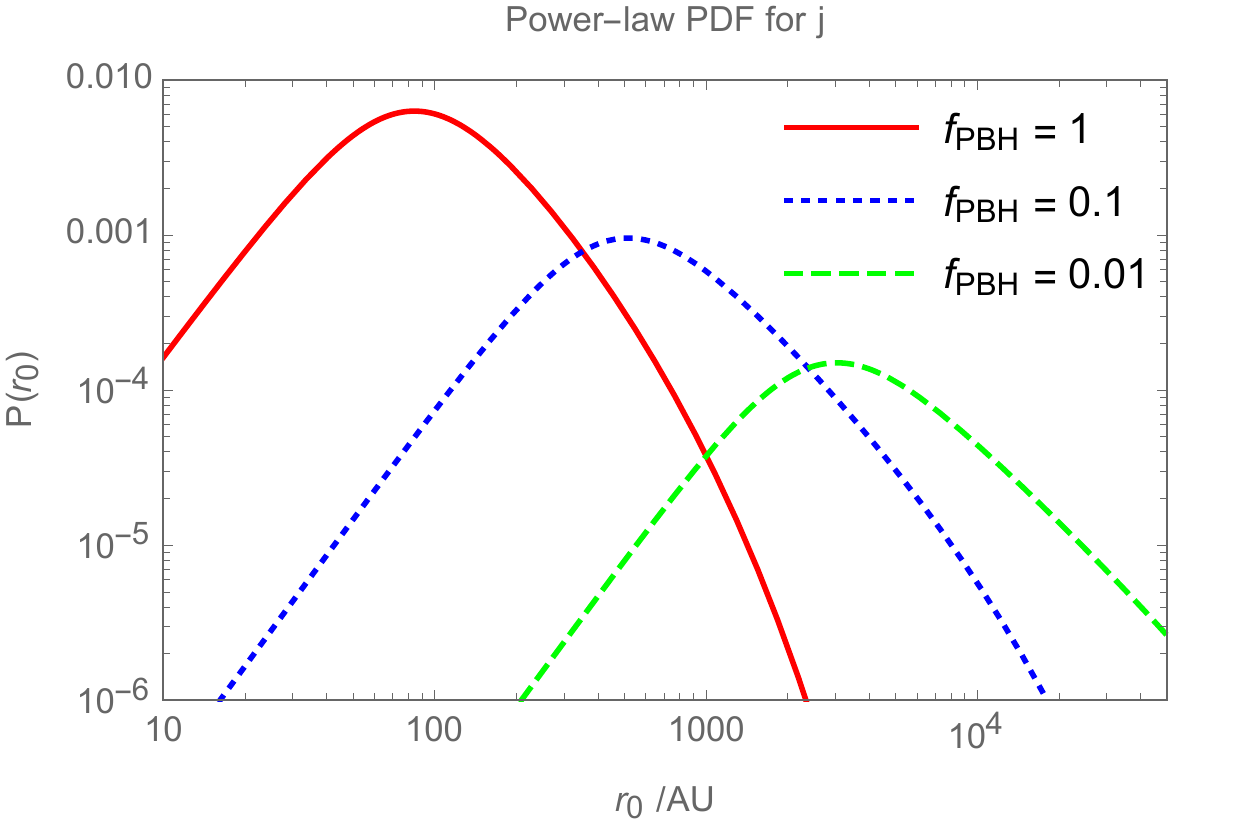} 
  \includegraphics[width=0.49\textwidth]{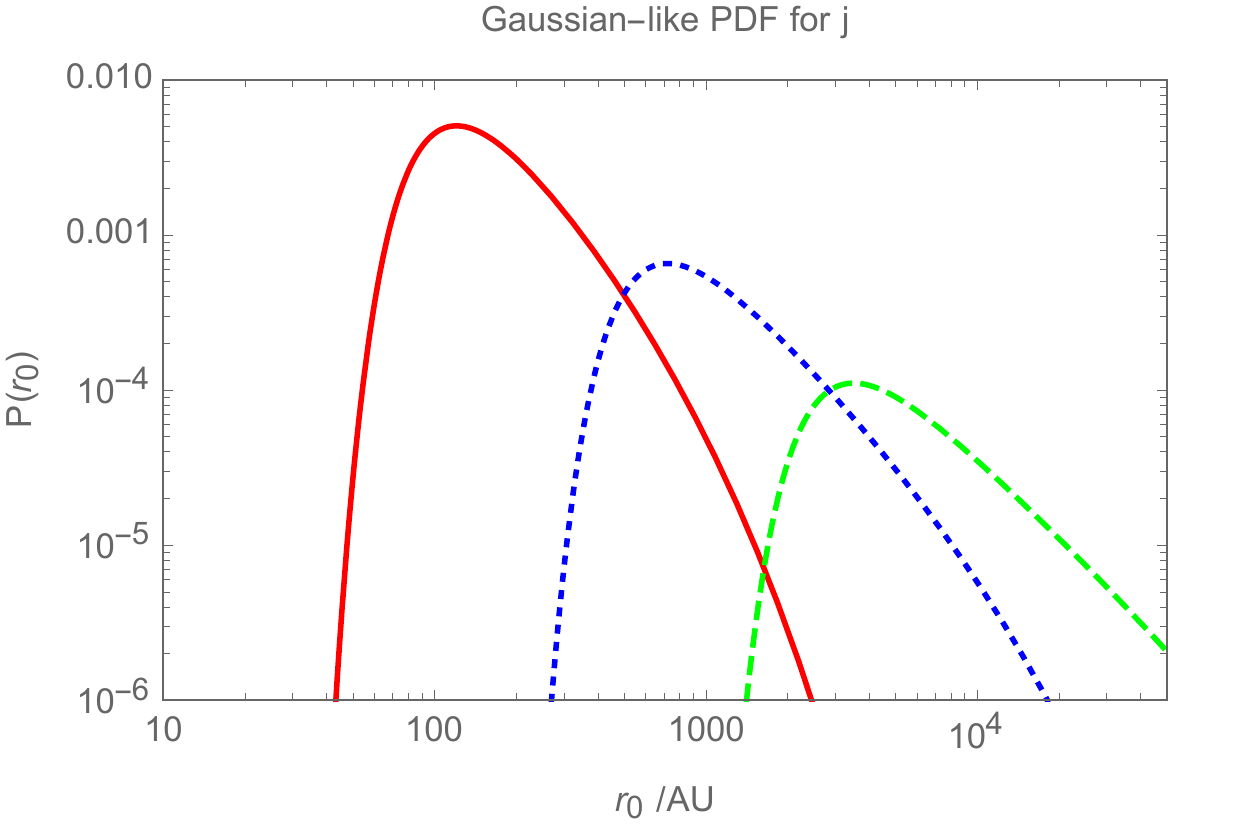} 
 % \vspace{-2.0cm}
  \caption{ The distribution of semi-major axes $r_0$ for binaries merging today, assuming either a power-law distribution, equation \eqref{eqn:powerLaw}, or a Gaussian-like distribution, equation \eqref{eqn:gaussLike}. We have assumed the following choices for the parameters: masses $m_1 = m_2 = m_\mathrm{c} = 20\,\msun$, mass function width $\sigma_m = 0.05$, and an exclusion zone $A=2$ times greater than the initial PBH separation. The abundance of PBHs is described by $\fPBH$, the fraction of dark matter composed of PBHs. Both distributions peak at similar $r_0$, but the power-law distribution has a significantly larger tail at small $r_0$.}
  \label{fig:a_dist}
\end{figure*}

\begin{figure*}[t!]
%\vspace{-1.5cm}
 \centering
  \includegraphics[width=0.49\textwidth]{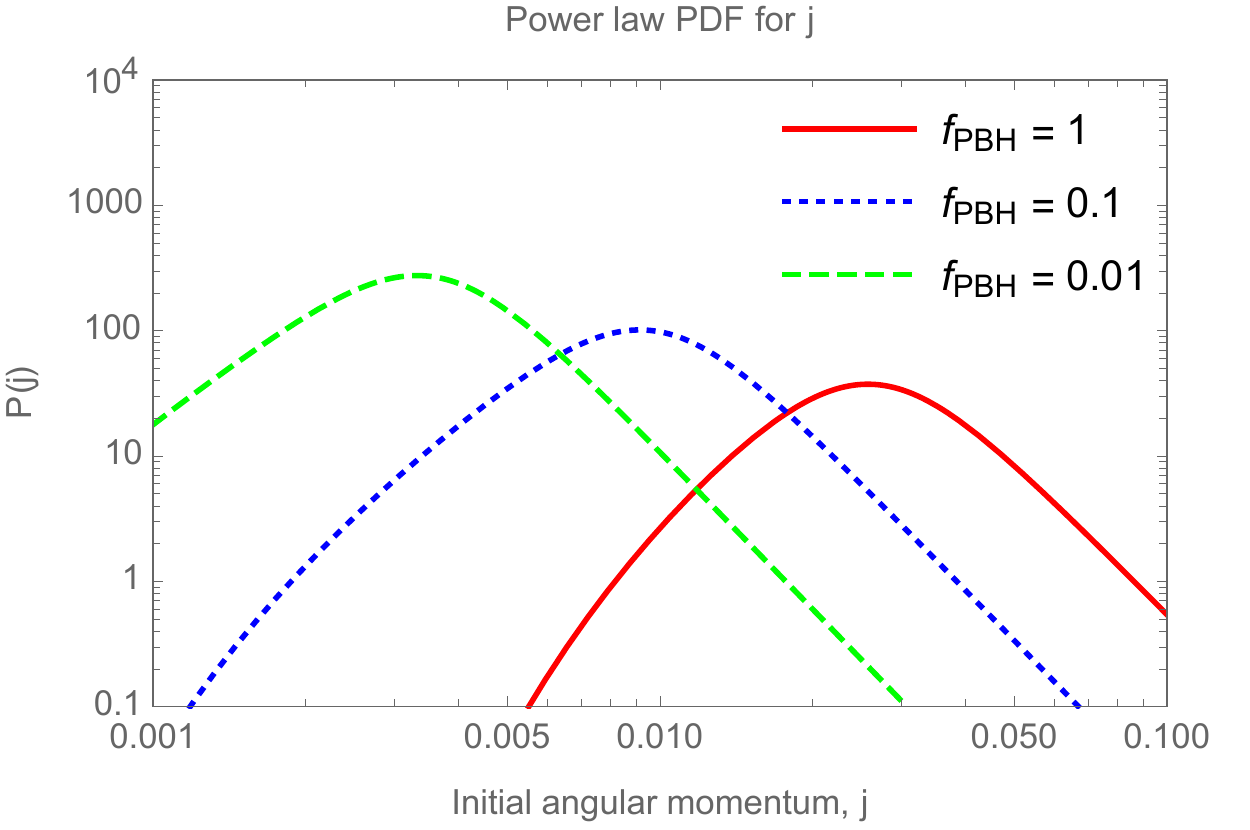} 
  \includegraphics[width=0.49\textwidth]{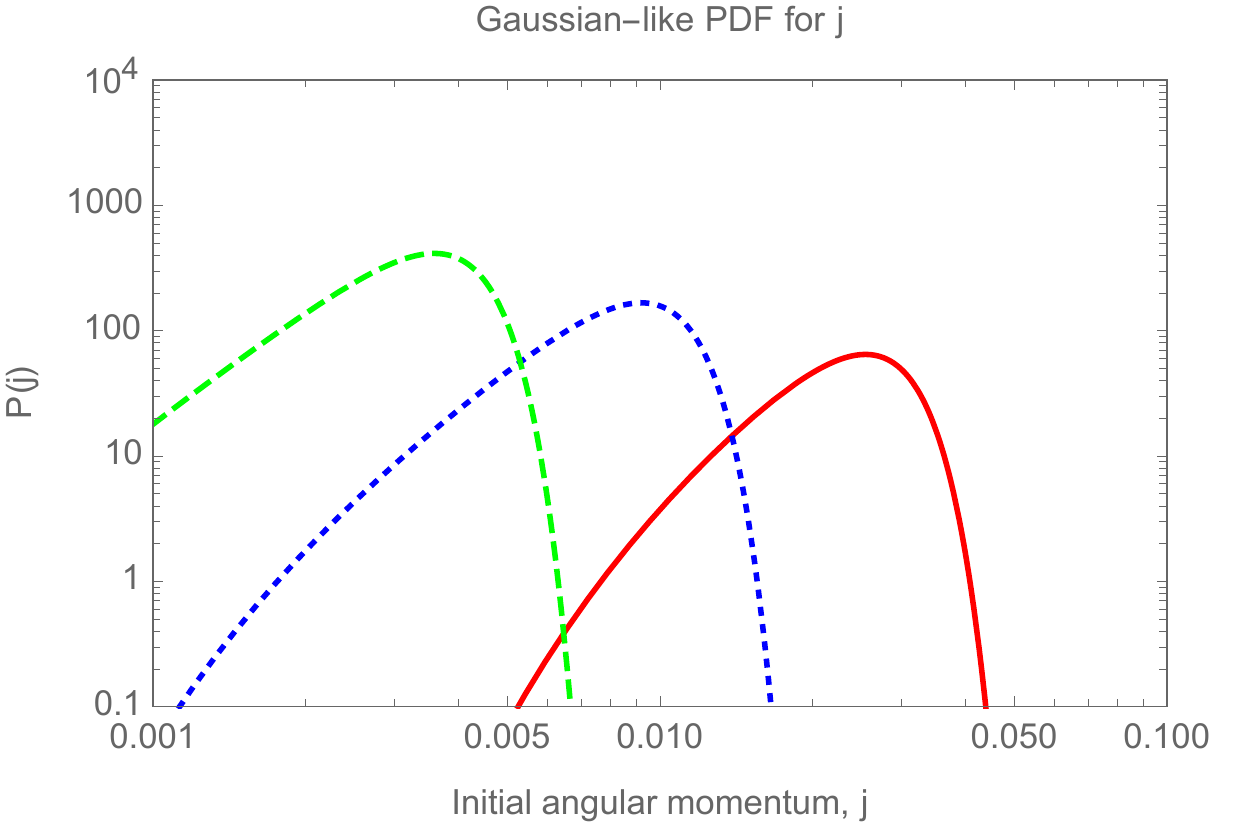} 
 % \vspace{-2.0cm}
  \caption{ The distribution of initial angular momentum $j$ for binaries merging today, again using either a power-law distribution, equation \eqref{eqn:powerLaw}, or a Gaussian-like distribution, equation \eqref{eqn:gaussLike}. We have used the same parameter choices as previously: masses $m_1 = m_2 = m_\mathrm{c} = 20\,\msun$, mass function width $\sigma_m = 0.05$, and an exclusion zone $A=2$ times greater than the initial PBH separation, with the abundance of PBHs is described by $\fPBH$, the fraction of dark matter composed of PBHs. Again, we see that both distributions predict a similar characteristic $j$, and we see a larger tail for high $j$ for the power-law distribution (corresponding to the low $r_0$ tail in figure \ref{fig:a_dist}). } 
  \label{fig:j_dist}
\end{figure*}

Depending on the choices of these parameters, typical semi-major axes for PBHs expecting to merge today can vary from tens to tens of thousands of AUs, whilst the initial angular momentum can vary from $\mathcal{O}(10^{-3})$ to $\mathcal{O}(10^{-2})$ --- meaning extremely high eccentricities, $e \gtrsim 0.999$. Figure \ref{fig:a_dist} shows the PDF of initial semi-major axes for PBHs expected to merge today ($13.7\,\gyr$ after formation of the binary) for different values of $\fPBH$, whilst figure \ref{fig:j_dist} shows the same for the PDF of the initial angular momentum. We have assumed the following choices for the parameters: masses $m_1 = m_2 = m_\mathrm{c} = 20\,\msun$, mass function width $\sigma_m = 0.05$, and an exclusion zone $A=2$ times greater than the initial PBH separation.

\section{Fly-bys in the secular regime}
\label{sect:sec}

Any object with mass $\mper$ passing by the PBH binary with mass $m\equiv m_1+m_2$ will affect the binary system, potentially breaking it up (e.g., \citep{1983ApJ...268..319H,1983ApJ...268..342H,1991MNRAS.250..555H,1993ApJS...85..347H,1993ApJ...403..256H,1993ApJ...403..271G,1993ApJ...415..631S,1993ApJ...411..285D,1996ApJ...467..348M,1996ApJ...467..359H,2012PhRvD..85l3005K,2018arXiv180208654S}). In a sub-type of interactions, the third object passes in a wide orbit relative to the binary, which conserves the binary's semimajor axis $r_a$, but induces changes to the angular-momentum and eccentricity vectors. The latter case, known as the secular regime (e.g., \citep{1975MNRAS.173..729H,1996MNRAS.282.1064H,2018MNRAS.476.4139H,2019MNRAS.487.5630H}), is characterised by the dimensionless quantity (also known as the `adiabatic ratio')
\be
\label{eq:R_def}
\Rsec = \left [ \left ( 1 + \frac{\mper}{m} \right ) \left (\frac{r_a}{\qper} \right )^3 \left ( 1 + \eper \right ) \right ]^{1/2},
\ee
which is the ratio of the perturber's angular speed at periapsis to the binary's mean motion. Here, $\qper$ is the perturber's periapsis distance to the binary's center of mass, and $\eper$ is the eccentricity of the perturber's orbit ($\eper\geq1$). Note that the perturber eccentricity $\eper$ can be written in terms of the velocity at infinity $v$ as
\be 
\label{eq:eper}
E = 1 + \frac{Q v^2}{G(\mper+m)} \approx 1 + 8 \times 10^5 \left (\frac{\qper}{10^6\,\au} \right ) \left ( \frac{v}{200\,\kms} \right )^2 \left ( \frac{M+m}{60\,\msun} \right )^{-1}.
\ee
In our model, encounters are typically highly hyperbolic ($\eper \gg 1$). If $\Rsec \ll1$, this indicates that the secular regime applies, and it is appropriate to average the equations of motion over the binary's orbital phase. 

\begin{comment}
To estimate the typical $\Rsec$ in our scenario, we first consider the rate $\Gamma$ of perturbers approaching the binary within a distance $\renc$, which is given approximately by (e.g., \citep{2017AJ....154..272H})
\be
\Gamma \approx 2 \sqrt{2\pi} \renc^2 n \sigma_v,
\label{eqn:gamma}
\ee
where $\sigma_v$ is the velocity dispersion. We require that the number of encounters within a time $\Delta t= 1\,\gyr$ is equal to $N_\mathrm{enc} = \Gamma \Delta t = 10$, giving
\be
\renc = \sqrt{\frac{N_\mathrm{enc}}{2\sqrt{2\pi} n \sigma_v \Delta t}} \approx 10^6\,\au \, \left (\frac{N_\mathrm{enc}}{10} \right )^{1/2} \left ( \frac{n}{9 \times 10^{-6} \, \pc^{-3}} \right )^{-1/2} \left ( \frac{\sigma_v}{6\,\kms} \right )^{-1/2} \left ( \frac{\Delta t}{1\,\gyr} \right )^{-1/2}.
\ee
With $\renc \sim 10^6\,\au$, the adiabatic ratio is estimated to be
\end{comment}

For our fiducial model, the typical adiabatic ratio is
\be
\Rsec \sim 10^{-3} \, \left ( \frac{m}{40\,\msun} \right )^{-1/2} \left ( \frac{r_a}{100\,\au} \right )^{3/2} \left ( \frac{\qper}{10^6\,\au} \right )^{-1} \left (\frac{v}{200\,\kms} \right ),
\ee
where we used that $\eper \gg 1$. Since $\Rsec \ll1$, we can safely assume that the overwhelming majority of perturbers are within the secular regime.

In the secular regime, the changes to the orbital parameters can be computed analytically. In particular, in the limit of parabolic encounters ($\eper=1$), the change in the scalar eccentricity $\Delta e$ can be expressed relatively compactly according to \citep{2019MNRAS.487.5630H}
\begin{align}
\label{eq:Deltae_sec}
\nonumber \Delta e &= \epssa \frac{15\pi}{4} e \sqrt{1-e^2} \sin 2\omega \sin^2i + \epssa^2 \frac{3}{512} \pi e \biggl [ +4 \cos 2 i \biggl \{3 \pi  \left(81 e^2-56\right)+200 \left(1-e^2\right) \cos 2 \omega  \sin 2 \Omega \biggl \} \\
\nonumber &\qquad  +3 \pi  \biggl \{ 200 e^2 \sin ^4i \cos 4 \omega +8 \left(16 e^2+9\right) \sin ^2 2 i \cos 2 \omega +\left(39 e^2+36\right) \cos 4 i-299 e^2+124\biggl \} \\
&\qquad + 100 \left(1-e^2\right) \sin 2 \omega  \biggl \{ \left (5 \cos i+3 \cos 3 i\right ) \cos 2 \Omega +6 \sin i \sin 2 i\biggl \} \biggl ] + \mathcal{O} \left (  \epssa^3\right ).
\end{align}
Here, 
\begin{align}
\label{eq:epssa}
\nonumber \epssa &\equiv \left [ \frac{\mper^2}{m(m+\mper)} \left ( \frac{r_a}{\qper} \right )^3 \left(1+\eper \right )^{-3} \right ]^{1/2} \\
&\sim 6 \times 10^{-16} \, \left( \frac{m+M}{60\,\msun} \right ) \left ( \frac{m}{40\,\msun} \right )^{-1/2} \left ( \frac{M}{20\,\msun} \right ) \left ( \frac{r_a}{100\,\au} \right )^{3/2} \left ( \frac{\qper}{10^6\,\au} \right )^{-3} \left ( \frac{v}{200\,\kms} \right )^{-3}
\end{align}
measures the strength of the perturbation (the second line assumes $\eper \gg 1$), and $i$ (inclination), $\omega$ (argument of periapsis), and $\Omega$ (longitude of the ascending node) quantify the binary's orbital orientation with respect to the perturber's orbital plane.  Equation~(\ref{eq:Deltae_sec}) is valid to second order in $\epssa$ and excludes octupole-order terms that arise if $m_1 \neq m_2$. If $m_1 \neq m_2$, then the octupole-order terms are non zero; they are smaller than the quadrupole-order terms by a factor which is on the order of \citep{2019MNRAS.488.5192H}
\be
\epsoct \equiv \frac{|m_1-m_2|}{m_1+m_2} \frac{a}{\qper} \frac{1}{1+\eper} \sim 10^{-10} \, \frac{|m_1-m_2|}{m_1+m_2} \left ( \frac{r_a}{100\,\au} \right ) \left ( \frac{\qper}{10^6\,\au} \right )^{-1}
\ee
(assuming $\eper\sim10^6$ for the numerical estimate). 

In our Monte Carlo calculations (section~\ref{sect:mc} below), we will calculate the effect of the perturbation in the secular approximation using the analytic expressions of \citep{2019MNRAS.487.5630H,2019MNRAS.488.5192H}. Specifically, we include terms of order $\epssa$ and $\epssa^2$ for a given $\eper$. Given the excessively large number of individual terms involved and their small values (see Table 1 of \citep{2019MNRAS.488.5192H}), we omit all octupole-order terms associated with $\epssa^2$ (the octupole-order terms associated with $\epssa$ are included).

\section{Estimated effect of fly-bys}
\label{sect:estimate}

In this section, we will provide an analytic estimate of the cumulative effect of nearby PBHs passing near a binary PBH. We will begin by estimating the number density of PBHs residing within a DM halo. For \emph{Milky Way}-type haloes, the extent of the halo is typically considered to be the region in which the density is 200 times the background matter density of the Universe, $\rho_\mathrm{halo} = 200 \, \Omega_\mathrm{M} \rho_\mathrm{c}$, where $\Omega_\mathrm{M}$ is the total matter density parameter, and $\rho_\mathrm{c}$ is the critical density of the universe today. We will take the numerical values, $\Omega_\mathrm{M} = 0.315$ and $\rho_\mathrm{c} = 1.68\times10^{-23}\,\mathrm{M_\odot \,\au}^{-3}$. For generality, and accounting for the fact that binary PBHs may be found predominantly in PBH clusters with a higher average density, we will consider the halo density to be $X_\mathrm{halo}$ times greater than the background matter density:
\be
n_\mathrm{PBH}=\frac{X_\mathrm{halo}\fPBH\rho_\mathrm{c}}{\bar{m}_\mathrm{PBH}},
\label{eqn:rate}
\ee
where $\bar{m}_\mathrm{PBH}$ is the average PBH mass, and recall that $\fPBH$ is the fraction of DM composed of PBHs. For the remainder of this section, we will consider a monochromatic mass function of PBHs, and thus drop the bar notation, such that all PBHs have mass $m_\mathrm{PBH}$.

The encounter rate is given, as a function of the encounter radius, by
\be
\Gamma \sim \frac{X_\mathrm{halo}\fPBH\rho_\mathrm{c}}{\bar{m}_\mathrm{PBH}}\sigma_v \int \mathrm{d}r 2r,
\ee
where $\sigma_v$ is the velocity dispersion (which in this section we take to be constant, and all perturbers will move at this speed relative to the binary). The reason for not performing the integral will become apparent soon. Note that we are here ignoring the effect of gravitational focusing, which, in the secular regime being considered, has a negligible effect upon the impact parameter and the distribution of encounters at different radii.

The expected number of encounters is then given by $N = \Gamma\tau$, where $\tau$ is the time for which the system is observed, we will use $\tau=13\,\gyr$ as a fiducial value: 
\be
N \approx 10^{-10} \left( \frac{X_\mathrm{halo}}{200} \right) \left( \frac{\sigma_v}{200\,\kms} \right) \left( \frac{\tau}{13\,\gyr} \right) \left( \frac{m_\mathrm{PBH}}{20 \,\msun} \right)^{-1}\fPBH \int \frac{\mathrm{d}r 2r}{\au^2}.
\label{eqn:N}
\ee

Assuming that all fly-bys are well described by the secular regime, we will use equation \eqref{eq:epssa} above to provide an order of magnitude estimate for the effect of an individual fly-by on the eccentricity of a binary:

\be
\Delta e \sim \epssa \approx 6\times 10^{-16}\left( \frac{r_a}{10^2 \,\au} \right)^{3/2}\left( \frac{Q}{10^6 \,\au} \right)^{-3}\left( \frac{\sigma_v}{200\,\kms} \right)^{-3},
\ee
and note that the effect on the semi-major axis is negligible in the secular regime. Also recall that in this section, we assume all PBHs to have mass $m_\mathrm{PBH}$ (including the perturbers).

Analytically calculating the total expected change to the eccentricity combined with orbital evolution due to GW emission is complex due to the cumulative nature of the interactions\footnote{However, in idealised cases and without GW emission, the steady-state due to secular encounters can be computed analytically, see \citep{2019MNRAS.488.5192H}.}, and so in order to provide a simple order-of-magnitude estimate, we will simply consider the eccentricity to change by a positive $\epssa$ in each encounter, and ignore the time-evolution of the orbital parameters. Thus, by equating the impact parameter $Q$ with the radius of an encounter $r$, an upper limit for the total change in eccentricity (by taking the sum of $\Delta e$ from each fly-by, $\Sigma \Delta e$) is given by
\begin{align}
\nonumber x &\equiv \Sigma \Delta e \approx 6 \times 10^{-26} \left( \frac{r_a}{10^2 \,\au} \right)^{3/2} \left( \frac{\sigma_v}{200\,\kms} \right)^{-2} \left( \frac{X_\mathrm{halo}}{200} \right) \left( \frac{\tau}{13\,\gyr} \right) \left( \frac{m_\mathrm{PBH}}{20 \,\msun} \right)^{-1}\fPBH \\
&\qquad \qquad \times \int\limits_{r_\mathrm{min}}^{r_\mathrm{max}} \frac{\mathrm{d}r 2r}{\au^2}\left( \frac{r}{10^6 \,\au} \right)^{-3}.
\label{eqn:sigmaDeltae}
\end{align}
The integral in the number of encounters now becomes important, because encounters at different radii have a different effect on the eccentricity. We can see that the integral will diverge as $r\rightarrow 0$. This is due to the fact that, as the encounter radius decreases, the chance of an encounter within this radius decreases as $r^{-2}$, but the effect increases as $r^3$ --- so even though one does not expect any encounters at small $r$, the expectation is a large change to the eccentricity from such encounters (although at this point the encounter is no longer secular). 

We therefore implement a minimum radius for encounters, such that the expected number of encounters given by equation \eqref{eqn:N}, between 0 and $r_\mathrm{min}$, is $N=0.5$ --- such that most PBHs do not experience a closer encounter than this.
\be
r_\mathrm{min} \approx \sqrt{2}\times 10^{5}\left( \frac{X_\mathrm{halo}}{200} \right)^{-1/2} \left( \frac{\sigma_v}{200\,\kms} \right)^{-1/2} \left( \frac{\tau}{13\,\gyr} \right)^{-1/2} \left( \frac{m_\mathrm{PBH}}{20\,\msun} \right)^{1/2}\fPBH^{-1/2}.
\ee
In the Monte Carlo simulations, we require a maximum encounter radius $R_\mathrm{enc}$ in order to have a finite number of encounters (see section \ref{sect:mc}). However, this is larger, by many orders of magnitude, than $r_\mathrm{min}$, and so here we simply take $r_\mathrm{max}\rightarrow \infty$. Accounting for these limits in the integration gives
\be
x \approx 10^{-12} \left( \frac{r_a}{10^2 \,\au} \right)^{3/2} \left( \frac{\sigma_v}{200\,\kms} \right)^{-3/2} \left( \frac{X_\mathrm{halo}}{200} \right)^{3/2} \left( \frac{\tau}{13\,\gyr} \right)^{3/2} \left( \frac{m_\mathrm{PBH}}{20 \,\msun} \right)^{-3/2}\fPBH^{3/2}.
\ee

The coalescence time for a highly eccentric binary PBH system is expected to be proportional to $j^7 = (1-e^2)^{7/2}$. Assuming that the change in eccentricity is small, we can write $e=e_0 + x$, where the subscript $0$ denotes the fiducial eccentricity, and treat $x$ as an expansion variable. To first order in $x$ then, the coalescence time changes by 
\be
\delta_t \equiv \frac{t-t_0}{t_0} = \frac{(1-(e_0+x)^2)^{7/2}-(1-e_0^2)^{7/2}}{(1-e_0^2)^{7/2}} \approx -\frac{7e_0}{1-e_0^2}x = -\frac{7}{j_0^2}x,
\ee
where in the last equality we have made the substitution $j_0 = \sqrt{1-e_0^2}$, and we have neglected the $e_0\approx1$ term in the numerator. This can be expressed in terms of the model parameters as
\begin{align}
\nonumber \delta_t \approx -6 \times 10^{-8} \left( \frac{j_0}{0.01} \right)^{-2} \left( \frac{r_a}{10^2 \,\au} \right)^{3/2} \left( \frac{\sigma_v}{200 \,\kms} \right)^{-3/2} \left( \frac{X_\mathrm{halo}}{200} \right)^{3/2} \\
\qquad \times \left( \frac{\tau}{13\,\gyr} \right)^{3/2} \left( \frac{m_\mathrm{PBH}}{20 \,\msun} \right)^{-3/2}\fPBH^{3/2}.
\label{eqn:deltaTestimate}
\end{align}
We can therefore expect that, for the fiducial model, the effect of fly-bys on the encounter rate is negligible. We note that this is only intended as an approximate number, and that many of the parameters are not independent. For example, we saw in section \ref{sect:formation} that the characteristic semi-major axis and eccentricity of the binary orbits are functions of the PBH abundance $\fPBH$ and mass $m_\mathrm{PBH}$.

We will further investigate the change in lifetime of binary systems by numerically evolving them over time, as well as considering variations from the fiducial model where a significant effect may be seen.

\section{Binary system evolution}
\label{sect:mc}
We simulate the evolution of a PBH binary after decoupling from the Hubble flow using a Monte Carlo approach. We take into account perturbations from passing PBHs in the secular approximation (see section~\ref{sect:sec}), and decay of the orbital energy and angular momentum due to GW emission. Our algorithm consists of the following steps.

\begin{itemize}
    \item We sample a next perturber given the encounter rate $\Gamma$ (computed from equation~\eqref{eqn:rate}), 
    i.e., the perturber will encounter the binary at a time delay $\Delta t$, where the probability that the time delay exceeds $\Delta t$ is given by $\exp(-\Gamma \Delta t)$. The impact parameter $b$ is sampled from a distribution $\mathrm{d}N/\mathrm{d}b\propto b$ with $0<b<\renc$. The perturber's mass $\mper$ is sampled from a lognormal mass function, described in equation \eqref{eqn:logNormalMassFn}
    
    \item In order for a finite number of encounters to be considered, it is necessary to define a maximum encounter radius $\renc$, above which we neglect the effect of fly-bys. Since we are considering the effect of an encounter between a binary and a single PBH, we set $\renc$ to a radius around the binary where there is a low probability of finding multiple PBHs: $\renc = 0.1 \left( 4\pi n/3 \right)^{-1/3}$, where $n$ is the number density of PBHs in a DM halo. We do not expect a cut-off in the impact parameter to have a significant effect, as fly-bys with a smaller impact parameter have a larger cumulative effect (see section \ref{sect:estimate}), and we have verified that increasing/decreasing the factor 0.1 by an order of magnitude does not affect the results.
    
    \item We apply the effects of the perturber's passage on the binary given $\mper$ and $b$ using the analytic expressions for $\Delta e$ from \citep{2019MNRAS.487.5630H,2019MNRAS.488.5192H}. Here, we assume that the binary's orbital orientation is random (i.e., flat distributions in $\cos i$, $\omega$, and $\Omega$). Evidently, the binary's orientation actually remains fixed between encounters whereas the perturbers plausibly approach from random orientations. However, for the actual computation of $\Delta e$, only the relative orientation matters, so this distinction is unimportant as long as the orientation of the perturbers is isotropic, which is what we assume. 
    \item In-between the current time and the time of the next perturber, we take into account the decay of the orbit due to GW emission by numerically solving the set of ordinary differential equations (ODEs) from \citep{1964PhRv..136.1224P}, i.e.,
    \begin{subequations}
    \begin{align}
        \frac{\mathrm{d} r_a}{\mathrm{d} t} &= - \frac{64}{5} \frac{G^3 m_1 m_2 m}{c^5 r_a^3 \left(1-e^2\right)^{7/2}} \left (1 + \frac{73}{24} e^2 + \frac{37}{96} e^4 \right ); \\
        \frac{\mathrm{d} e}{\mathrm{d} t} &= - \frac{304}{15} e \frac{G^3 m_1 m_2 m}{c^5 r_a^4 \left(1-e^2\right)^{5/2}} \left (1 + \frac{121}{304} e^2 \right ).
    \end{align}
    \label{eqn:PetersEqns}
    \end{subequations}
    We integrate the above set of equations using the \textsc{CVODE} library \citep{1996ComPh..10..138C} in \textsc{C}, and continue until $r_\mathrm{p} = r_a (1-e) < 100 \, r_\mathrm{g}$, where $r_\mathrm{g} \equiv G m/c^2$ is the binary's gravitational radius. When this condition has been satisfied, we consider the binary to have merged. We note that, by the time of having reached $r_\mathrm{p}= \alpha \,r_\mathrm{g} = 100\,r_\mathrm{g}$, the binary has mostly circularised; the remaining merger time is \citep{1964PhRv..136.1224P}
    \be
    T_\mathrm{c} = \frac{5}{256} \alpha^4 \frac{m^2}{m_1 m_2} \frac{Gm}{c^3} \simeq 0.4\,\mathrm{hr} \, \left ( \frac{\alpha}{100} \right )^4 \left ( \frac{m}{40\,\msun} \right )^3 \left ( \frac{m_1}{20 \, \msun} \right )^{-1} \left ( \frac{m_2}{20 \, \msun} \right )^{-1}.
    \ee
\end{itemize}

\section{Results}
\label{sect:results}

In this section, we will discuss the results of the simulations of PBH binary evolution. Firstly, we will investigate the merger time calculated by a numeric integration of the equations governing the binary evolution, equation \eqref{eqn:PetersEqns}, in the absence of encounters which can perturb the system, and then including the effect of perturbers.

\subsection{Numerical evolution of binary primordial black holes}

In the limit of high eccentricity, $e\rightarrow 1$, an analytic expression for the coalescence time $\tau_\mathrm{A}$ for binary systems is given by \cite{1964PhRv..136.1224P}
\be
\tau_\mathrm{A} = \frac{3}{85}\frac{c^5}{G^3}\frac{r_0^4(1-e_0^2)^{7/2}}{m_1 m_2 (m_1+m_2)}.
\label{eqn:approxTime}
\ee
Since binary PBHs themselves are typically highly eccentric, this equation has typically been used to calculate the coalescence time of binary PBH systems (i.e. \cite{Ali-Haimoud:2017rtz,Raidal:2018bbj}), and is a key component in predicting the merger rate observable today.

In this section, we test the accuracy of this approximation by numerically integrating equation \eqref{eqn:PetersEqns}, as described in section \ref{sect:mc}, for PBH binaries with random initial conditions to give a numeric value for the coalescence time $\tau_\mathrm{N}$, and comparing this to the time $\tau_\mathrm{A}$ calculated from equation \eqref{eqn:approxTime}. We then consider the effect this might have on the observable merger rate today, finding that, whilst equation \eqref{eqn:approxTime} can be inaccurate to $\mathcal{O}(10\%)$, this is likely to have a negligible effect on the merger rate.

Our initial PBH binaries are randomly generated using the methods described in section \ref{sect:formation}. The relevant parameters which will affect the distribution of initial conditions are: the peak mass of the PBH mass function $m_\mathrm{c}$, the width of the mass function $\sigma_m$, the size of the exclusion zone required for a binary to form $A$, and the total abundance of PBHs, parameterized by $\fPBH$. The great majority of binary systems merge early during the history of the universe, $\tau<1\,\gyr$. Since such systems are not relevant for the observation of mergers happening today, we limit our selection of binary PBHs to those with a lifetime greater than $1\,\gyr$. 

If PBHs make up the entirety of DM, $\fPBH=1$, we find that the analytic expressions only matches the numeric results to $\mathcal{O}(10\%)$, but matches much more closely, to within $\mathcal{O}(1\%)$ when PBHs are less abundant, $\fPBH\approx 0.01$. This is likely due to the fact that, whilst all long-lived PBH binaries are highly eccentric, they are significantly more eccentric for low $\fPBH$ (see section \ref{sect:formation} for more discussion). We find that changing $m_\mathrm{c}$, $\sigma_m$ and $A$ does have an effect, this effect is subdominant to the effect of changing $\fPBH$. We therefore fix $\mu=20\,\msun$, $\sigma_m=0.5$ and $A=2$ for the remainder of the discussion in this section. 

\begin{figure*}[t!]
%\vspace{-1.5cm}
 \centering
  \includegraphics[width=0.49\textwidth]{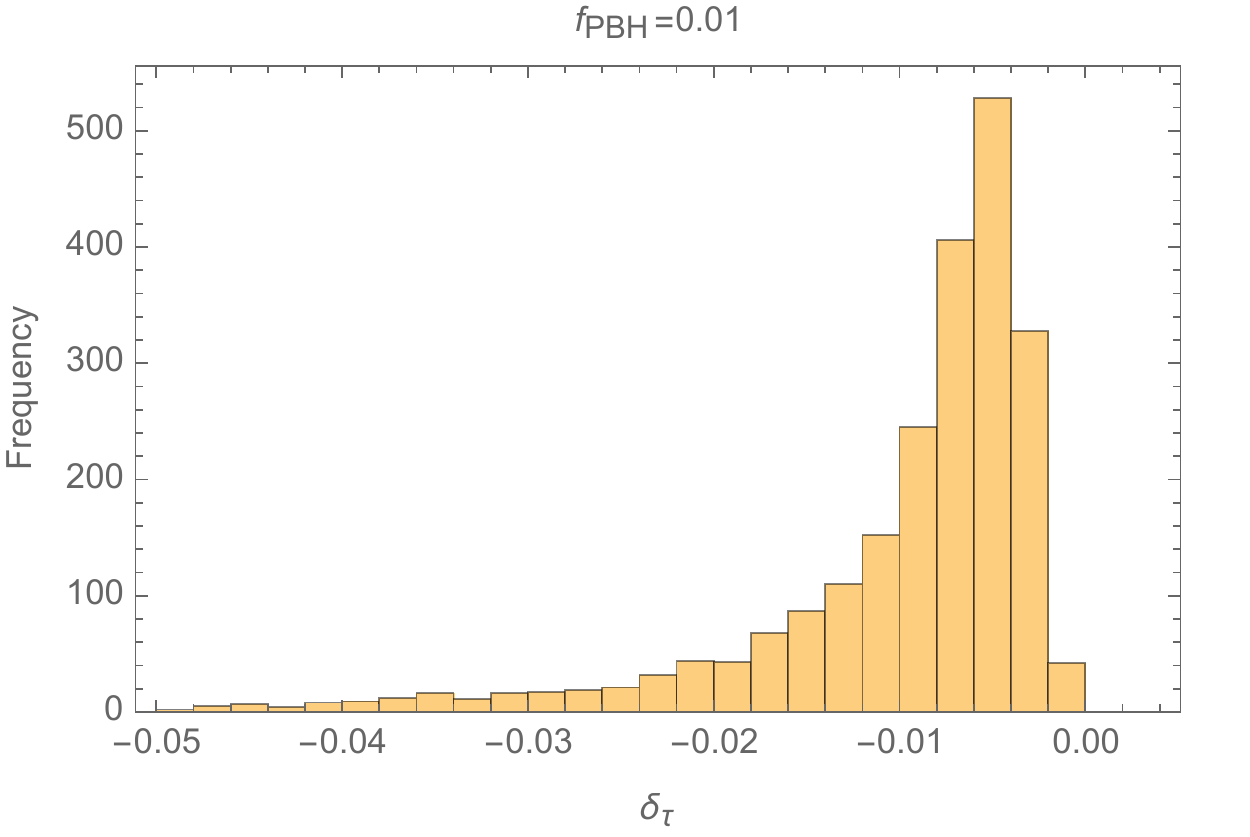} 
  \includegraphics[width=0.49\textwidth]{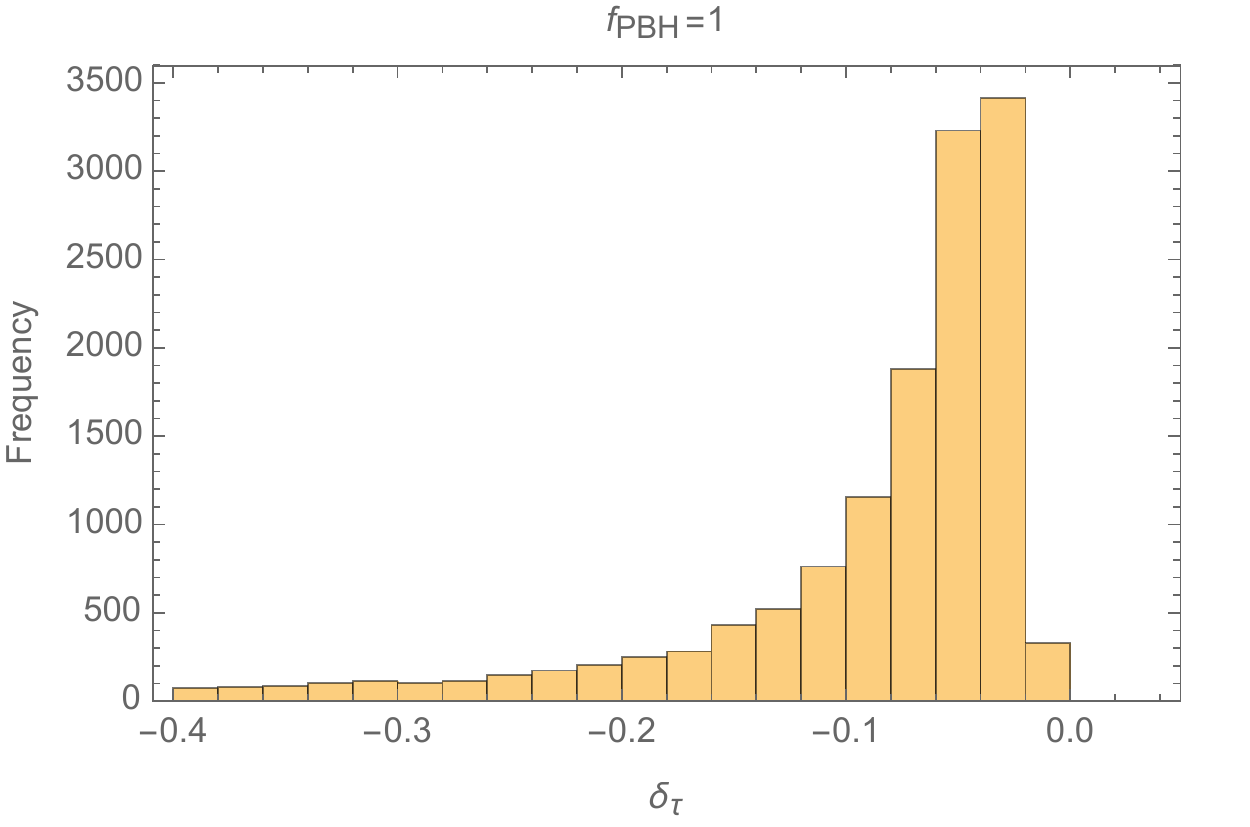} 
 % \vspace{-2.0cm}
  \caption{ We show the fractional error in the coalescence time for binary systems with an expected lifetime between 1 and $20\,\gyr$, for $\fPBH=0.01$ and $1$. For small $\fPBH$, the analytic expression is accurate to order $1\%$. For large PBH abundance, $\fPBH=1$, the analytic expression significantly overestimate the coalescence time, by order $10\%$, owing to the lower typical binary eccentricity in such cases. } 
  \label{fig:analyticVSnumeric}
\end{figure*}

We define the quantity $\delta_\tau$ as the relative change in $\tau$
\be
\delta_\tau = \frac{\tau_\mathrm{N}-\tau_\mathrm{A}}{\tau_\mathrm{A}}.
\ee
Figure \ref{fig:analyticVSnumeric} shows a histogram of $\delta_\tau$, for 100 000 different random binary PBHs with an expected lifetime $\tau_\mathrm{A}>1\,\gyr$, for $\fPBH = 0.01, 1$ (only PBHs with a coalescence time between 1 and $20\,\gyr$ are plotted). We can see that equation \eqref{eqn:approxTime} overpredicts the coalescence time, finding $\langle \delta_\tau \rangle = -0.097$ for $\fPBH=1$, and $\langle \delta_\tau \rangle = -0.015$ for $\fPBH=0.01$.

At first glance, this may be expected to have a significant effect on the merger rate observable today, since the merger rate is expected to decrease over time. To estimate how much the merger rate is affected, we will compare the number of binaries merging over time when the coalescence times is calculated numerically or analytically. To achieve this, we use 100 000 initial binaries (with a lifetime greater than 1 Gyr) and calculate their coalescence time, and sort the coalescence times into 1 Gyr bins.

\begin{figure*}[t!]

 \centering
  \includegraphics[width=0.49\textwidth]{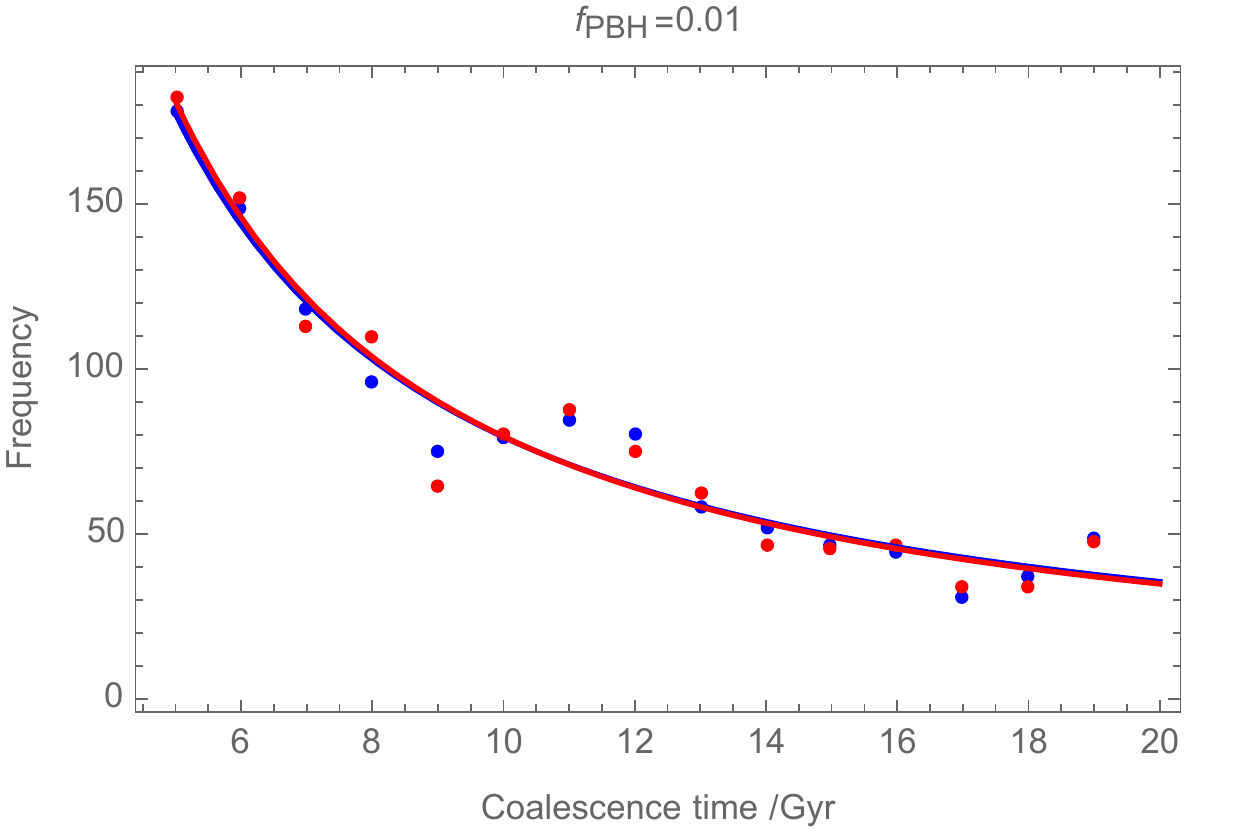} 
  \includegraphics[width=0.49\textwidth]{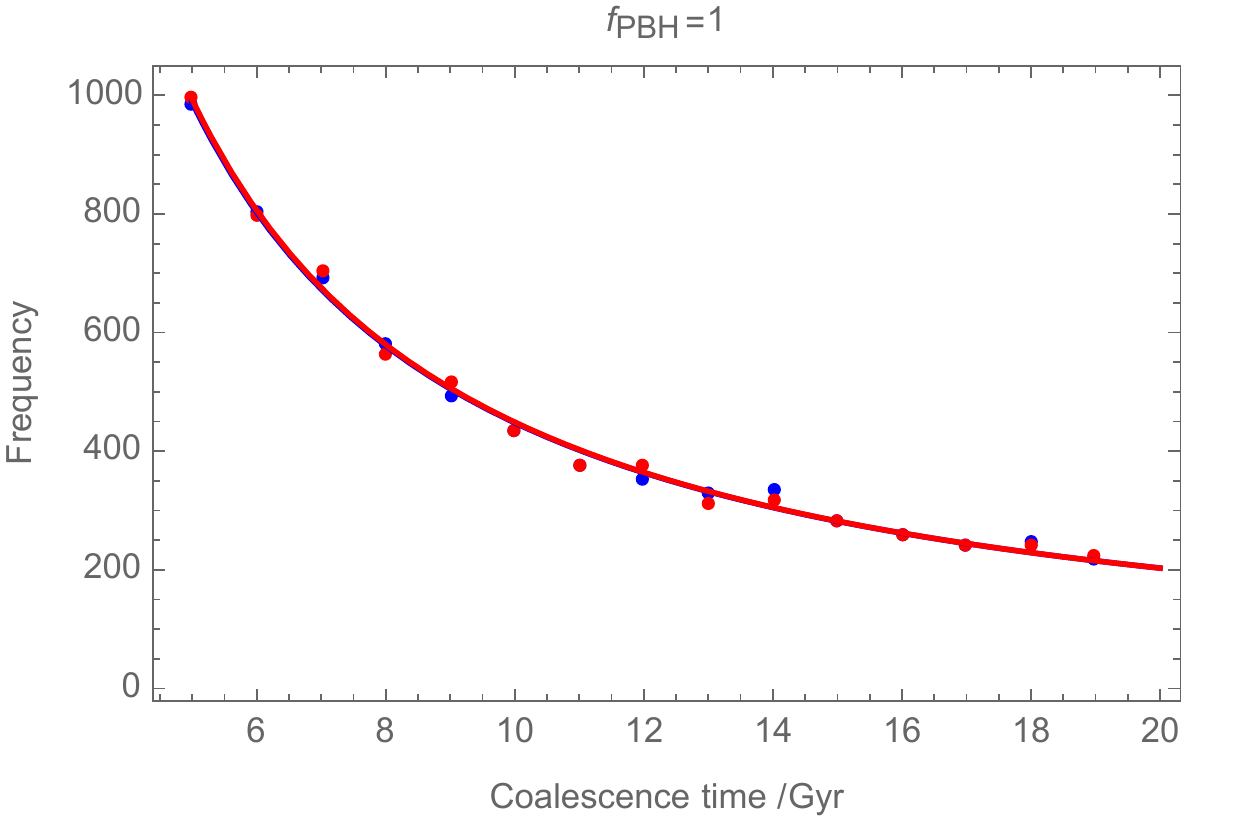} 

  \caption{ We show the number of binaries (out of 100 000 initial binaries) merging in each $1\,\gyr$ bin, for $\fPBH=0.01$ and 1. The blue points show the coalescence times as calculated with equation \eqref{eqn:approxTime}, and the red points show coalescence times calculated numerically. The blue/red lines show a power-law line of best fit. Even though there can be significant error in the coalescence times arising from the use of equation \eqref{eqn:approxTime}, this has little to no effect on the number of PBHs merging at any given time. Note that the larger scatter in the left-hand plot is due to the smaller number of PBHs merging in the selection for the $\fPBH=0.01$ model.} 
  \label{fig:mergeRate}
\end{figure*}

Figure \ref{fig:mergeRate} shows the number of binaries calculated merging in each 1 Gyr bin, for 100 000 initial binaries each for $\fPBH=0.01,1$. The red (blue) points show the number of binaries merging in each 1 Gyr bin, calculated from equation \eqref{eqn:approxTime} (calculated numerically). \citet{Raidal:2018bbj} predicts that the merger rate $R$ follows a power-law with respect to time $t$, $R\propto t^{-34/37}$, so we fit a power-law to the data points, shown with a red (blue) lines. %Using randomly generated binaries to calculate the merger rate dependance on time, we find a marginally stronger dependence than found in \citet{Raidal:2018bbj}, $R \propto \frac{1}{t^{\sim 1.1}}$ - although this depends on how the binaries are b.

 We see only a small deviation in the number of PBHs merging at any given time --- which is consistent with the random errors caused by the finite sample size. We therefore conclude that, whilst the analytic expression for the merger time can have a significant error (especially for individual binaries), this has a negligible impact on the merger rate predicted --- and may therefore be safely ignored.

Finally, we turn our attention to the amplitude of the numerical error made when integrating \eqref{eqn:PetersEqns}. Numerical errors made in the ODE integration can lead to errors in the coalescence time compared to the true solution. It is necessary to quantify these errors when considering the effects of fly-bys as well, in order to be able to properly distinguish the physical effects of fly-bys from unphysical numerical noise. We find that the largest numerical error made in our integrations gives rise to $\delta_\tau$ of order $10^{-8}$, which, as we will see, is safely several orders of magnitude smaller than the observed signal due to fly-bys.

\subsection{The effect of fly-bys}

In this section we will discuss the effect of distant encounters between binary PBH systems and individual PBHs --- referred to as fly-bys. We begin with a random sample of PBH binaries, before submitting them to random encounters with other PBHs. To make the task manageable, we make the simple assumption that such binary PBHs are part of a DM halo (and/or PBH cluster) from shortly after the time of formation until today --- and neglect any time evolution of such haloes/clusters. A more complete investigation of the evolution of binary systems from the early universe, through structure formation, up until the present epoch will require the use of $N$-body simulations with PBH DM.

There are a large number of parameters which can affect the typical coalescence times of binaries, and so we will investigate a fiducial model with fixed values for these parameters, before discussing the effect of individual parameters. For the fiducial model, we take the following values for the required parameters:
\begin{itemize}
    \item $m_\mathrm{c} =20\,\msun$, $\sigma_m = 0.5$, close to the values given by \citet{Raidal:2018bbj} as a best-fit model to explain the black holes coalescence events observed by LIGO.
    \item $\fPBH=1$ (although we note that this is expected to produce a larger frequency of coalescence events than are observed).
    \item $A=2$, throughout we will make the simple assumption that if two PBHs form within a distance $x$ of each other, they will eventually form a binary if there are no other PBHs within a distance $2x$. Changing this by order unity has a similar order effect on the distribution of initial semi-major axes, and a negligible effect on our final results.
    \item $X_\mathrm{halo} = 200$, corresponding to the usual definition of DM halos. The great majority of DM in the universe is expected to be found inside halos with a density greater than 200 times the background density of the universe --- though we note that binary PBHs may be found in denser sub-haloes, and the effect of mass segregation may mean that the heavier binary systems migrate towards the denser cores of DM halos.
    \item $\sigma_v = 200\,\kms$, corresponding approximately to the velocity dispersion for a Milky Way-type halo calculated using virial theorem. We take a zeroth order estimate for $\sigma$, but the actual velocity dispersion within a DM halo is position dependant, and calculating the relative velocity of binaries with respect to nearby PBHs goes beyond the scope of this work.
\end{itemize}

We will compare the coalescence time calculated with and without encounters from the Monte Carlo simulations of binary PBHs. We will again use $\delta_\tau$ to represent the fractional change in the merger time for PBHs,
\be
\delta_\tau = \frac{\tau_\mathrm{MC}-\tau_\mathrm{N}}{\tau_\mathrm{N}},
\ee
where $\tau_\mathrm{MC}$ is the coalescence time given by the Monte Carlo simulations including fly-bys, and $\tau_\mathrm{N}$ is the coalescence time predicted from numerically solving equation \eqref{eqn:PetersEqns} from the initial conditions of the binary. 

\begin{figure*}[t!]
%\vspace{-1.5cm}
 \centering
  \includegraphics[width=0.6\textwidth]{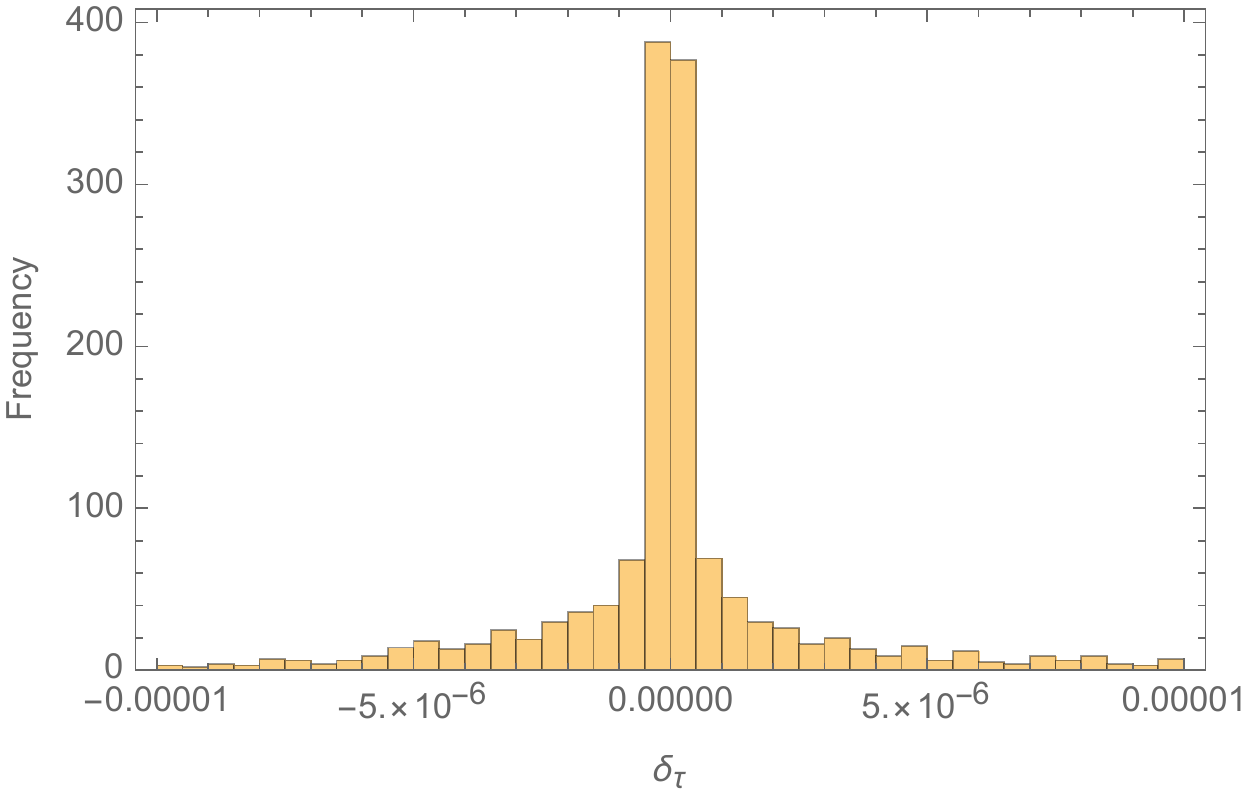} 
 % \vspace{-2.0cm}
  \caption{ The fractional change in coalescence times due to the effect of fly-bys, for the fiducial model described above.} 
  \label{fig:deltaThistogram}
\end{figure*}

Figure \ref{fig:deltaThistogram} shows a histogram for $\delta_\tau$ with a coalescence time between $10-15\,\gyr$, starting from 100 000 initial binaries\footnote{Instead including all binaries which merge between 1 and 20 Gyr has a small effect, and would give  $\sigma_\tau = 1.72\times 10^{-4}$.}. The mean is $\langle \delta_\tau \rangle = -3.26\times 10^{-6}$, which is negligible compared to the variance of the distribution. It can be seen that most binaries experience a change in coalescence time of order $10^{-6}$. However, the standard deviation is orders of magnitude larger, $\sigma_\tau = 2.31\times 10^{-4}$.

As may have been expected from the divergence in the integral in equation \eqref{eqn:sigmaDeltae}, the standard deviation is dominated by outliers in the distribution, a small number of binaries experience a change in merger time orders of magnitude larger than typical (up to $\delta_\tau=\mathcal{O}(0.01)$ for the binaries considered). Thus, in order to describe the typical effect of fly-bys with a more representative number, and to investigate the effect of changing the model parameters, we will neglect the tails of the distribution, retaining only the central $90\%$ of the data\footnote{Another approach is to simply run more initial binaries, but in this case, $\sigma_{\tau}$ instead becomes dominated by the cut-off for strong encounters.}. In this case, we find a significantly smaller standard deviation $\sigma^*_{\tau} = 4.64\times 10^{-6}$ (and the mean is $\langle \delta^*_{\tau} \rangle = -5.81\times 10^{-8}$).

\subsubsection{Changing the model parameters}

\begin{figure*}[t!]
%\vspace{-1.5cm}
 \centering
  \includegraphics[width=0.49\textwidth]{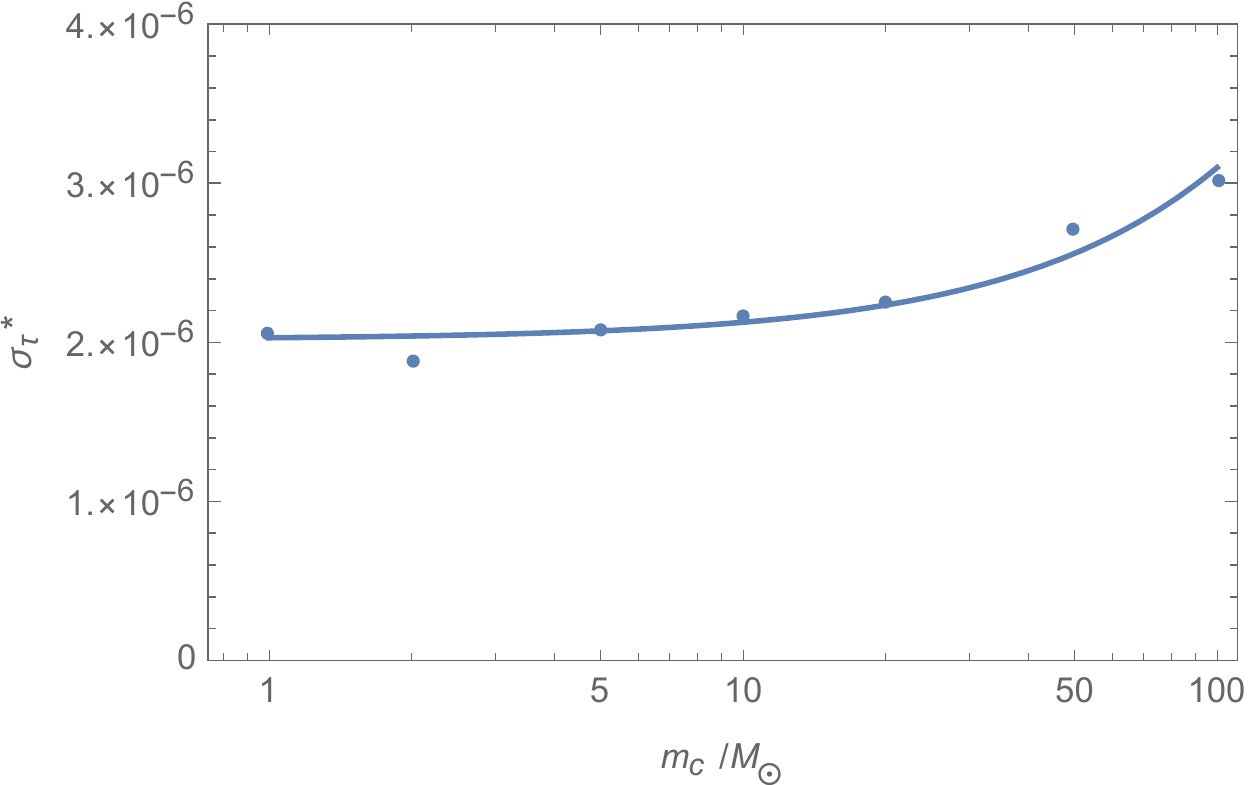} 
  \includegraphics[width=0.49\textwidth]{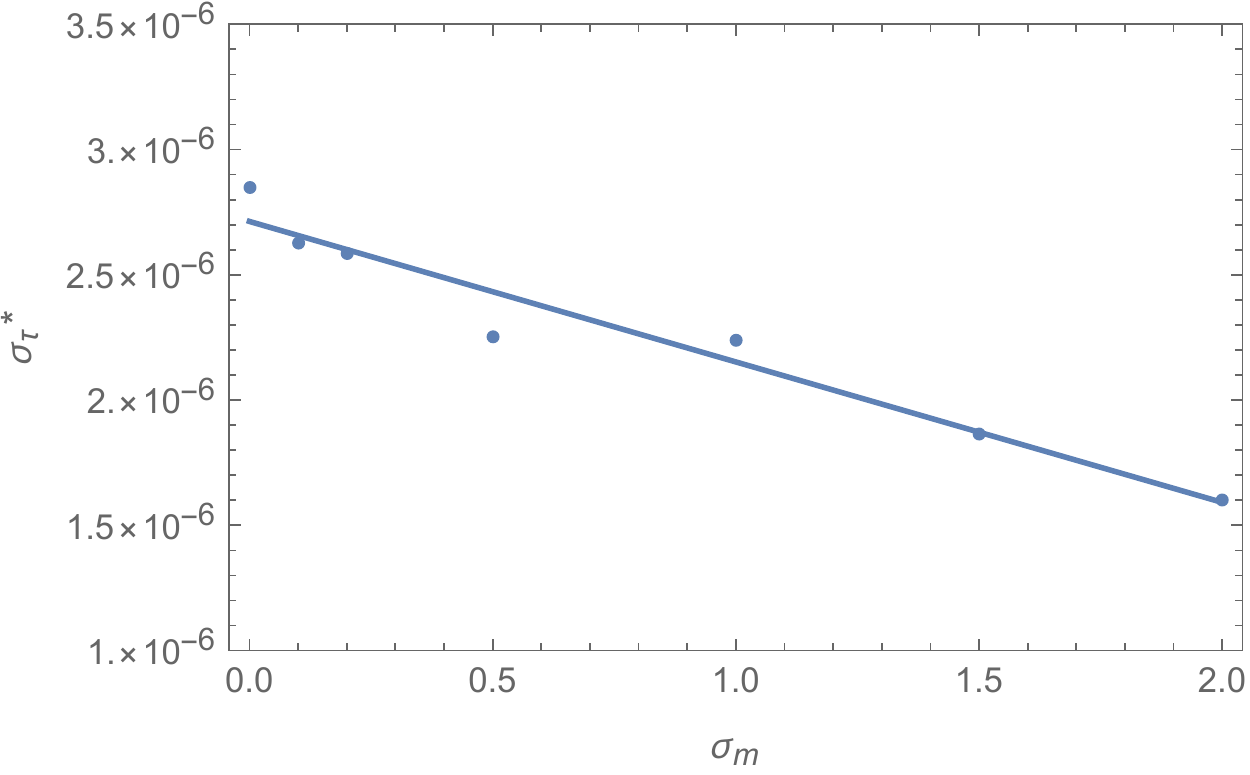}
  \includegraphics[width=0.49\textwidth]{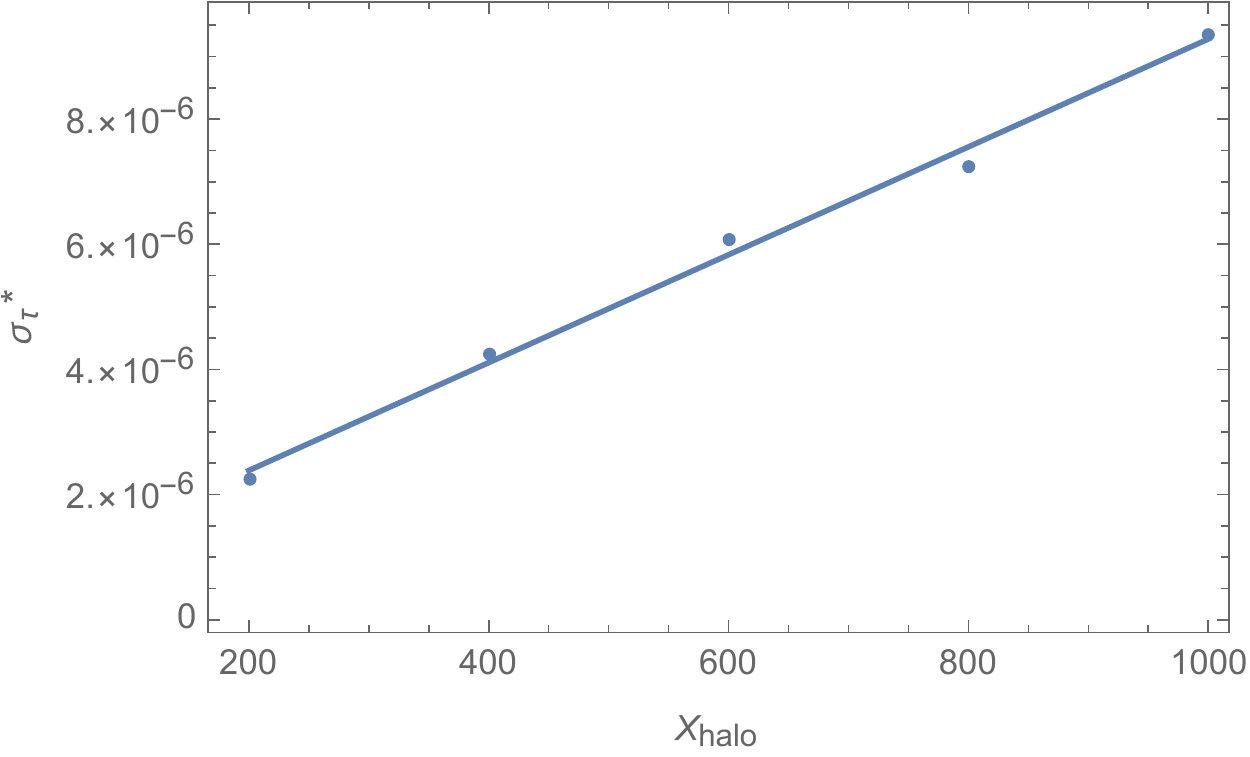} 
  \includegraphics[width=0.49\textwidth]{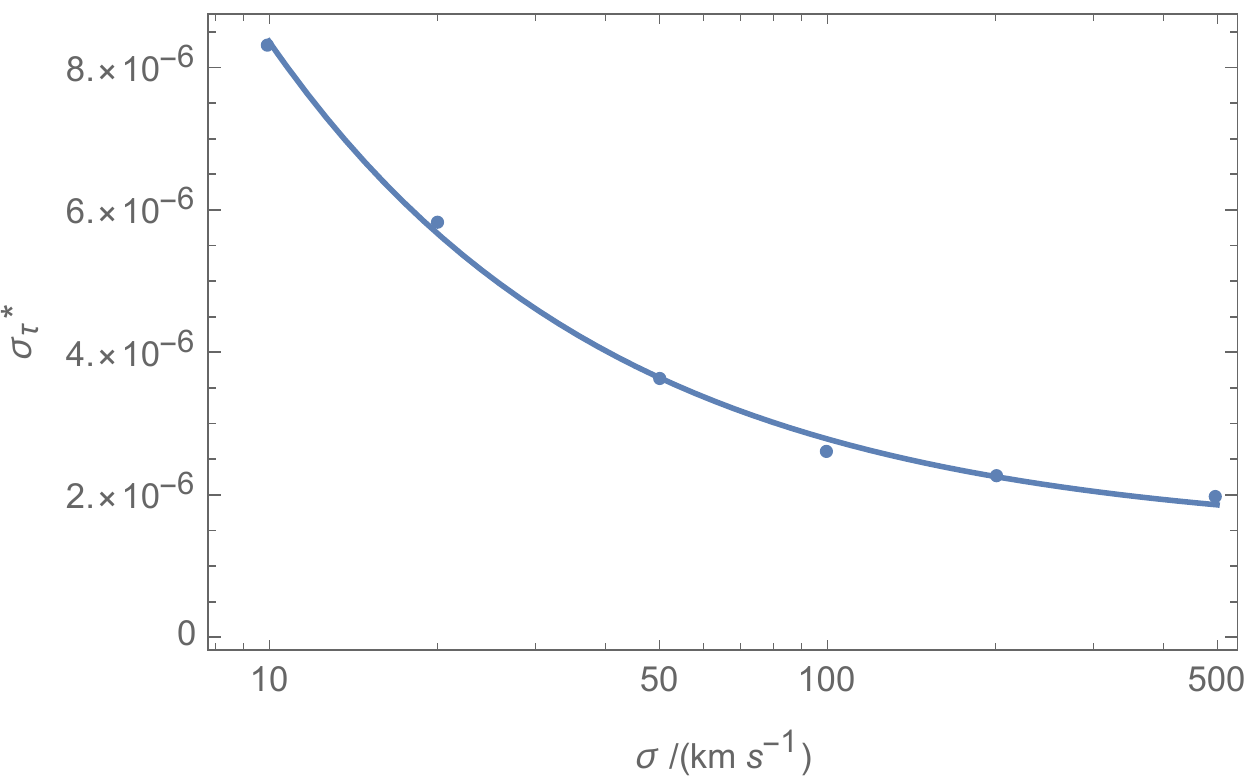} 
  \includegraphics[width=0.49\textwidth]{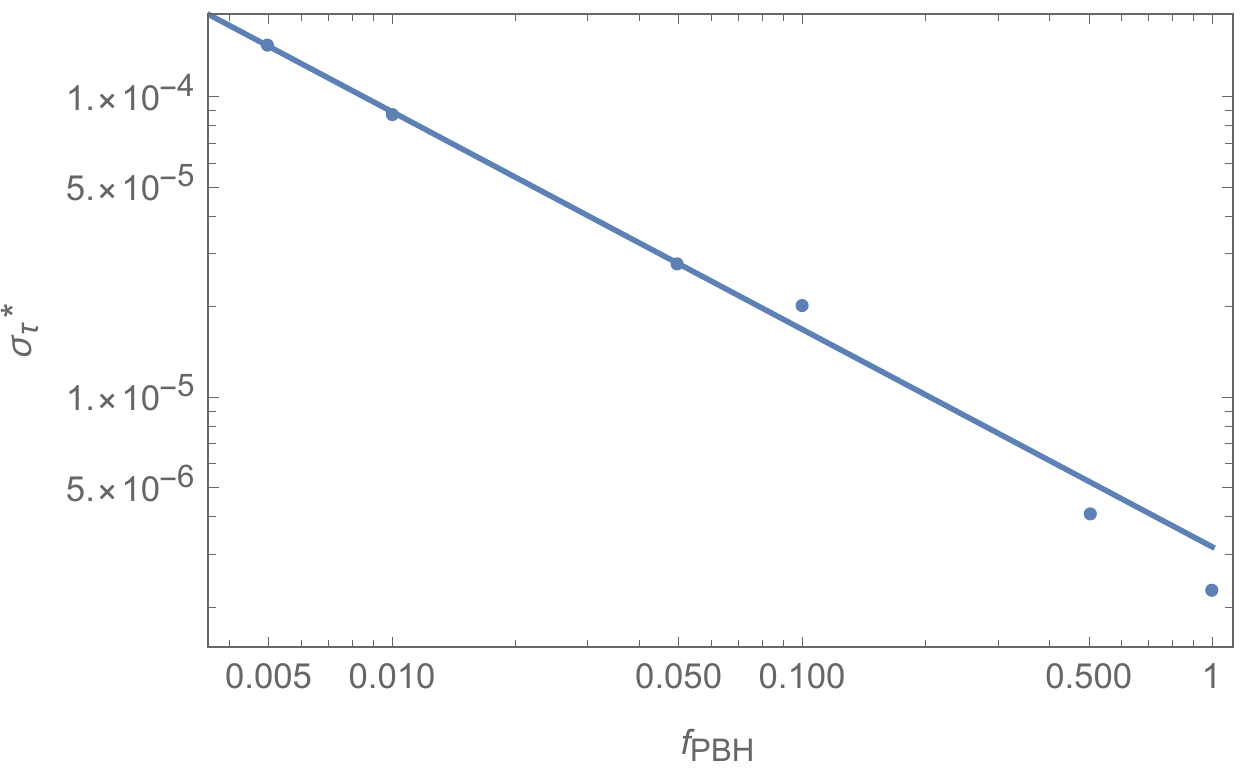} 
 % \vspace{-2.0cm}
  \caption{ The effect of changing the model parameters on the standard deviation of the change in coalescence time $\sigma^*_\tau$. The top row shows the effect of changing the mass function parameters, $m_\mathrm{c}$ and $\sigma_m$, which have a small effect. The second row shows how the halo parameters affect the binary coalescence times, with denser haloes (larger $x_\mathrm{halo}$) and lower velocity dispersion $\sigma$ resulting in fly-bys having a larger effect. Finally, we consider the PBH abundance $\fPBH$, seeing that fly-bys have a larger effect for smaller $\fPBH$, due to the increased characteristic semi-major axis of binaries in that case. For illustration purposes, we have plotted linear lines of best fit for $m_\mathrm{c}$, $\sigma_m$, and $X_\mathrm{halo}$, and power-law lines of best fit for $\sigma_v$ and $\fPBH$.} 
  \label{fig:effectOfParameters}
\end{figure*}

We now ask the question of how the coalescence time is affected for different parameter choices. Starting from 10 000 initial binaries in each case, figure \ref{fig:effectOfParameters} shows how $\sigma^*_\tau$ is affected by changing the model parameters. Here, to increase the sample size, we consider all binaries which merge within $1-20\,\gyr$ after formation, which has a negligible effect on the final values for $\sigma^*_\tau$. The top row shows that $\sigma^*_\tau$ only has a weak dependence on the mass function\footnote{We note that, due to the effects of critical collapse, the width of the mass function is not expected to be smaller than $\sigma_m\approx 0.1$ \cite{GCBY,Young:2019gfc}}. On the other hand, changing the parameters $\fPBH$, $X_\mathrm{halo}$ or $\sigma_v$ has a relatively strong effect.

Decreasing $\fPBH$ might be expected to result in fly-bys having a smaller effect, since there are less PBHs in a DM halo to interact with a given binary. However, as described in section \ref{sect:formation}, a smaller $\fPBH$ also implies that binary PBHs will form with significantly larger semi-major axes (and those merging by today also had a higher initial eccentricity) --- meaning that encounters have a larger effect on the orbital parameters. 

Changing $X_\mathrm{halo}$ is found to have a significant effect on $\sigma^*_\tau$. If we consider that PBH binaries might be found within denser DM haloes (and neglecting the fact that this is likely to change the velocity dispersion), this has the simple effect of increasing the number of encounters that occur --- without changing the nature of the encounters. We find the expected result, therefore, that $\sigma^*_\tau$ is approximately proportionate to $X_\mathrm{halo}$ --- the effect is not exactly proportionate due to the cut-off in the maximum encounter radius considered.

It might naively be expected that decreasing the velocity dispersion $\sigma_v$ will lead to a smaller effect on the coalescence time --- since this implies there will be fewer encounters. However, each fly-by takes a lot longer to occur, and thus has an overall larger impact as $\sigma_v$ decreases (see equation \eqref{eqn:deltaTestimate}) --- and we therefore see that decreasing $\sigma_v$ from $200\,\kms$ to $10\,\kms$ makes the effect of fly-bys significantly larger.

We now consider the largest value for $\sigma_\tau$ which might be obtained for reasonable choices of the model parameters. We have seen that changing the mass function parameters does not strongly affect our results, so we will keep $m_c=20\,\msun$ and $\sigma_m=0.5$ fixed. Constraints on PBH abundance arising from GW signals from merging PBHs are of order $10^{-3}$, so we will take $\fPBH=5\times 10^{-3}$. For the parameters related to DM haloes, we will consider that binaries might be more likely to be found in denser regions (for example, due to the formation of PBH clusters \cite{Inman:2019wvr}), where not only is the density higher, but the velocity dispersion is likely to be lower \cite{Hoeft:2003ea}. We will therefore consider $X_\mathrm{halo}=1000$ and $\sigma_v=10\,\kms$. With these choices of parameters, we find $\sigma_\tau = 8.60\times 10^{-2}$ (and $\sigma^*_\tau = 2.62\times 10^{-4}$).

\subsubsection{Non-secular encounters in the low $\fPBH$ regime}

As $\fPBH$ becomes small, we find that fly-bys have a larger effect on the coalescence time. This is due to the fact that, while typical semi-major axes of binaries and impact parameters of encounters both increase, the semi-major axes of binaries typically grow by more --- resulting in fly-bys having a larger effect. 

This also means that the chance of non-secular encounters ('strong encounters') increases, i.e. encounters with an adiabatic ratio $\mathcal{R}\gtrsim 1$. Our current formalism is not capable of dealing with such ``strong encounters'', and when a strong encounter does occur in the Monte Carlo simulation, this is recorded, and evolution of that binary is halted. For $\fPBH=\mathcal{O}(1)$, the number of binaries experiencing a strong encounter is negligible. However, for $\fPBH=\mathcal{O}(10^{-3})$, we find that $\sim 10\%$ of binaries experience a strong encounter --- which is actually a larger fraction than the number of binaries merging between $10-15\,\gyr$.

\begin{figure*}[t!]
%\vspace{-1.5cm}
 \centering
  \includegraphics[width=0.6\textwidth]{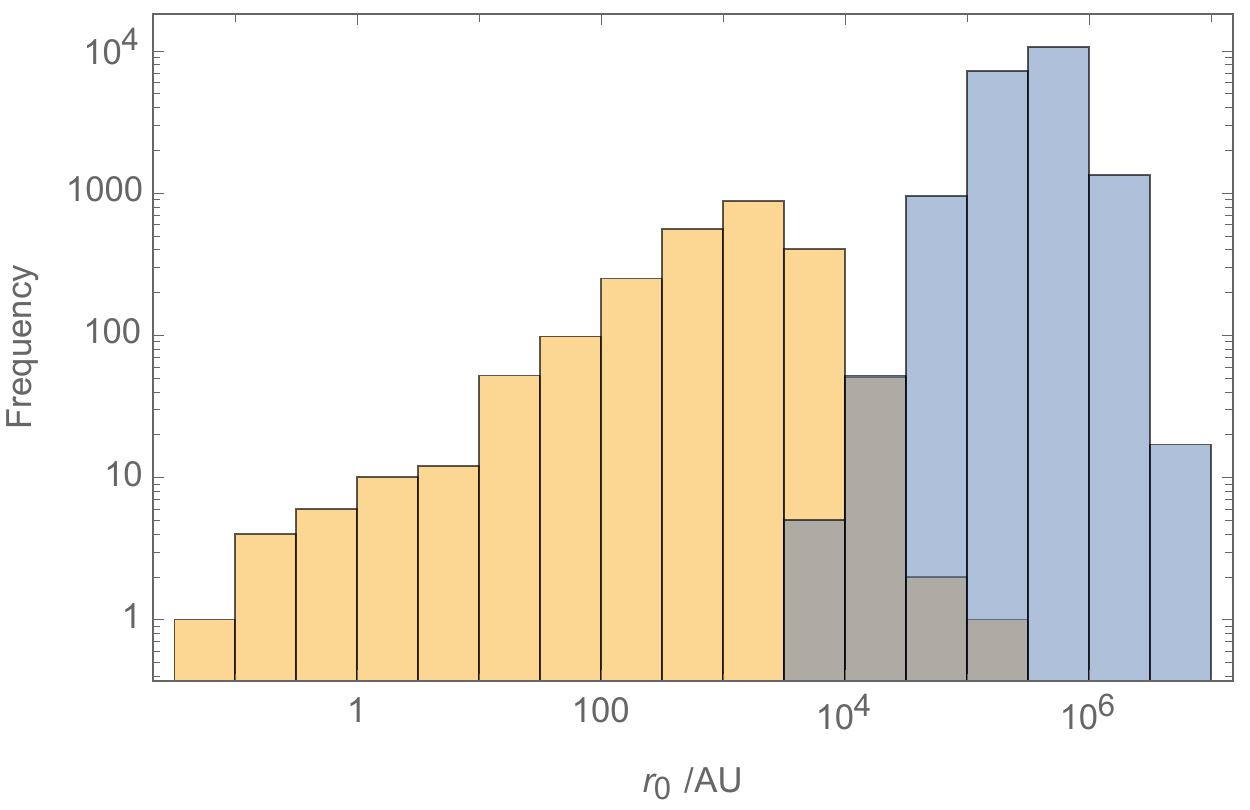} 
  %\includegraphics[width=0.49\textwidth]{semiMajorMergedBinaries.pdf} 
  %\includegraphics[width=0.49\textwidth]{semiMajorStrongEncounters.pdf} 
 % \vspace{-2.0cm}
  \caption{ The frequency of binaries with initial semi-major axis $r_0$ of binaries which coalesce (orange) and those which experience a strong encounter (blue), during the time of the simulation. We see that PBHs which merge during the lifetime of the simulation typically have a much smaller $r_0$ than those which experience a strong encounter, with a small overlap.} 
  \label{fig:semiMajorComparison}
\end{figure*}

However, the semi-major axes of binaries experiencing a strong encounter are orders of magnitude higher than the semi-major axes of binaries which merge within the simulation time. Figure \ref{fig:semiMajorComparison} shows a comparison of the semi-major axes of binaries which merge in the simulation, compared to those which experience a strong encounter (for the parameter choices given at the end of the previous subsection). For binaries which are expected to merge in the current lifetime of the universe, the semi-major axes are of order $10^3\,\au$, whilst for those experience strong encounters, the semi-major axes are of order $10^6\,\au$. 

Neglecting such binaries is therefore unlikely to affect the result and conclusions presented in this paper --- although it is conceivable that strong encounters may drive binary systems to merge much earlier, thereby increasing the merger rate observed today. We also note that such wide binaries may have been disrupted in the early universe shortly after formation. We leave further consideration of such binaries for future work.

\section{Discussion}
\label{sect:discuss}

We have considered PBH binaries which form in the early universe, shortly after the formation of the PBHs themselves. Utilising a Monte Carlo approach, the evolution of a large sample of initial binaries are numerically evolved forwards through time in order to determine their coalescence time. 

We account for the impact of encounters of fly-bys in DM haloes (whilst binary-binary encounters are rarer and have a negligible effect in the secular limit compared to binary-single encounters \citep{2020MNRAS.494..850H}). In section \ref{sect:estimate} we also developed an analytic estimate for the change in coalescence time due to the effect of fly-bys, and in later sections we investigated the effect with a Monte Carlo approach. We typically find that binary-single encounters have a small effect on the coalescence, changing the lifetime by order $10^{-6}$, although a small number of binaries experience a much larger effect. Considering a more extreme model, we find that the typical binary lifetime is unlikely to change by more than $10^{-2}$. 

We note that we have neglected the formation of dense DM haloes (``spikes'') around PBHs in models where PBHs only constitute a small fraction of dark matter (i.e.  \cite{Boucenna:2017ghj,Adamek:2019gns}). In addition to increasing the effective mass of PBHs, these DM spikes will also affect the in-spiral of binary PBHs and their gravitational waveforms \cite{Kavanagh:2020cfn}. A PBH (with a DM spike) flying by a binary system is far enough away to be considered as a point mass, and since the important factor is the ratio of perturber mass to binary masses, the change in mass due to the DM spikes is unlikely to have a signficant effect on the outcome of the perturber. Neglecting the DM spikes surrounding PBHs is therefore expected to have a negligible effect on our conclusions.

We also numerically solving the evolution of binary orbits to an analytic approximation used in previous studies, equation \eqref{eqn:approxTime}. By using a numeric method we obtain a more accurate value for the coalescence time than using the analytic expression. Whilst the analytic expression generally overestimates the coalescence time for individual systems by around $10\%$, we find that it has a negligible impact on the merger rate of binaries, and so may safely be neglected.

We therefore conclude that neglecting the impact of binary-single encounters after formation of a binary is unlikely to result in significant error to the coalescence time, placing the constraints on PBH abundance arising from the detected GW signals from merging PBHs on a more secure footing. However, we note that further work may be necessary before this can be stated with certainty --- our results show a need to study the evolution of PBH binaries in the early universe, during the structure formation, and the dynamics of PBHs within DM halos.

A recent paper by \citet{Jedamzik:2020ypm} performed a similar analysis to the one presented here, analysing the effect of interactions between a binary PBH and a third by-passing PBH. The results presented there are complimentary to those which we present here, and concern with the evolution of binaries in extremely dense clusters (which are denser by many orders of magnitude than the DM haloes considered here), which form at high redshifts but evaporate by lower redshifts. The conclusion reached in that paper is that many binaries in such clusters are disrupted, and that the merger rate observed today is consistent with PBHs composing the entirety of dark matter.

\section*{Acknowledgements}

SY is supported by a Humboldt Research Fellowship. We would like to thank Eiichiro Komatsu, Fabian Schmidt, Thorsten Naab, Antti Rantala and Wilma Trick for helpful discussion related to this paper.

\bibliography{bibfile}

%merlin.mbs apsrev4-1.bst 2010-07-25 4.21a (PWD, AO, DPC) hacked
%Control: key (0)
%Control: author (8) initials jnrlst
%Control: editor formatted (1) identically to author
%Control: production of article title (-1) disabled
%Control: page (0) single
%Control: year (1) truncated
%Control: production of eprint (0) enabled
\begin{thebibliography}{133}%
\makeatletter
\providecommand \@ifxundefined [1]{%
 \@ifx{#1\undefined}
}%
\providecommand \@ifnum [1]{%
 \ifnum #1\expandafter \@firstoftwo
 \else \expandafter \@secondoftwo
 \fi
}%
\providecommand \@ifx [1]{%
 \ifx #1\expandafter \@firstoftwo
 \else \expandafter \@secondoftwo
 \fi
}%
\providecommand \natexlab [1]{#1}%
\providecommand \enquote  [1]{``#1''}%
\providecommand \bibnamefont  [1]{#1}%
\providecommand \bibfnamefont [1]{#1}%
\providecommand \citenamefont [1]{#1}%
\providecommand \href@noop [0]{\@secondoftwo}%
\providecommand \href [0]{\begingroup \@sanitize@url \@href}%
\providecommand \@href[1]{\@@startlink{#1}\@@href}%
\providecommand \@@href[1]{\endgroup#1\@@endlink}%
\providecommand \@sanitize@url [0]{\catcode `\\12\catcode `\$12\catcode
  `\&12\catcode `\#12\catcode `\^12\catcode `\_12\catcode `\%12\relax}%
\providecommand \@@startlink[1]{}%
\providecommand \@@endlink[0]{}%
\providecommand \url  [0]{\begingroup\@sanitize@url \@url }%
\providecommand \@url [1]{\endgroup\@href {#1}{\urlprefix }}%
\providecommand \urlprefix  [0]{URL }%
\providecommand \Eprint [0]{\href }%
\providecommand \doibase [0]{http://dx.doi.org/}%
\providecommand \selectlanguage [0]{\@gobble}%
\providecommand \bibinfo  [0]{\@secondoftwo}%
\providecommand \bibfield  [0]{\@secondoftwo}%
\providecommand \translation [1]{[#1]}%
\providecommand \BibitemOpen [0]{}%
\providecommand \bibitemStop [0]{}%
\providecommand \bibitemNoStop [0]{.\EOS\space}%
\providecommand \EOS [0]{\spacefactor3000\relax}%
\providecommand \BibitemShut  [1]{\csname bibitem#1\endcsname}%
\let\auto@bib@innerbib\@empty
%</preamble>
\bibitem [{\citenamefont {Abbott}\ \emph
  {et~al.}(2016{\natexlab{a}})\citenamefont {Abbott} \emph
  {et~al.}}]{Abbott:2016blz}%
  \BibitemOpen
  \bibfield  {author} {\bibinfo {author} {\bibfnamefont {B.}~\bibnamefont
  {Abbott}} \emph {et~al.} (\bibinfo {collaboration} {LIGO Scientific,
  Virgo}),\ }\href {\doibase 10.1103/PhysRevLett.116.061102} {\bibfield
  {journal} {\bibinfo  {journal} {Phys. Rev. Lett.}\ }\textbf {\bibinfo
  {volume} {116}},\ \bibinfo {pages} {061102} (\bibinfo {year}
  {2016}{\natexlab{a}})},\ \Eprint {http://arxiv.org/abs/1602.03837}
  {arXiv:1602.03837 [gr-qc]} \BibitemShut {NoStop}%
\bibitem [{\citenamefont {Abbott}\ \emph
  {et~al.}(2016{\natexlab{b}})\citenamefont {Abbott} \emph
  {et~al.}}]{Abbott:2016nmj}%
  \BibitemOpen
  \bibfield  {author} {\bibinfo {author} {\bibfnamefont {B.~P.}\ \bibnamefont
  {Abbott}} \emph {et~al.} (\bibinfo {collaboration} {LIGO Scientific,
  Virgo}),\ }\href {\doibase 10.1103/PhysRevLett.116.241103} {\bibfield
  {journal} {\bibinfo  {journal} {Phys. Rev. Lett.}\ }\textbf {\bibinfo
  {volume} {116}},\ \bibinfo {pages} {241103} (\bibinfo {year}
  {2016}{\natexlab{b}})},\ \Eprint {http://arxiv.org/abs/1606.04855}
  {arXiv:1606.04855 [gr-qc]} \BibitemShut {NoStop}%
\bibitem [{\citenamefont {Abbott}\ \emph
  {et~al.}(2017{\natexlab{a}})\citenamefont {Abbott} \emph
  {et~al.}}]{Abbott:2017vtc}%
  \BibitemOpen
  \bibfield  {author} {\bibinfo {author} {\bibfnamefont {B.~P.}\ \bibnamefont
  {Abbott}} \emph {et~al.} (\bibinfo {collaboration} {LIGO Scientific,
  VIRGO}),\ }\href {\doibase 10.1103/PhysRevLett.118.221101} {\bibfield
  {journal} {\bibinfo  {journal} {Phys. Rev. Lett.}\ }\textbf {\bibinfo
  {volume} {118}},\ \bibinfo {pages} {221101} (\bibinfo {year}
  {2017}{\natexlab{a}})},\ \bibinfo {note} {[Erratum: Phys.Rev.Lett. 121,
  129901 (2018)]},\ \Eprint {http://arxiv.org/abs/1706.01812} {arXiv:1706.01812
  [gr-qc]} \BibitemShut {NoStop}%
\bibitem [{\citenamefont {Abbott}\ \emph
  {et~al.}(2017{\natexlab{b}})\citenamefont {Abbott} \emph
  {et~al.}}]{Abbott:2017gyy}%
  \BibitemOpen
  \bibfield  {author} {\bibinfo {author} {\bibfnamefont {B.~P.}\ \bibnamefont
  {Abbott}} \emph {et~al.} (\bibinfo {collaboration} {LIGO Scientific,
  Virgo}),\ }\href {\doibase 10.3847/2041-8213/aa9f0c} {\bibfield  {journal}
  {\bibinfo  {journal} {Astrophys. J.}\ }\textbf {\bibinfo {volume} {851}},\
  \bibinfo {pages} {L35} (\bibinfo {year} {2017}{\natexlab{b}})},\ \Eprint
  {http://arxiv.org/abs/1711.05578} {arXiv:1711.05578 [astro-ph.HE]}
  \BibitemShut {NoStop}%
\bibitem [{\citenamefont {Abbott}\ \emph
  {et~al.}(2017{\natexlab{c}})\citenamefont {Abbott} \emph
  {et~al.}}]{Abbott:2017oio}%
  \BibitemOpen
  \bibfield  {author} {\bibinfo {author} {\bibfnamefont {B.}~\bibnamefont
  {Abbott}} \emph {et~al.} (\bibinfo {collaboration} {LIGO Scientific,
  Virgo}),\ }\href {\doibase 10.1103/PhysRevLett.119.141101} {\bibfield
  {journal} {\bibinfo  {journal} {Phys. Rev. Lett.}\ }\textbf {\bibinfo
  {volume} {119}},\ \bibinfo {pages} {141101} (\bibinfo {year}
  {2017}{\natexlab{c}})},\ \Eprint {http://arxiv.org/abs/1709.09660}
  {arXiv:1709.09660 [gr-qc]} \BibitemShut {NoStop}%
\bibitem [{\citenamefont {Abbott}\ \emph {et~al.}(2019)\citenamefont {Abbott}
  \emph {et~al.}}]{LIGOScientific:2018mvr}%
  \BibitemOpen
  \bibfield  {author} {\bibinfo {author} {\bibfnamefont {B.}~\bibnamefont
  {Abbott}} \emph {et~al.} (\bibinfo {collaboration} {LIGO Scientific,
  Virgo}),\ }\href {\doibase 10.1103/PhysRevX.9.031040} {\bibfield  {journal}
  {\bibinfo  {journal} {Phys. Rev. X}\ }\textbf {\bibinfo {volume} {9}},\
  \bibinfo {pages} {031040} (\bibinfo {year} {2019})},\ \Eprint
  {http://arxiv.org/abs/1811.12907} {arXiv:1811.12907 [astro-ph.HE]}
  \BibitemShut {NoStop}%
\bibitem [{\citenamefont {Abbott}\ \emph
  {et~al.}(2020{\natexlab{a}})\citenamefont {Abbott} \emph
  {et~al.}}]{Abbott:2020uma}%
  \BibitemOpen
  \bibfield  {author} {\bibinfo {author} {\bibfnamefont {B.}~\bibnamefont
  {Abbott}} \emph {et~al.} (\bibinfo {collaboration} {LIGO Scientific,
  Virgo}),\ }\href {\doibase 10.3847/2041-8213/ab75f5} {\bibfield  {journal}
  {\bibinfo  {journal} {Astrophys. J. Lett.}\ }\textbf {\bibinfo {volume}
  {892}},\ \bibinfo {pages} {L3} (\bibinfo {year} {2020}{\natexlab{a}})},\
  \Eprint {http://arxiv.org/abs/2001.01761} {arXiv:2001.01761 [astro-ph.HE]}
  \BibitemShut {NoStop}%
\bibitem [{\citenamefont {Abbott}\ \emph
  {et~al.}(2020{\natexlab{b}})\citenamefont {Abbott} \emph
  {et~al.}}]{LIGOScientific:2020stg}%
  \BibitemOpen
  \bibfield  {author} {\bibinfo {author} {\bibfnamefont {R.}~\bibnamefont
  {Abbott}} \emph {et~al.} (\bibinfo {collaboration} {LIGO Scientific,
  Virgo}),\ }\href@noop {} {\  (\bibinfo {year} {2020}{\natexlab{b}})},\
  \Eprint {http://arxiv.org/abs/2004.08342} {arXiv:2004.08342 [astro-ph.HE]}
  \BibitemShut {NoStop}%
\bibitem [{\citenamefont {{Tutukov}}\ and\ \citenamefont
  {{Yungelson}}(1973)}]{1973NInfo..27...70T}%
  \BibitemOpen
  \bibfield  {author} {\bibinfo {author} {\bibfnamefont {A.}~\bibnamefont
  {{Tutukov}}}\ and\ \bibinfo {author} {\bibfnamefont {L.}~\bibnamefont
  {{Yungelson}}},\ }\href@noop {} {\bibfield  {journal} {\bibinfo  {journal}
  {Nauchnye Informatsii}\ }\textbf {\bibinfo {volume} {27}},\ \bibinfo {pages}
  {70} (\bibinfo {year} {1973})}\BibitemShut {NoStop}%
\bibitem [{\citenamefont {{Lipunov}}\ \emph {et~al.}(1997)\citenamefont
  {{Lipunov}}, \citenamefont {{Postnov}},\ and\ \citenamefont
  {{Prokhorov}}}]{1997MNRAS.288..245L}%
  \BibitemOpen
  \bibfield  {author} {\bibinfo {author} {\bibfnamefont {V.~M.}\ \bibnamefont
  {{Lipunov}}}, \bibinfo {author} {\bibfnamefont {K.~A.}\ \bibnamefont
  {{Postnov}}}, \ and\ \bibinfo {author} {\bibfnamefont {M.~E.}\ \bibnamefont
  {{Prokhorov}}},\ }\href {\doibase 10.1093/mnras/288.1.245} {\bibfield
  {journal} {\bibinfo  {journal} {\mnras}\ }\textbf {\bibinfo {volume} {288}},\
  \bibinfo {pages} {245} (\bibinfo {year} {1997})},\ \Eprint
  {http://arxiv.org/abs/astro-ph/9702060} {arXiv:astro-ph/9702060 [astro-ph]}
  \BibitemShut {NoStop}%
\bibitem [{\citenamefont {{Tutukov}}\ and\ \citenamefont
  {{Yungelson}}(1993)}]{1993MNRAS.260..675T}%
  \BibitemOpen
  \bibfield  {author} {\bibinfo {author} {\bibfnamefont {A.~V.}\ \bibnamefont
  {{Tutukov}}}\ and\ \bibinfo {author} {\bibfnamefont {L.~R.}\ \bibnamefont
  {{Yungelson}}},\ }\href {\doibase 10.1093/mnras/260.3.675} {\bibfield
  {journal} {\bibinfo  {journal} {\mnras}\ }\textbf {\bibinfo {volume} {260}},\
  \bibinfo {pages} {675} (\bibinfo {year} {1993})}\BibitemShut {NoStop}%
\bibitem [{\citenamefont {{Belczynski}}\ \emph {et~al.}(2002)\citenamefont
  {{Belczynski}}, \citenamefont {{Kalogera}},\ and\ \citenamefont
  {{Bulik}}}]{2002ApJ...572..407B}%
  \BibitemOpen
  \bibfield  {author} {\bibinfo {author} {\bibfnamefont {K.}~\bibnamefont
  {{Belczynski}}}, \bibinfo {author} {\bibfnamefont {V.}~\bibnamefont
  {{Kalogera}}}, \ and\ \bibinfo {author} {\bibfnamefont {T.}~\bibnamefont
  {{Bulik}}},\ }\href {\doibase 10.1086/340304} {\bibfield  {journal} {\bibinfo
   {journal} {\apj}\ }\textbf {\bibinfo {volume} {572}},\ \bibinfo {pages}
  {407} (\bibinfo {year} {2002})},\ \Eprint
  {http://arxiv.org/abs/astro-ph/0111452} {arXiv:astro-ph/0111452 [astro-ph]}
  \BibitemShut {NoStop}%
\bibitem [{\citenamefont {{Voss}}\ and\ \citenamefont
  {{Tauris}}(2003)}]{2003MNRAS.342.1169V}%
  \BibitemOpen
  \bibfield  {author} {\bibinfo {author} {\bibfnamefont {R.}~\bibnamefont
  {{Voss}}}\ and\ \bibinfo {author} {\bibfnamefont {T.~M.}\ \bibnamefont
  {{Tauris}}},\ }\href {\doibase 10.1046/j.1365-8711.2003.06616.x} {\bibfield
  {journal} {\bibinfo  {journal} {\mnras}\ }\textbf {\bibinfo {volume} {342}},\
  \bibinfo {pages} {1169} (\bibinfo {year} {2003})},\ \Eprint
  {http://arxiv.org/abs/astro-ph/0303227} {astro-ph/0303227} \BibitemShut
  {NoStop}%
\bibitem [{\citenamefont {{Kalogera}}\ \emph {et~al.}(2007)\citenamefont
  {{Kalogera}}, \citenamefont {{Belczynski}}, \citenamefont {{Kim}},
  \citenamefont {{O'Shaughnessy}},\ and\ \citenamefont
  {{Willems}}}]{2007PhR...442...75K}%
  \BibitemOpen
  \bibfield  {author} {\bibinfo {author} {\bibfnamefont {V.}~\bibnamefont
  {{Kalogera}}}, \bibinfo {author} {\bibfnamefont {K.}~\bibnamefont
  {{Belczynski}}}, \bibinfo {author} {\bibfnamefont {C.}~\bibnamefont {{Kim}}},
  \bibinfo {author} {\bibfnamefont {R.}~\bibnamefont {{O'Shaughnessy}}}, \ and\
  \bibinfo {author} {\bibfnamefont {B.}~\bibnamefont {{Willems}}},\ }\href
  {\doibase 10.1016/j.physrep.2007.02.008} {\bibfield  {journal} {\bibinfo
  {journal} {\physrep}\ }\textbf {\bibinfo {volume} {442}},\ \bibinfo {pages}
  {75} (\bibinfo {year} {2007})},\ \Eprint
  {http://arxiv.org/abs/astro-ph/0612144} {astro-ph/0612144} \BibitemShut
  {NoStop}%
\bibitem [{\citenamefont {{Dominik}}\ \emph {et~al.}(2012)\citenamefont
  {{Dominik}}, \citenamefont {{Belczynski}}, \citenamefont {{Fryer}},
  \citenamefont {{Holz}}, \citenamefont {{Berti}}, \citenamefont {{Bulik}},
  \citenamefont {{Mandel}},\ and\ \citenamefont
  {{O'Shaughnessy}}}]{2012ApJ...759...52D}%
  \BibitemOpen
  \bibfield  {author} {\bibinfo {author} {\bibfnamefont {M.}~\bibnamefont
  {{Dominik}}}, \bibinfo {author} {\bibfnamefont {K.}~\bibnamefont
  {{Belczynski}}}, \bibinfo {author} {\bibfnamefont {C.}~\bibnamefont
  {{Fryer}}}, \bibinfo {author} {\bibfnamefont {D.~E.}\ \bibnamefont {{Holz}}},
  \bibinfo {author} {\bibfnamefont {E.}~\bibnamefont {{Berti}}}, \bibinfo
  {author} {\bibfnamefont {T.}~\bibnamefont {{Bulik}}}, \bibinfo {author}
  {\bibfnamefont {I.}~\bibnamefont {{Mandel}}}, \ and\ \bibinfo {author}
  {\bibfnamefont {R.}~\bibnamefont {{O'Shaughnessy}}},\ }\href {\doibase
  10.1088/0004-637X/759/1/52} {\bibfield  {journal} {\bibinfo  {journal}
  {\apj}\ }\textbf {\bibinfo {volume} {759}},\ \bibinfo {eid} {52} (\bibinfo
  {year} {2012})},\ \Eprint {http://arxiv.org/abs/1202.4901} {arXiv:1202.4901
  [astro-ph.HE]} \BibitemShut {NoStop}%
\bibitem [{\citenamefont {{Dominik}}\ \emph {et~al.}(2013)\citenamefont
  {{Dominik}}, \citenamefont {{Belczynski}}, \citenamefont {{Fryer}},
  \citenamefont {{Holz}}, \citenamefont {{Berti}}, \citenamefont {{Bulik}},
  \citenamefont {{Mandel}},\ and\ \citenamefont
  {{O'Shaughnessy}}}]{2013ApJ...779...72D}%
  \BibitemOpen
  \bibfield  {author} {\bibinfo {author} {\bibfnamefont {M.}~\bibnamefont
  {{Dominik}}}, \bibinfo {author} {\bibfnamefont {K.}~\bibnamefont
  {{Belczynski}}}, \bibinfo {author} {\bibfnamefont {C.}~\bibnamefont
  {{Fryer}}}, \bibinfo {author} {\bibfnamefont {D.~E.}\ \bibnamefont {{Holz}}},
  \bibinfo {author} {\bibfnamefont {E.}~\bibnamefont {{Berti}}}, \bibinfo
  {author} {\bibfnamefont {T.}~\bibnamefont {{Bulik}}}, \bibinfo {author}
  {\bibfnamefont {I.}~\bibnamefont {{Mandel}}}, \ and\ \bibinfo {author}
  {\bibfnamefont {R.}~\bibnamefont {{O'Shaughnessy}}},\ }\href {\doibase
  10.1088/0004-637X/779/1/72} {\bibfield  {journal} {\bibinfo  {journal}
  {\apj}\ }\textbf {\bibinfo {volume} {779}},\ \bibinfo {eid} {72} (\bibinfo
  {year} {2013})},\ \Eprint {http://arxiv.org/abs/1308.1546} {arXiv:1308.1546
  [astro-ph.HE]} \BibitemShut {NoStop}%
\bibitem [{\citenamefont {{Belczynski}}\ \emph {et~al.}(2014)\citenamefont
  {{Belczynski}}, \citenamefont {{Buonanno}}, \citenamefont {{Cantiello}},
  \citenamefont {{Fryer}}, \citenamefont {{Holz}}, \citenamefont {{Mandel}},
  \citenamefont {{Miller}},\ and\ \citenamefont
  {{Walczak}}}]{2014ApJ...789..120B}%
  \BibitemOpen
  \bibfield  {author} {\bibinfo {author} {\bibfnamefont {K.}~\bibnamefont
  {{Belczynski}}}, \bibinfo {author} {\bibfnamefont {A.}~\bibnamefont
  {{Buonanno}}}, \bibinfo {author} {\bibfnamefont {M.}~\bibnamefont
  {{Cantiello}}}, \bibinfo {author} {\bibfnamefont {C.~L.}\ \bibnamefont
  {{Fryer}}}, \bibinfo {author} {\bibfnamefont {D.~E.}\ \bibnamefont {{Holz}}},
  \bibinfo {author} {\bibfnamefont {I.}~\bibnamefont {{Mandel}}}, \bibinfo
  {author} {\bibfnamefont {M.~C.}\ \bibnamefont {{Miller}}}, \ and\ \bibinfo
  {author} {\bibfnamefont {M.}~\bibnamefont {{Walczak}}},\ }\href {\doibase
  10.1088/0004-637X/789/2/120} {\bibfield  {journal} {\bibinfo  {journal}
  {\apj}\ }\textbf {\bibinfo {volume} {789}},\ \bibinfo {eid} {120} (\bibinfo
  {year} {2014})},\ \Eprint {http://arxiv.org/abs/1403.0677} {arXiv:1403.0677
  [astro-ph.HE]} \BibitemShut {NoStop}%
\bibitem [{\citenamefont {{Belczynski}}\ \emph {et~al.}(2016)\citenamefont
  {{Belczynski}}, \citenamefont {{Holz}}, \citenamefont {{Bulik}},\ and\
  \citenamefont {{O'Shaughnessy}}}]{2016Natur.534..512B}%
  \BibitemOpen
  \bibfield  {author} {\bibinfo {author} {\bibfnamefont {K.}~\bibnamefont
  {{Belczynski}}}, \bibinfo {author} {\bibfnamefont {D.~E.}\ \bibnamefont
  {{Holz}}}, \bibinfo {author} {\bibfnamefont {T.}~\bibnamefont {{Bulik}}}, \
  and\ \bibinfo {author} {\bibfnamefont {R.}~\bibnamefont {{O'Shaughnessy}}},\
  }\href {\doibase 10.1038/nature18322} {\bibfield  {journal} {\bibinfo
  {journal} {\nat}\ }\textbf {\bibinfo {volume} {534}},\ \bibinfo {pages} {512}
  (\bibinfo {year} {2016})},\ \Eprint {http://arxiv.org/abs/1602.04531}
  {arXiv:1602.04531 [astro-ph.HE]} \BibitemShut {NoStop}%
\bibitem [{\citenamefont {{Zaldarriaga}}\ \emph {et~al.}(2018)\citenamefont
  {{Zaldarriaga}}, \citenamefont {{Kushnir}},\ and\ \citenamefont
  {{Kollmeier}}}]{2018MNRAS.473.4174Z}%
  \BibitemOpen
  \bibfield  {author} {\bibinfo {author} {\bibfnamefont {M.}~\bibnamefont
  {{Zaldarriaga}}}, \bibinfo {author} {\bibfnamefont {D.}~\bibnamefont
  {{Kushnir}}}, \ and\ \bibinfo {author} {\bibfnamefont {J.~A.}\ \bibnamefont
  {{Kollmeier}}},\ }\href {\doibase 10.1093/mnras/stx2577} {\bibfield
  {journal} {\bibinfo  {journal} {\mnras}\ }\textbf {\bibinfo {volume} {473}},\
  \bibinfo {pages} {4174} (\bibinfo {year} {2018})},\ \Eprint
  {http://arxiv.org/abs/1702.00885} {arXiv:1702.00885 [astro-ph.HE]}
  \BibitemShut {NoStop}%
\bibitem [{\citenamefont {{Gerosa}}\ \emph {et~al.}(2018)\citenamefont
  {{Gerosa}}, \citenamefont {{Berti}}, \citenamefont {{O'Shaughnessy}},
  \citenamefont {{Belczynski}}, \citenamefont {{Kesden}}, \citenamefont
  {{Wysocki}},\ and\ \citenamefont {{Gladysz}}}]{2018PhRvD..98h4036G}%
  \BibitemOpen
  \bibfield  {author} {\bibinfo {author} {\bibfnamefont {D.}~\bibnamefont
  {{Gerosa}}}, \bibinfo {author} {\bibfnamefont {E.}~\bibnamefont {{Berti}}},
  \bibinfo {author} {\bibfnamefont {R.}~\bibnamefont {{O'Shaughnessy}}},
  \bibinfo {author} {\bibfnamefont {K.}~\bibnamefont {{Belczynski}}}, \bibinfo
  {author} {\bibfnamefont {M.}~\bibnamefont {{Kesden}}}, \bibinfo {author}
  {\bibfnamefont {D.}~\bibnamefont {{Wysocki}}}, \ and\ \bibinfo {author}
  {\bibfnamefont {W.}~\bibnamefont {{Gladysz}}},\ }\href {\doibase
  10.1103/PhysRevD.98.084036} {\bibfield  {journal} {\bibinfo  {journal}
  {\prd}\ }\textbf {\bibinfo {volume} {98}},\ \bibinfo {eid} {084036} (\bibinfo
  {year} {2018})},\ \Eprint {http://arxiv.org/abs/1808.02491} {arXiv:1808.02491
  [astro-ph.HE]} \BibitemShut {NoStop}%
\bibitem [{\citenamefont {{Qin}}\ \emph {et~al.}(2019)\citenamefont {{Qin}},
  \citenamefont {{Marchant}}, \citenamefont {{Fragos}}, \citenamefont
  {{Meynet}},\ and\ \citenamefont {{Kalogera}}}]{2019ApJ...870L..18Q}%
  \BibitemOpen
  \bibfield  {author} {\bibinfo {author} {\bibfnamefont {Y.}~\bibnamefont
  {{Qin}}}, \bibinfo {author} {\bibfnamefont {P.}~\bibnamefont {{Marchant}}},
  \bibinfo {author} {\bibfnamefont {T.}~\bibnamefont {{Fragos}}}, \bibinfo
  {author} {\bibfnamefont {G.}~\bibnamefont {{Meynet}}}, \ and\ \bibinfo
  {author} {\bibfnamefont {V.}~\bibnamefont {{Kalogera}}},\ }\href {\doibase
  10.3847/2041-8213/aaf97b} {\bibfield  {journal} {\bibinfo  {journal} {\apjl}\
  }\textbf {\bibinfo {volume} {870}},\ \bibinfo {eid} {L18} (\bibinfo {year}
  {2019})},\ \Eprint {http://arxiv.org/abs/1810.13016} {arXiv:1810.13016
  [astro-ph.SR]} \BibitemShut {NoStop}%
\bibitem [{\citenamefont {{Bavera}}\ \emph {et~al.}(2020)\citenamefont
  {{Bavera}}, \citenamefont {{Fragos}}, \citenamefont {{Qin}}, \citenamefont
  {{Zapartas}}, \citenamefont {{Neijssel}}, \citenamefont {{Mandel}},
  \citenamefont {{Batta}}, \citenamefont {{Gaebel}}, \citenamefont
  {{Kimball}},\ and\ \citenamefont {{Stevenson}}}]{2020A&A...635A..97B}%
  \BibitemOpen
  \bibfield  {author} {\bibinfo {author} {\bibfnamefont {S.~S.}\ \bibnamefont
  {{Bavera}}}, \bibinfo {author} {\bibfnamefont {T.}~\bibnamefont {{Fragos}}},
  \bibinfo {author} {\bibfnamefont {Y.}~\bibnamefont {{Qin}}}, \bibinfo
  {author} {\bibfnamefont {E.}~\bibnamefont {{Zapartas}}}, \bibinfo {author}
  {\bibfnamefont {C.~J.}\ \bibnamefont {{Neijssel}}}, \bibinfo {author}
  {\bibfnamefont {I.}~\bibnamefont {{Mandel}}}, \bibinfo {author}
  {\bibfnamefont {A.}~\bibnamefont {{Batta}}}, \bibinfo {author} {\bibfnamefont
  {S.~M.}\ \bibnamefont {{Gaebel}}}, \bibinfo {author} {\bibfnamefont
  {C.}~\bibnamefont {{Kimball}}}, \ and\ \bibinfo {author} {\bibfnamefont
  {S.}~\bibnamefont {{Stevenson}}},\ }\href {\doibase
  10.1051/0004-6361/201936204} {\bibfield  {journal} {\bibinfo  {journal}
  {\aap}\ }\textbf {\bibinfo {volume} {635}},\ \bibinfo {eid} {A97} (\bibinfo
  {year} {2020})},\ \Eprint {http://arxiv.org/abs/1906.12257} {arXiv:1906.12257
  [astro-ph.HE]} \BibitemShut {NoStop}%
\bibitem [{\citenamefont {{Belczynski}}\ \emph {et~al.}(2020)\citenamefont
  {{Belczynski}}, \citenamefont {{Klencki}}, \citenamefont {{Fields}},
  \citenamefont {{Olejak}}, \citenamefont {{Berti}}, \citenamefont {{Meynet}},
  \citenamefont {{Fryer}}, \citenamefont {{Holz}}, \citenamefont
  {{O'Shaughnessy}}, \citenamefont {{Brown}}, \citenamefont {{Bulik}},
  \citenamefont {{Leung}}, \citenamefont {{Nomoto}}, \citenamefont {{Madau}},
  \citenamefont {{Hirschi}}, \citenamefont {{Kaiser}}, \citenamefont {{Jones}},
  \citenamefont {{Mondal}}, \citenamefont {{Chruslinska}}, \citenamefont
  {{Drozda}}, \citenamefont {{Gerosa}}, \citenamefont {{Doctor}}, \citenamefont
  {{Giersz}}, \citenamefont {{Ekstrom}}, \citenamefont {{Georgy}},
  \citenamefont {{Askar}}, \citenamefont {{Baibhav}}, \citenamefont
  {{Wysocki}}, \citenamefont {{Natan}}, \citenamefont {{Farr}}, \citenamefont
  {{Wiktorowicz}}, \citenamefont {{Coleman Miller}}, \citenamefont {{Farr}},\
  and\ \citenamefont {{Lasota}}}]{2020A&A...636A.104B}%
  \BibitemOpen
  \bibfield  {author} {\bibinfo {author} {\bibfnamefont {K.}~\bibnamefont
  {{Belczynski}}}, \bibinfo {author} {\bibfnamefont {J.}~\bibnamefont
  {{Klencki}}}, \bibinfo {author} {\bibfnamefont {C.~E.}\ \bibnamefont
  {{Fields}}}, \bibinfo {author} {\bibfnamefont {A.}~\bibnamefont {{Olejak}}},
  \bibinfo {author} {\bibfnamefont {E.}~\bibnamefont {{Berti}}}, \bibinfo
  {author} {\bibfnamefont {G.}~\bibnamefont {{Meynet}}}, \bibinfo {author}
  {\bibfnamefont {C.~L.}\ \bibnamefont {{Fryer}}}, \bibinfo {author}
  {\bibfnamefont {D.~E.}\ \bibnamefont {{Holz}}}, \bibinfo {author}
  {\bibfnamefont {R.}~\bibnamefont {{O'Shaughnessy}}}, \bibinfo {author}
  {\bibfnamefont {D.~A.}\ \bibnamefont {{Brown}}}, \bibinfo {author}
  {\bibfnamefont {T.}~\bibnamefont {{Bulik}}}, \bibinfo {author} {\bibfnamefont
  {S.~C.}\ \bibnamefont {{Leung}}}, \bibinfo {author} {\bibfnamefont
  {K.}~\bibnamefont {{Nomoto}}}, \bibinfo {author} {\bibfnamefont
  {P.}~\bibnamefont {{Madau}}}, \bibinfo {author} {\bibfnamefont
  {R.}~\bibnamefont {{Hirschi}}}, \bibinfo {author} {\bibfnamefont
  {E.}~\bibnamefont {{Kaiser}}}, \bibinfo {author} {\bibfnamefont
  {S.}~\bibnamefont {{Jones}}}, \bibinfo {author} {\bibfnamefont
  {S.}~\bibnamefont {{Mondal}}}, \bibinfo {author} {\bibfnamefont
  {M.}~\bibnamefont {{Chruslinska}}}, \bibinfo {author} {\bibfnamefont
  {P.}~\bibnamefont {{Drozda}}}, \bibinfo {author} {\bibfnamefont
  {D.}~\bibnamefont {{Gerosa}}}, \bibinfo {author} {\bibfnamefont
  {Z.}~\bibnamefont {{Doctor}}}, \bibinfo {author} {\bibfnamefont
  {M.}~\bibnamefont {{Giersz}}}, \bibinfo {author} {\bibfnamefont
  {S.}~\bibnamefont {{Ekstrom}}}, \bibinfo {author} {\bibfnamefont
  {C.}~\bibnamefont {{Georgy}}}, \bibinfo {author} {\bibfnamefont
  {A.}~\bibnamefont {{Askar}}}, \bibinfo {author} {\bibfnamefont
  {V.}~\bibnamefont {{Baibhav}}}, \bibinfo {author} {\bibfnamefont
  {D.}~\bibnamefont {{Wysocki}}}, \bibinfo {author} {\bibfnamefont
  {T.}~\bibnamefont {{Natan}}}, \bibinfo {author} {\bibfnamefont {W.~M.}\
  \bibnamefont {{Farr}}}, \bibinfo {author} {\bibfnamefont {G.}~\bibnamefont
  {{Wiktorowicz}}}, \bibinfo {author} {\bibfnamefont {M.}~\bibnamefont
  {{Coleman Miller}}}, \bibinfo {author} {\bibfnamefont {B.}~\bibnamefont
  {{Farr}}}, \ and\ \bibinfo {author} {\bibfnamefont {J.~P.}\ \bibnamefont
  {{Lasota}}},\ }\href {\doibase 10.1051/0004-6361/201936528ARXIV: 1706.07053}
  {\bibfield  {journal} {\bibinfo  {journal} {\aap}\ }\textbf {\bibinfo
  {volume} {636}},\ \bibinfo {eid} {A104} (\bibinfo {year} {2020})},\ \Eprint
  {http://arxiv.org/abs/1706.07053} {arXiv:1706.07053 [astro-ph.HE]}
  \BibitemShut {NoStop}%
\bibitem [{\citenamefont {{Sigurdsson}}\ and\ \citenamefont
  {{Hernquist}}(1993)}]{1993Natur.364..423S}%
  \BibitemOpen
  \bibfield  {author} {\bibinfo {author} {\bibfnamefont {S.}~\bibnamefont
  {{Sigurdsson}}}\ and\ \bibinfo {author} {\bibfnamefont {L.}~\bibnamefont
  {{Hernquist}}},\ }\href {\doibase 10.1038/364423a0} {\bibfield  {journal}
  {\bibinfo  {journal} {\nat}\ }\textbf {\bibinfo {volume} {364}},\ \bibinfo
  {pages} {423} (\bibinfo {year} {1993})}\BibitemShut {NoStop}%
\bibitem [{\citenamefont {{Portegies Zwart}}\ and\ \citenamefont
  {{McMillan}}(2000)}]{2000ApJ...528L..17P}%
  \BibitemOpen
  \bibfield  {author} {\bibinfo {author} {\bibfnamefont {S.~F.}\ \bibnamefont
  {{Portegies Zwart}}}\ and\ \bibinfo {author} {\bibfnamefont {S.~L.~W.}\
  \bibnamefont {{McMillan}}},\ }\href {\doibase 10.1086/312422} {\bibfield
  {journal} {\bibinfo  {journal} {\apjl}\ }\textbf {\bibinfo {volume} {528}},\
  \bibinfo {pages} {L17} (\bibinfo {year} {2000})},\ \Eprint
  {http://arxiv.org/abs/astro-ph/9910061} {astro-ph/9910061} \BibitemShut
  {NoStop}%
\bibitem [{\citenamefont {{Portegies Zwart}}\ \emph {et~al.}(2004)\citenamefont
  {{Portegies Zwart}}, \citenamefont {{Baumgardt}}, \citenamefont {{Hut}},
  \citenamefont {{Makino}},\ and\ \citenamefont
  {{McMillan}}}]{2004Natur.428..724P}%
  \BibitemOpen
  \bibfield  {author} {\bibinfo {author} {\bibfnamefont {S.~F.}\ \bibnamefont
  {{Portegies Zwart}}}, \bibinfo {author} {\bibfnamefont {H.}~\bibnamefont
  {{Baumgardt}}}, \bibinfo {author} {\bibfnamefont {P.}~\bibnamefont {{Hut}}},
  \bibinfo {author} {\bibfnamefont {J.}~\bibnamefont {{Makino}}}, \ and\
  \bibinfo {author} {\bibfnamefont {S.~L.~W.}\ \bibnamefont {{McMillan}}},\
  }\href {\doibase 10.1038/nature02448} {\bibfield  {journal} {\bibinfo
  {journal} {\nat}\ }\textbf {\bibinfo {volume} {428}},\ \bibinfo {pages} {724}
  (\bibinfo {year} {2004})},\ \Eprint {http://arxiv.org/abs/astro-ph/0402622}
  {arXiv:astro-ph/0402622 [astro-ph]} \BibitemShut {NoStop}%
\bibitem [{\citenamefont {{O'Leary}}\ \emph {et~al.}(2006)\citenamefont
  {{O'Leary}}, \citenamefont {{Rasio}}, \citenamefont {{Fregeau}},
  \citenamefont {{Ivanova}},\ and\ \citenamefont
  {{O'Shaughnessy}}}]{2006ApJ...637..937O}%
  \BibitemOpen
  \bibfield  {author} {\bibinfo {author} {\bibfnamefont {R.~M.}\ \bibnamefont
  {{O'Leary}}}, \bibinfo {author} {\bibfnamefont {F.~A.}\ \bibnamefont
  {{Rasio}}}, \bibinfo {author} {\bibfnamefont {J.~M.}\ \bibnamefont
  {{Fregeau}}}, \bibinfo {author} {\bibfnamefont {N.}~\bibnamefont
  {{Ivanova}}}, \ and\ \bibinfo {author} {\bibfnamefont {R.}~\bibnamefont
  {{O'Shaughnessy}}},\ }\href {\doibase 10.1086/498446} {\bibfield  {journal}
  {\bibinfo  {journal} {\apj}\ }\textbf {\bibinfo {volume} {637}},\ \bibinfo
  {pages} {937} (\bibinfo {year} {2006})},\ \Eprint
  {http://arxiv.org/abs/astro-ph/0508224} {astro-ph/0508224} \BibitemShut
  {NoStop}%
\bibitem [{\citenamefont {{Antonini}}\ and\ \citenamefont
  {{Perets}}(2012)}]{2012ApJ...757...27A}%
  \BibitemOpen
  \bibfield  {author} {\bibinfo {author} {\bibfnamefont {F.}~\bibnamefont
  {{Antonini}}}\ and\ \bibinfo {author} {\bibfnamefont {H.~B.}\ \bibnamefont
  {{Perets}}},\ }\href {\doibase 10.1088/0004-637X/757/1/27} {\bibfield
  {journal} {\bibinfo  {journal} {\apj}\ }\textbf {\bibinfo {volume} {757}},\
  \bibinfo {eid} {27} (\bibinfo {year} {2012})},\ \Eprint
  {http://arxiv.org/abs/1203.2938} {arXiv:1203.2938} \BibitemShut {NoStop}%
\bibitem [{\citenamefont {{Thompson}}(2011)}]{2011ApJ...741...82T}%
  \BibitemOpen
  \bibfield  {author} {\bibinfo {author} {\bibfnamefont {T.~A.}\ \bibnamefont
  {{Thompson}}},\ }\href {\doibase 10.1088/0004-637X/741/2/82} {\bibfield
  {journal} {\bibinfo  {journal} {\apj}\ }\textbf {\bibinfo {volume} {741}},\
  \bibinfo {eid} {82} (\bibinfo {year} {2011})},\ \Eprint
  {http://arxiv.org/abs/1011.4322} {arXiv:1011.4322 [astro-ph.HE]} \BibitemShut
  {NoStop}%
\bibitem [{\citenamefont {{Antonini}}\ \emph {et~al.}(2014)\citenamefont
  {{Antonini}}, \citenamefont {{Murray}},\ and\ \citenamefont
  {{Mikkola}}}]{2014ApJ...781...45A}%
  \BibitemOpen
  \bibfield  {author} {\bibinfo {author} {\bibfnamefont {F.}~\bibnamefont
  {{Antonini}}}, \bibinfo {author} {\bibfnamefont {N.}~\bibnamefont
  {{Murray}}}, \ and\ \bibinfo {author} {\bibfnamefont {S.}~\bibnamefont
  {{Mikkola}}},\ }\href {\doibase 10.1088/0004-637X/781/1/45} {\bibfield
  {journal} {\bibinfo  {journal} {\apj}\ }\textbf {\bibinfo {volume} {781}},\
  \bibinfo {eid} {45} (\bibinfo {year} {2014})},\ \Eprint
  {http://arxiv.org/abs/1308.3674} {arXiv:1308.3674 [astro-ph.HE]} \BibitemShut
  {NoStop}%
\bibitem [{\citenamefont {{Ziosi}}\ \emph {et~al.}(2014)\citenamefont
  {{Ziosi}}, \citenamefont {{Mapelli}}, \citenamefont {{Branchesi}},\ and\
  \citenamefont {{Tormen}}}]{2014MNRAS.441.3703Z}%
  \BibitemOpen
  \bibfield  {author} {\bibinfo {author} {\bibfnamefont {B.~M.}\ \bibnamefont
  {{Ziosi}}}, \bibinfo {author} {\bibfnamefont {M.}~\bibnamefont {{Mapelli}}},
  \bibinfo {author} {\bibfnamefont {M.}~\bibnamefont {{Branchesi}}}, \ and\
  \bibinfo {author} {\bibfnamefont {G.}~\bibnamefont {{Tormen}}},\ }\href
  {\doibase 10.1093/mnras/stu824} {\bibfield  {journal} {\bibinfo  {journal}
  {\mnras}\ }\textbf {\bibinfo {volume} {441}},\ \bibinfo {pages} {3703}
  (\bibinfo {year} {2014})},\ \Eprint {http://arxiv.org/abs/1404.7147}
  {arXiv:1404.7147} \BibitemShut {NoStop}%
\bibitem [{\citenamefont {{Prodan}}\ \emph {et~al.}(2015)\citenamefont
  {{Prodan}}, \citenamefont {{Antonini}},\ and\ \citenamefont
  {{Perets}}}]{2015ApJ...799..118P}%
  \BibitemOpen
  \bibfield  {author} {\bibinfo {author} {\bibfnamefont {S.}~\bibnamefont
  {{Prodan}}}, \bibinfo {author} {\bibfnamefont {F.}~\bibnamefont
  {{Antonini}}}, \ and\ \bibinfo {author} {\bibfnamefont {H.~B.}\ \bibnamefont
  {{Perets}}},\ }\href {\doibase 10.1088/0004-637X/799/2/118} {\bibfield
  {journal} {\bibinfo  {journal} {\apj}\ }\textbf {\bibinfo {volume} {799}},\
  \bibinfo {eid} {118} (\bibinfo {year} {2015})},\ \Eprint
  {http://arxiv.org/abs/1405.6029} {arXiv:1405.6029} \BibitemShut {NoStop}%
\bibitem [{\citenamefont {{Rodriguez}}\ \emph {et~al.}(2015)\citenamefont
  {{Rodriguez}}, \citenamefont {{Morscher}}, \citenamefont {{Pattabiraman}},
  \citenamefont {{Chatterjee}}, \citenamefont {{Haster}},\ and\ \citenamefont
  {{Rasio}}}]{2015PhRvL.115e1101R}%
  \BibitemOpen
  \bibfield  {author} {\bibinfo {author} {\bibfnamefont {C.~L.}\ \bibnamefont
  {{Rodriguez}}}, \bibinfo {author} {\bibfnamefont {M.}~\bibnamefont
  {{Morscher}}}, \bibinfo {author} {\bibfnamefont {B.}~\bibnamefont
  {{Pattabiraman}}}, \bibinfo {author} {\bibfnamefont {S.}~\bibnamefont
  {{Chatterjee}}}, \bibinfo {author} {\bibfnamefont {C.-J.}\ \bibnamefont
  {{Haster}}}, \ and\ \bibinfo {author} {\bibfnamefont {F.~A.}\ \bibnamefont
  {{Rasio}}},\ }\href {\doibase 10.1103/PhysRevLett.115.051101} {\bibfield
  {journal} {\bibinfo  {journal} {Physical Review Letters}\ }\textbf {\bibinfo
  {volume} {115}},\ \bibinfo {eid} {051101} (\bibinfo {year} {2015})},\ \Eprint
  {http://arxiv.org/abs/1505.00792} {arXiv:1505.00792 [astro-ph.HE]}
  \BibitemShut {NoStop}%
\bibitem [{\citenamefont {{Mapelli}}(2016)}]{2016MNRAS.459.3432M}%
  \BibitemOpen
  \bibfield  {author} {\bibinfo {author} {\bibfnamefont {M.}~\bibnamefont
  {{Mapelli}}},\ }\href {\doibase 10.1093/mnras/stw869} {\bibfield  {journal}
  {\bibinfo  {journal} {\mnras}\ }\textbf {\bibinfo {volume} {459}},\ \bibinfo
  {pages} {3432} (\bibinfo {year} {2016})},\ \Eprint
  {http://arxiv.org/abs/1604.03559} {arXiv:1604.03559} \BibitemShut {NoStop}%
\bibitem [{\citenamefont {{Stephan}}\ \emph {et~al.}(2016)\citenamefont
  {{Stephan}}, \citenamefont {{Naoz}}, \citenamefont {{Ghez}}, \citenamefont
  {{Witzel}}, \citenamefont {{Sitarski}}, \citenamefont {{Do}},\ and\
  \citenamefont {{Kocsis}}}]{2016MNRAS.460.3494S}%
  \BibitemOpen
  \bibfield  {author} {\bibinfo {author} {\bibfnamefont {A.~P.}\ \bibnamefont
  {{Stephan}}}, \bibinfo {author} {\bibfnamefont {S.}~\bibnamefont {{Naoz}}},
  \bibinfo {author} {\bibfnamefont {A.~M.}\ \bibnamefont {{Ghez}}}, \bibinfo
  {author} {\bibfnamefont {G.}~\bibnamefont {{Witzel}}}, \bibinfo {author}
  {\bibfnamefont {B.~N.}\ \bibnamefont {{Sitarski}}}, \bibinfo {author}
  {\bibfnamefont {T.}~\bibnamefont {{Do}}}, \ and\ \bibinfo {author}
  {\bibfnamefont {B.}~\bibnamefont {{Kocsis}}},\ }\href {\doibase
  10.1093/mnras/stw1220} {\bibfield  {journal} {\bibinfo  {journal} {\mnras}\
  }\textbf {\bibinfo {volume} {460}},\ \bibinfo {pages} {3494} (\bibinfo {year}
  {2016})},\ \Eprint {http://arxiv.org/abs/1603.02709} {arXiv:1603.02709
  [astro-ph.SR]} \BibitemShut {NoStop}%
\bibitem [{\citenamefont {{Kimpson}}\ \emph {et~al.}(2016)\citenamefont
  {{Kimpson}}, \citenamefont {{Spera}}, \citenamefont {{Mapelli}},\ and\
  \citenamefont {{Ziosi}}}]{2016MNRAS.463.2443K}%
  \BibitemOpen
  \bibfield  {author} {\bibinfo {author} {\bibfnamefont {T.~O.}\ \bibnamefont
  {{Kimpson}}}, \bibinfo {author} {\bibfnamefont {M.}~\bibnamefont {{Spera}}},
  \bibinfo {author} {\bibfnamefont {M.}~\bibnamefont {{Mapelli}}}, \ and\
  \bibinfo {author} {\bibfnamefont {B.~M.}\ \bibnamefont {{Ziosi}}},\ }\href
  {\doibase 10.1093/mnras/stw2085} {\bibfield  {journal} {\bibinfo  {journal}
  {\mnras}\ }\textbf {\bibinfo {volume} {463}},\ \bibinfo {pages} {2443}
  (\bibinfo {year} {2016})},\ \Eprint {http://arxiv.org/abs/1608.05422}
  {arXiv:1608.05422} \BibitemShut {NoStop}%
\bibitem [{\citenamefont {{Antonini}}\ and\ \citenamefont
  {{Rasio}}(2016)}]{2016ApJ...831..187A}%
  \BibitemOpen
  \bibfield  {author} {\bibinfo {author} {\bibfnamefont {F.}~\bibnamefont
  {{Antonini}}}\ and\ \bibinfo {author} {\bibfnamefont {F.~A.}\ \bibnamefont
  {{Rasio}}},\ }\href {\doibase 10.3847/0004-637X/831/2/187} {\bibfield
  {journal} {\bibinfo  {journal} {\apj}\ }\textbf {\bibinfo {volume} {831}},\
  \bibinfo {eid} {187} (\bibinfo {year} {2016})},\ \Eprint
  {http://arxiv.org/abs/1606.04889} {arXiv:1606.04889 [astro-ph.HE]}
  \BibitemShut {NoStop}%
\bibitem [{\citenamefont {{Rodriguez}}\ \emph
  {et~al.}(2016{\natexlab{a}})\citenamefont {{Rodriguez}}, \citenamefont
  {{Zevin}}, \citenamefont {{Pankow}}, \citenamefont {{Kalogera}},\ and\
  \citenamefont {{Rasio}}}]{2016ApJ...832L...2R}%
  \BibitemOpen
  \bibfield  {author} {\bibinfo {author} {\bibfnamefont {C.~L.}\ \bibnamefont
  {{Rodriguez}}}, \bibinfo {author} {\bibfnamefont {M.}~\bibnamefont
  {{Zevin}}}, \bibinfo {author} {\bibfnamefont {C.}~\bibnamefont {{Pankow}}},
  \bibinfo {author} {\bibfnamefont {V.}~\bibnamefont {{Kalogera}}}, \ and\
  \bibinfo {author} {\bibfnamefont {F.~A.}\ \bibnamefont {{Rasio}}},\ }\href
  {\doibase 10.3847/2041-8205/832/1/L2} {\bibfield  {journal} {\bibinfo
  {journal} {\apjl}\ }\textbf {\bibinfo {volume} {832}},\ \bibinfo {eid} {L2}
  (\bibinfo {year} {2016}{\natexlab{a}})},\ \Eprint
  {http://arxiv.org/abs/1609.05916} {arXiv:1609.05916 [astro-ph.HE]}
  \BibitemShut {NoStop}%
\bibitem [{\citenamefont {{Rodriguez}}\ \emph
  {et~al.}(2016{\natexlab{b}})\citenamefont {{Rodriguez}}, \citenamefont
  {{Chatterjee}},\ and\ \citenamefont {{Rasio}}}]{2016PhRvD..93h4029R}%
  \BibitemOpen
  \bibfield  {author} {\bibinfo {author} {\bibfnamefont {C.~L.}\ \bibnamefont
  {{Rodriguez}}}, \bibinfo {author} {\bibfnamefont {S.}~\bibnamefont
  {{Chatterjee}}}, \ and\ \bibinfo {author} {\bibfnamefont {F.~A.}\
  \bibnamefont {{Rasio}}},\ }\href {\doibase 10.1103/PhysRevD.93.084029}
  {\bibfield  {journal} {\bibinfo  {journal} {\prd}\ }\textbf {\bibinfo
  {volume} {93}},\ \bibinfo {eid} {084029} (\bibinfo {year}
  {2016}{\natexlab{b}})},\ \Eprint {http://arxiv.org/abs/1602.02444}
  {arXiv:1602.02444 [astro-ph.HE]} \BibitemShut {NoStop}%
\bibitem [{\citenamefont {{Silsbee}}\ and\ \citenamefont
  {{Tremaine}}(2017)}]{2017ApJ...836...39S}%
  \BibitemOpen
  \bibfield  {author} {\bibinfo {author} {\bibfnamefont {K.}~\bibnamefont
  {{Silsbee}}}\ and\ \bibinfo {author} {\bibfnamefont {S.}~\bibnamefont
  {{Tremaine}}},\ }\href {\doibase 10.3847/1538-4357/aa5729} {\bibfield
  {journal} {\bibinfo  {journal} {\apj}\ }\textbf {\bibinfo {volume} {836}},\
  \bibinfo {eid} {39} (\bibinfo {year} {2017})},\ \Eprint
  {http://arxiv.org/abs/1608.07642} {arXiv:1608.07642 [astro-ph.HE]}
  \BibitemShut {NoStop}%
\bibitem [{\citenamefont {{Chatterjee}}\ \emph {et~al.}(2017)\citenamefont
  {{Chatterjee}}, \citenamefont {{Rodriguez}}, \citenamefont {{Kalogera}},\
  and\ \citenamefont {{Rasio}}}]{2017ApJ...836L..26C}%
  \BibitemOpen
  \bibfield  {author} {\bibinfo {author} {\bibfnamefont {S.}~\bibnamefont
  {{Chatterjee}}}, \bibinfo {author} {\bibfnamefont {C.~L.}\ \bibnamefont
  {{Rodriguez}}}, \bibinfo {author} {\bibfnamefont {V.}~\bibnamefont
  {{Kalogera}}}, \ and\ \bibinfo {author} {\bibfnamefont {F.~A.}\ \bibnamefont
  {{Rasio}}},\ }\href {\doibase 10.3847/2041-8213/aa5caa} {\bibfield  {journal}
  {\bibinfo  {journal} {\apjl}\ }\textbf {\bibinfo {volume} {836}},\ \bibinfo
  {eid} {L26} (\bibinfo {year} {2017})},\ \Eprint
  {http://arxiv.org/abs/1609.06689} {arXiv:1609.06689 [astro-ph.GA]}
  \BibitemShut {NoStop}%
\bibitem [{\citenamefont {{Samsing}}\ and\ \citenamefont
  {{Ramirez-Ruiz}}(2017)}]{2017ApJ...840L..14S}%
  \BibitemOpen
  \bibfield  {author} {\bibinfo {author} {\bibfnamefont {J.}~\bibnamefont
  {{Samsing}}}\ and\ \bibinfo {author} {\bibfnamefont {E.}~\bibnamefont
  {{Ramirez-Ruiz}}},\ }\href {\doibase 10.3847/2041-8213/aa6f0b} {\bibfield
  {journal} {\bibinfo  {journal} {\apjl}\ }\textbf {\bibinfo {volume} {840}},\
  \bibinfo {eid} {L14} (\bibinfo {year} {2017})},\ \Eprint
  {http://arxiv.org/abs/1703.09703} {arXiv:1703.09703 [astro-ph.HE]}
  \BibitemShut {NoStop}%
\bibitem [{\citenamefont {{Antonini}}\ \emph {et~al.}(2017)\citenamefont
  {{Antonini}}, \citenamefont {{Toonen}},\ and\ \citenamefont
  {{Hamers}}}]{2017ApJ...841...77A}%
  \BibitemOpen
  \bibfield  {author} {\bibinfo {author} {\bibfnamefont {F.}~\bibnamefont
  {{Antonini}}}, \bibinfo {author} {\bibfnamefont {S.}~\bibnamefont
  {{Toonen}}}, \ and\ \bibinfo {author} {\bibfnamefont {A.~S.}\ \bibnamefont
  {{Hamers}}},\ }\href {\doibase 10.3847/1538-4357/aa6f5e} {\bibfield
  {journal} {\bibinfo  {journal} {\apj}\ }\textbf {\bibinfo {volume} {841}},\
  \bibinfo {eid} {77} (\bibinfo {year} {2017})},\ \Eprint
  {http://arxiv.org/abs/1703.06614} {arXiv:1703.06614 [astro-ph.GA]}
  \BibitemShut {NoStop}%
\bibitem [{\citenamefont {{Petrovich}}\ and\ \citenamefont
  {{Antonini}}(2017)}]{2017ApJ...846..146P}%
  \BibitemOpen
  \bibfield  {author} {\bibinfo {author} {\bibfnamefont {C.}~\bibnamefont
  {{Petrovich}}}\ and\ \bibinfo {author} {\bibfnamefont {F.}~\bibnamefont
  {{Antonini}}},\ }\href {\doibase 10.3847/1538-4357/aa8628} {\bibfield
  {journal} {\bibinfo  {journal} {\apj}\ }\textbf {\bibinfo {volume} {846}},\
  \bibinfo {eid} {146} (\bibinfo {year} {2017})},\ \Eprint
  {http://arxiv.org/abs/1705.05848} {arXiv:1705.05848 [astro-ph.HE]}
  \BibitemShut {NoStop}%
\bibitem [{\citenamefont {{Antonini}}\ \emph {et~al.}(2018)\citenamefont
  {{Antonini}}, \citenamefont {{Rodriguez}}, \citenamefont {{Petrovich}},\ and\
  \citenamefont {{Fischer}}}]{2018MNRAS.480L..58A}%
  \BibitemOpen
  \bibfield  {author} {\bibinfo {author} {\bibfnamefont {F.}~\bibnamefont
  {{Antonini}}}, \bibinfo {author} {\bibfnamefont {C.~L.}\ \bibnamefont
  {{Rodriguez}}}, \bibinfo {author} {\bibfnamefont {C.}~\bibnamefont
  {{Petrovich}}}, \ and\ \bibinfo {author} {\bibfnamefont {C.~L.}\ \bibnamefont
  {{Fischer}}},\ }\href {\doibase 10.1093/mnrasl/sly126} {\bibfield  {journal}
  {\bibinfo  {journal} {\mnras}\ }\textbf {\bibinfo {volume} {480}},\ \bibinfo
  {pages} {L58} (\bibinfo {year} {2018})},\ \Eprint
  {http://arxiv.org/abs/1711.07142} {arXiv:1711.07142 [astro-ph.HE]}
  \BibitemShut {NoStop}%
\bibitem [{\citenamefont {{Rodriguez}}\ \emph
  {et~al.}(2018{\natexlab{a}})\citenamefont {{Rodriguez}}, \citenamefont
  {{Amaro-Seoane}}, \citenamefont {{Chatterjee}}, \citenamefont {{Kremer}},
  \citenamefont {{Rasio}}, \citenamefont {{Samsing}}, \citenamefont {{Ye}},\
  and\ \citenamefont {{Zevin}}}]{2018PhRvD..98l3005R}%
  \BibitemOpen
  \bibfield  {author} {\bibinfo {author} {\bibfnamefont {C.~L.}\ \bibnamefont
  {{Rodriguez}}}, \bibinfo {author} {\bibfnamefont {P.}~\bibnamefont
  {{Amaro-Seoane}}}, \bibinfo {author} {\bibfnamefont {S.}~\bibnamefont
  {{Chatterjee}}}, \bibinfo {author} {\bibfnamefont {K.}~\bibnamefont
  {{Kremer}}}, \bibinfo {author} {\bibfnamefont {F.~A.}\ \bibnamefont
  {{Rasio}}}, \bibinfo {author} {\bibfnamefont {J.}~\bibnamefont {{Samsing}}},
  \bibinfo {author} {\bibfnamefont {C.~S.}\ \bibnamefont {{Ye}}}, \ and\
  \bibinfo {author} {\bibfnamefont {M.}~\bibnamefont {{Zevin}}},\ }\href
  {\doibase 10.1103/PhysRevD.98.123005} {\bibfield  {journal} {\bibinfo
  {journal} {\prd}\ }\textbf {\bibinfo {volume} {98}},\ \bibinfo {eid} {123005}
  (\bibinfo {year} {2018}{\natexlab{a}})},\ \Eprint
  {http://arxiv.org/abs/1811.04926} {arXiv:1811.04926 [astro-ph.HE]}
  \BibitemShut {NoStop}%
\bibitem [{\citenamefont {{Samsing}}\ \emph
  {et~al.}(2018{\natexlab{a}})\citenamefont {{Samsing}}, \citenamefont
  {{Askar}},\ and\ \citenamefont {{Giersz}}}]{2018ApJ...855..124S}%
  \BibitemOpen
  \bibfield  {author} {\bibinfo {author} {\bibfnamefont {J.}~\bibnamefont
  {{Samsing}}}, \bibinfo {author} {\bibfnamefont {A.}~\bibnamefont {{Askar}}},
  \ and\ \bibinfo {author} {\bibfnamefont {M.}~\bibnamefont {{Giersz}}},\
  }\href {\doibase 10.3847/1538-4357/aaab52} {\bibfield  {journal} {\bibinfo
  {journal} {\apj}\ }\textbf {\bibinfo {volume} {855}},\ \bibinfo {eid} {124}
  (\bibinfo {year} {2018}{\natexlab{a}})},\ \Eprint
  {http://arxiv.org/abs/1712.06186} {arXiv:1712.06186 [astro-ph.HE]}
  \BibitemShut {NoStop}%
\bibitem [{\citenamefont {{Samsing}}\ \emph
  {et~al.}(2018{\natexlab{b}})\citenamefont {{Samsing}}, \citenamefont
  {{MacLeod}},\ and\ \citenamefont {{Ramirez-Ruiz}}}]{2018ApJ...853..140S}%
  \BibitemOpen
  \bibfield  {author} {\bibinfo {author} {\bibfnamefont {J.}~\bibnamefont
  {{Samsing}}}, \bibinfo {author} {\bibfnamefont {M.}~\bibnamefont
  {{MacLeod}}}, \ and\ \bibinfo {author} {\bibfnamefont {E.}~\bibnamefont
  {{Ramirez-Ruiz}}},\ }\href {\doibase 10.3847/1538-4357/aaa715} {\bibfield
  {journal} {\bibinfo  {journal} {\apj}\ }\textbf {\bibinfo {volume} {853}},\
  \bibinfo {eid} {140} (\bibinfo {year} {2018}{\natexlab{b}})},\ \Eprint
  {http://arxiv.org/abs/1706.03776} {arXiv:1706.03776 [astro-ph.HE]}
  \BibitemShut {NoStop}%
\bibitem [{\citenamefont {{Hamers}}\ \emph {et~al.}(2018)\citenamefont
  {{Hamers}}, \citenamefont {{Bar-Or}}, \citenamefont {{Petrovich}},\ and\
  \citenamefont {{Antonini}}}]{2018ApJ...865....2H}%
  \BibitemOpen
  \bibfield  {author} {\bibinfo {author} {\bibfnamefont {A.~S.}\ \bibnamefont
  {{Hamers}}}, \bibinfo {author} {\bibfnamefont {B.}~\bibnamefont {{Bar-Or}}},
  \bibinfo {author} {\bibfnamefont {C.}~\bibnamefont {{Petrovich}}}, \ and\
  \bibinfo {author} {\bibfnamefont {F.}~\bibnamefont {{Antonini}}},\ }\href
  {\doibase 10.3847/1538-4357/aadae2} {\bibfield  {journal} {\bibinfo
  {journal} {\apj}\ }\textbf {\bibinfo {volume} {865}},\ \bibinfo {eid} {2}
  (\bibinfo {year} {2018})},\ \Eprint {http://arxiv.org/abs/1805.10313}
  {arXiv:1805.10313 [astro-ph.HE]} \BibitemShut {NoStop}%
\bibitem [{\citenamefont {{Hoang}}\ \emph {et~al.}(2018)\citenamefont
  {{Hoang}}, \citenamefont {{Naoz}}, \citenamefont {{Kocsis}}, \citenamefont
  {{Rasio}},\ and\ \citenamefont {{Dosopoulou}}}]{2018ApJ...856..140H}%
  \BibitemOpen
  \bibfield  {author} {\bibinfo {author} {\bibfnamefont {B.-M.}\ \bibnamefont
  {{Hoang}}}, \bibinfo {author} {\bibfnamefont {S.}~\bibnamefont {{Naoz}}},
  \bibinfo {author} {\bibfnamefont {B.}~\bibnamefont {{Kocsis}}}, \bibinfo
  {author} {\bibfnamefont {F.~A.}\ \bibnamefont {{Rasio}}}, \ and\ \bibinfo
  {author} {\bibfnamefont {F.}~\bibnamefont {{Dosopoulou}}},\ }\href {\doibase
  10.3847/1538-4357/aaafce} {\bibfield  {journal} {\bibinfo  {journal} {\apj}\
  }\textbf {\bibinfo {volume} {856}},\ \bibinfo {eid} {140} (\bibinfo {year}
  {2018})},\ \Eprint {http://arxiv.org/abs/1706.09896} {arXiv:1706.09896
  [astro-ph.HE]} \BibitemShut {NoStop}%
\bibitem [{\citenamefont {{Samsing}}(2018)}]{2018PhRvD..97j3014S}%
  \BibitemOpen
  \bibfield  {author} {\bibinfo {author} {\bibfnamefont {J.}~\bibnamefont
  {{Samsing}}},\ }\href {\doibase 10.1103/PhysRevD.97.103014} {\bibfield
  {journal} {\bibinfo  {journal} {\prd}\ }\textbf {\bibinfo {volume} {97}},\
  \bibinfo {eid} {103014} (\bibinfo {year} {2018})},\ \Eprint
  {http://arxiv.org/abs/1711.07452} {arXiv:1711.07452 [astro-ph.HE]}
  \BibitemShut {NoStop}%
\bibitem [{\citenamefont {{Arca-Sedda}}\ and\ \citenamefont
  {{Gualandris}}(2018)}]{2018MNRAS.477.4423A}%
  \BibitemOpen
  \bibfield  {author} {\bibinfo {author} {\bibfnamefont {M.}~\bibnamefont
  {{Arca-Sedda}}}\ and\ \bibinfo {author} {\bibfnamefont {A.}~\bibnamefont
  {{Gualandris}}},\ }\href {\doibase 10.1093/mnras/sty922} {\bibfield
  {journal} {\bibinfo  {journal} {\mnras}\ }\textbf {\bibinfo {volume} {477}},\
  \bibinfo {pages} {4423} (\bibinfo {year} {2018})},\ \Eprint
  {http://arxiv.org/abs/1804.06116} {arXiv:1804.06116} \BibitemShut {NoStop}%
\bibitem [{\citenamefont {{Rodriguez}}\ \emph
  {et~al.}(2018{\natexlab{b}})\citenamefont {{Rodriguez}}, \citenamefont
  {{Amaro-Seoane}}, \citenamefont {{Chatterjee}},\ and\ \citenamefont
  {{Rasio}}}]{2018PhRvL.120o1101R}%
  \BibitemOpen
  \bibfield  {author} {\bibinfo {author} {\bibfnamefont {C.~L.}\ \bibnamefont
  {{Rodriguez}}}, \bibinfo {author} {\bibfnamefont {P.}~\bibnamefont
  {{Amaro-Seoane}}}, \bibinfo {author} {\bibfnamefont {S.}~\bibnamefont
  {{Chatterjee}}}, \ and\ \bibinfo {author} {\bibfnamefont {F.~A.}\
  \bibnamefont {{Rasio}}},\ }\href {\doibase 10.1103/PhysRevLett.120.151101}
  {\bibfield  {journal} {\bibinfo  {journal} {Physical Review Letters}\
  }\textbf {\bibinfo {volume} {120}},\ \bibinfo {eid} {151101} (\bibinfo {year}
  {2018}{\natexlab{b}})},\ \Eprint {http://arxiv.org/abs/1712.04937}
  {arXiv:1712.04937 [astro-ph.HE]} \BibitemShut {NoStop}%
\bibitem [{\citenamefont {{Randall}}\ and\ \citenamefont
  {{Xianyu}}(2018{\natexlab{a}})}]{2018ApJ...853...93R}%
  \BibitemOpen
  \bibfield  {author} {\bibinfo {author} {\bibfnamefont {L.}~\bibnamefont
  {{Randall}}}\ and\ \bibinfo {author} {\bibfnamefont {Z.-Z.}\ \bibnamefont
  {{Xianyu}}},\ }\href {\doibase 10.3847/1538-4357/aaa1a2} {\bibfield
  {journal} {\bibinfo  {journal} {\apj}\ }\textbf {\bibinfo {volume} {853}},\
  \bibinfo {eid} {93} (\bibinfo {year} {2018}{\natexlab{a}})}\BibitemShut
  {NoStop}%
\bibitem [{\citenamefont {{Gond{\'a}n}}\ \emph {et~al.}(2018)\citenamefont
  {{Gond{\'a}n}}, \citenamefont {{Kocsis}}, \citenamefont {{Raffai}},\ and\
  \citenamefont {{Frei}}}]{2018ApJ...860....5G}%
  \BibitemOpen
  \bibfield  {author} {\bibinfo {author} {\bibfnamefont {L.}~\bibnamefont
  {{Gond{\'a}n}}}, \bibinfo {author} {\bibfnamefont {B.}~\bibnamefont
  {{Kocsis}}}, \bibinfo {author} {\bibfnamefont {P.}~\bibnamefont {{Raffai}}},
  \ and\ \bibinfo {author} {\bibfnamefont {Z.}~\bibnamefont {{Frei}}},\ }\href
  {\doibase 10.3847/1538-4357/aabfee} {\bibfield  {journal} {\bibinfo
  {journal} {\apj}\ }\textbf {\bibinfo {volume} {860}},\ \bibinfo {eid} {5}
  (\bibinfo {year} {2018})},\ \Eprint {http://arxiv.org/abs/1711.09989}
  {arXiv:1711.09989 [astro-ph.HE]} \BibitemShut {NoStop}%
\bibitem [{\citenamefont {{Randall}}\ and\ \citenamefont
  {{Xianyu}}(2018{\natexlab{b}})}]{2018ApJ...864..134R}%
  \BibitemOpen
  \bibfield  {author} {\bibinfo {author} {\bibfnamefont {L.}~\bibnamefont
  {{Randall}}}\ and\ \bibinfo {author} {\bibfnamefont {Z.-Z.}\ \bibnamefont
  {{Xianyu}}},\ }\href {\doibase 10.3847/1538-4357/aad7fe} {\bibfield
  {journal} {\bibinfo  {journal} {\apj}\ }\textbf {\bibinfo {volume} {864}},\
  \bibinfo {eid} {134} (\bibinfo {year} {2018}{\natexlab{b}})},\ \Eprint
  {http://arxiv.org/abs/1802.05718} {arXiv:1802.05718 [gr-qc]} \BibitemShut
  {NoStop}%
\bibitem [{\citenamefont {{Arca-Sedda}}\ and\ \citenamefont
  {{Capuzzo-Dolcetta}}(2019)}]{2019MNRAS.483..152A}%
  \BibitemOpen
  \bibfield  {author} {\bibinfo {author} {\bibfnamefont {M.}~\bibnamefont
  {{Arca-Sedda}}}\ and\ \bibinfo {author} {\bibfnamefont {R.}~\bibnamefont
  {{Capuzzo-Dolcetta}}},\ }\href {\doibase 10.1093/mnras/sty3096} {\bibfield
  {journal} {\bibinfo  {journal} {\mnras}\ }\textbf {\bibinfo {volume} {483}},\
  \bibinfo {pages} {152} (\bibinfo {year} {2019})},\ \Eprint
  {http://arxiv.org/abs/1709.05567} {arXiv:1709.05567 [astro-ph.GA]}
  \BibitemShut {NoStop}%
\bibitem [{\citenamefont {{Fragione}}\ and\ \citenamefont
  {{Loeb}}(2019)}]{2019MNRAS.486.4443F}%
  \BibitemOpen
  \bibfield  {author} {\bibinfo {author} {\bibfnamefont {G.}~\bibnamefont
  {{Fragione}}}\ and\ \bibinfo {author} {\bibfnamefont {A.}~\bibnamefont
  {{Loeb}}},\ }\href {\doibase 10.1093/mnras/stz1131} {\bibfield  {journal}
  {\bibinfo  {journal} {\mnras}\ }\textbf {\bibinfo {volume} {486}},\ \bibinfo
  {pages} {4443} (\bibinfo {year} {2019})},\ \Eprint
  {http://arxiv.org/abs/1903.10511} {arXiv:1903.10511 [astro-ph.GA]}
  \BibitemShut {NoStop}%
\bibitem [{\citenamefont {{Hamers}}\ and\ \citenamefont
  {{Samsing}}(2019{\natexlab{a}})}]{2019MNRAS.487.5630H}%
  \BibitemOpen
  \bibfield  {author} {\bibinfo {author} {\bibfnamefont {A.~S.}\ \bibnamefont
  {{Hamers}}}\ and\ \bibinfo {author} {\bibfnamefont {J.}~\bibnamefont
  {{Samsing}}},\ }\href {\doibase 10.1093/mnras/stz1646} {\bibfield  {journal}
  {\bibinfo  {journal} {\mnras}\ }\textbf {\bibinfo {volume} {487}},\ \bibinfo
  {pages} {5630} (\bibinfo {year} {2019}{\natexlab{a}})},\ \Eprint
  {http://arxiv.org/abs/1904.09624} {arXiv:1904.09624 [astro-ph.SR]}
  \BibitemShut {NoStop}%
\bibitem [{\citenamefont {{Samsing}}\ \emph {et~al.}(2019)\citenamefont
  {{Samsing}}, \citenamefont {{Hamers}},\ and\ \citenamefont
  {{Tyles}}}]{2019PhRvD.100d3010S}%
  \BibitemOpen
  \bibfield  {author} {\bibinfo {author} {\bibfnamefont {J.}~\bibnamefont
  {{Samsing}}}, \bibinfo {author} {\bibfnamefont {A.~S.}\ \bibnamefont
  {{Hamers}}}, \ and\ \bibinfo {author} {\bibfnamefont {J.~G.}\ \bibnamefont
  {{Tyles}}},\ }\href {\doibase 10.1103/PhysRevD.100.043010} {\bibfield
  {journal} {\bibinfo  {journal} {\prd}\ }\textbf {\bibinfo {volume} {100}},\
  \bibinfo {eid} {043010} (\bibinfo {year} {2019})},\ \Eprint
  {http://arxiv.org/abs/1906.07189} {arXiv:1906.07189 [astro-ph.HE]}
  \BibitemShut {NoStop}%
\bibitem [{\citenamefont {{Fragione}}\ and\ \citenamefont
  {{Kocsis}}(2020)}]{2020MNRAS.493.3920F}%
  \BibitemOpen
  \bibfield  {author} {\bibinfo {author} {\bibfnamefont {G.}~\bibnamefont
  {{Fragione}}}\ and\ \bibinfo {author} {\bibfnamefont {B.}~\bibnamefont
  {{Kocsis}}},\ }\href {\doibase 10.1093/mnras/staa443} {\bibfield  {journal}
  {\bibinfo  {journal} {\mnras}\ }\textbf {\bibinfo {volume} {493}},\ \bibinfo
  {pages} {3920} (\bibinfo {year} {2020})},\ \Eprint
  {http://arxiv.org/abs/1910.00407} {arXiv:1910.00407 [astro-ph.GA]}
  \BibitemShut {NoStop}%
\bibitem [{\citenamefont {{Madau}}\ and\ \citenamefont
  {{Dickinson}}(2014)}]{2014ARA&A..52..415M}%
  \BibitemOpen
  \bibfield  {author} {\bibinfo {author} {\bibfnamefont {P.}~\bibnamefont
  {{Madau}}}\ and\ \bibinfo {author} {\bibfnamefont {M.}~\bibnamefont
  {{Dickinson}}},\ }\href {\doibase 10.1146/annurev-astro-081811-125615}
  {\bibfield  {journal} {\bibinfo  {journal} {\araa}\ }\textbf {\bibinfo
  {volume} {52}},\ \bibinfo {pages} {415} (\bibinfo {year} {2014})},\ \Eprint
  {http://arxiv.org/abs/1403.0007} {arXiv:1403.0007 [astro-ph.CO]} \BibitemShut
  {NoStop}%
\bibitem [{\citenamefont {{Zel'dovich}}\ and\ \citenamefont
  {{Novikov}}(1967)}]{1967SvA....10..602Z}%
  \BibitemOpen
  \bibfield  {author} {\bibinfo {author} {\bibfnamefont {Y.~B.}\ \bibnamefont
  {{Zel'dovich}}}\ and\ \bibinfo {author} {\bibfnamefont {I.~D.}\ \bibnamefont
  {{Novikov}}},\ }\href@noop {} {\bibfield  {journal} {\bibinfo  {journal}
  {\sovast}\ }\textbf {\bibinfo {volume} {10}},\ \bibinfo {pages} {602}
  (\bibinfo {year} {1967})}\BibitemShut {NoStop}%
\bibitem [{\citenamefont {Hawking}(1971)}]{Hawking:1971ei}%
  \BibitemOpen
  \bibfield  {author} {\bibinfo {author} {\bibfnamefont {S.}~\bibnamefont
  {Hawking}},\ }\href@noop {} {\bibfield  {journal} {\bibinfo  {journal} {Mon.
  Not. Roy. Astron. Soc.}\ }\textbf {\bibinfo {volume} {152}},\ \bibinfo
  {pages} {75} (\bibinfo {year} {1971})}\BibitemShut {NoStop}%
%%CITATION = MNRAA,152,75;%%
\bibitem [{\citenamefont {Carr}\ and\ \citenamefont
  {Hawking}(1974)}]{Carr:1974nx}%
  \BibitemOpen
  \bibfield  {author} {\bibinfo {author} {\bibfnamefont {B.~J.}\ \bibnamefont
  {Carr}}\ and\ \bibinfo {author} {\bibfnamefont {S.~W.}\ \bibnamefont
  {Hawking}},\ }\href@noop {} {\bibfield  {journal} {\bibinfo  {journal} {Mon.
  Not. Roy. Astron. Soc.}\ }\textbf {\bibinfo {volume} {168}},\ \bibinfo
  {pages} {399} (\bibinfo {year} {1974})}\BibitemShut {NoStop}%
%%CITATION = MNRAA,168,399;%%
\bibitem [{\citenamefont {Hawking}(1989)}]{Hawking:1987bn}%
  \BibitemOpen
  \bibfield  {author} {\bibinfo {author} {\bibfnamefont {S.~W.}\ \bibnamefont
  {Hawking}},\ }\href {\doibase 10.1016/0370-2693(89)90206-2} {\bibfield
  {journal} {\bibinfo  {journal} {Phys. Lett.}\ }\textbf {\bibinfo {volume}
  {B231}},\ \bibinfo {pages} {237} (\bibinfo {year} {1989})}\BibitemShut
  {NoStop}%
%%CITATION = PHLTA,B231,237;%%
\bibitem [{\citenamefont {Hawking}\ \emph {et~al.}(1982)\citenamefont
  {Hawking}, \citenamefont {Moss},\ and\ \citenamefont
  {Stewart}}]{Hawking:1982ga}%
  \BibitemOpen
  \bibfield  {author} {\bibinfo {author} {\bibfnamefont {S.~W.}\ \bibnamefont
  {Hawking}}, \bibinfo {author} {\bibfnamefont {I.~G.}\ \bibnamefont {Moss}}, \
  and\ \bibinfo {author} {\bibfnamefont {J.~M.}\ \bibnamefont {Stewart}},\
  }\href {\doibase 10.1103/PhysRevD.26.2681} {\bibfield  {journal} {\bibinfo
  {journal} {Phys. Rev.}\ }\textbf {\bibinfo {volume} {D26}},\ \bibinfo {pages}
  {2681} (\bibinfo {year} {1982})}\BibitemShut {NoStop}%
%%CITATION = PHRVA,D26,2681;%%
\bibitem [{\citenamefont {{Carr}}\ and\ \citenamefont
  {{Kuhnel}}(2020)}]{2020arXiv200602838C}%
  \BibitemOpen
  \bibfield  {author} {\bibinfo {author} {\bibfnamefont {B.}~\bibnamefont
  {{Carr}}}\ and\ \bibinfo {author} {\bibfnamefont {F.}~\bibnamefont
  {{Kuhnel}}},\ }\href@noop {} {\bibfield  {journal} {\bibinfo  {journal}
  {arXiv e-prints}\ ,\ \bibinfo {eid} {arXiv:2006.02838}} (\bibinfo {year}
  {2020})},\ \Eprint {http://arxiv.org/abs/2006.02838} {arXiv:2006.02838
  [astro-ph.CO]} \BibitemShut {NoStop}%
\bibitem [{\citenamefont {{Postnov}}\ and\ \citenamefont
  {{Kuranov}}(2019)}]{2019MNRAS.483.3288P}%
  \BibitemOpen
  \bibfield  {author} {\bibinfo {author} {\bibfnamefont {K.~A.}\ \bibnamefont
  {{Postnov}}}\ and\ \bibinfo {author} {\bibfnamefont {A.~G.}\ \bibnamefont
  {{Kuranov}}},\ }\href {\doibase 10.1093/mnras/sty3313} {\bibfield  {journal}
  {\bibinfo  {journal} {\mnras}\ }\textbf {\bibinfo {volume} {483}},\ \bibinfo
  {pages} {3288} (\bibinfo {year} {2019})},\ \Eprint
  {http://arxiv.org/abs/1706.00369} {arXiv:1706.00369 [astro-ph.HE]}
  \BibitemShut {NoStop}%
\bibitem [{\citenamefont {Clesse}\ and\ \citenamefont
  {Garcia-Bellido}(2018)}]{Clesse:2017bsw}%
  \BibitemOpen
  \bibfield  {author} {\bibinfo {author} {\bibfnamefont {S.}~\bibnamefont
  {Clesse}}\ and\ \bibinfo {author} {\bibfnamefont {J.}~\bibnamefont
  {Garcia-Bellido}},\ }\href {\doibase 10.1016/j.dark.2018.08.004} {\bibfield
  {journal} {\bibinfo  {journal} {Phys. Dark Univ.}\ }\textbf {\bibinfo
  {volume} {22}},\ \bibinfo {pages} {137} (\bibinfo {year} {2018})},\ \Eprint
  {http://arxiv.org/abs/1711.10458} {arXiv:1711.10458 [astro-ph.CO]}
  \BibitemShut {NoStop}%
%%CITATION = ARXIV:1711.10458;%%
\bibitem [{\citenamefont {Postnov}\ and\ \citenamefont
  {Mitichkin}(2019)}]{Postnov:2019tmw}%
  \BibitemOpen
  \bibfield  {author} {\bibinfo {author} {\bibfnamefont {K.}~\bibnamefont
  {Postnov}}\ and\ \bibinfo {author} {\bibfnamefont {N.}~\bibnamefont
  {Mitichkin}},\ }\href {\doibase 10.1088/1475-7516/2019/06/044} {\bibfield
  {journal} {\bibinfo  {journal} {JCAP}\ }\textbf {\bibinfo {volume} {1906}},\
  \bibinfo {pages} {044} (\bibinfo {year} {2019})},\ \Eprint
  {http://arxiv.org/abs/1904.00570} {arXiv:1904.00570 [astro-ph.HE]}
  \BibitemShut {NoStop}%
%%CITATION = ARXIV:1904.00570;%%
\bibitem [{\citenamefont {Fernandez}\ and\ \citenamefont
  {Profumo}(2019)}]{Fernandez:2019kyb}%
  \BibitemOpen
  \bibfield  {author} {\bibinfo {author} {\bibfnamefont {N.}~\bibnamefont
  {Fernandez}}\ and\ \bibinfo {author} {\bibfnamefont {S.}~\bibnamefont
  {Profumo}},\ }\href {\doibase 10.1088/1475-7516/2019/08/022} {\bibfield
  {journal} {\bibinfo  {journal} {JCAP}\ }\textbf {\bibinfo {volume} {1908}},\
  \bibinfo {pages} {022} (\bibinfo {year} {2019})},\ \Eprint
  {http://arxiv.org/abs/1905.13019} {arXiv:1905.13019 [astro-ph.HE]}
  \BibitemShut {NoStop}%
%%CITATION = ARXIV:1905.13019;%%
\bibitem [{\citenamefont {De~Luca}\ \emph
  {et~al.}(2019{\natexlab{a}})\citenamefont {De~Luca}, \citenamefont
  {Desjacques}, \citenamefont {Franciolini}, \citenamefont {Malhotra},\ and\
  \citenamefont {Riotto}}]{DeLuca:2019buf}%
  \BibitemOpen
  \bibfield  {author} {\bibinfo {author} {\bibfnamefont {V.}~\bibnamefont
  {De~Luca}}, \bibinfo {author} {\bibfnamefont {V.}~\bibnamefont {Desjacques}},
  \bibinfo {author} {\bibfnamefont {G.}~\bibnamefont {Franciolini}}, \bibinfo
  {author} {\bibfnamefont {A.}~\bibnamefont {Malhotra}}, \ and\ \bibinfo
  {author} {\bibfnamefont {A.}~\bibnamefont {Riotto}},\ }\href {\doibase
  10.1088/1475-7516/2019/05/018} {\bibfield  {journal} {\bibinfo  {journal}
  {JCAP}\ }\textbf {\bibinfo {volume} {05}},\ \bibinfo {pages} {018} (\bibinfo
  {year} {2019}{\natexlab{a}})},\ \Eprint {http://arxiv.org/abs/1903.01179}
  {arXiv:1903.01179 [astro-ph.CO]} \BibitemShut {NoStop}%
\bibitem [{\citenamefont {De~Luca}\ \emph
  {et~al.}(2020{\natexlab{a}})\citenamefont {De~Luca}, \citenamefont
  {Franciolini}, \citenamefont {Pani},\ and\ \citenamefont
  {Riotto}}]{DeLuca:2020bjf}%
  \BibitemOpen
  \bibfield  {author} {\bibinfo {author} {\bibfnamefont {V.}~\bibnamefont
  {De~Luca}}, \bibinfo {author} {\bibfnamefont {G.}~\bibnamefont
  {Franciolini}}, \bibinfo {author} {\bibfnamefont {P.}~\bibnamefont {Pani}}, \
  and\ \bibinfo {author} {\bibfnamefont {A.}~\bibnamefont {Riotto}},\ }\href
  {\doibase 10.1088/1475-7516/2020/04/052} {\bibfield  {journal} {\bibinfo
  {journal} {JCAP}\ }\textbf {\bibinfo {volume} {04}},\ \bibinfo {pages} {052}
  (\bibinfo {year} {2020}{\natexlab{a}})},\ \Eprint
  {http://arxiv.org/abs/2003.02778} {arXiv:2003.02778 [astro-ph.CO]}
  \BibitemShut {NoStop}%
\bibitem [{\citenamefont {Bird}\ \emph {et~al.}(2016)\citenamefont {Bird},
  \citenamefont {Cholis}, \citenamefont {Muñoz}, \citenamefont {Ali-Haimoud},
  \citenamefont {Kamionkowski}, \citenamefont {Kovetz}, \citenamefont
  {Raccanelli},\ and\ \citenamefont {Riess}}]{Bird:2016dcv}%
  \BibitemOpen
  \bibfield  {author} {\bibinfo {author} {\bibfnamefont {S.}~\bibnamefont
  {Bird}}, \bibinfo {author} {\bibfnamefont {I.}~\bibnamefont {Cholis}},
  \bibinfo {author} {\bibfnamefont {J.~B.}\ \bibnamefont {Muñoz}}, \bibinfo
  {author} {\bibfnamefont {Y.}~\bibnamefont {Ali-Haimoud}}, \bibinfo {author}
  {\bibfnamefont {M.}~\bibnamefont {Kamionkowski}}, \bibinfo {author}
  {\bibfnamefont {E.~D.}\ \bibnamefont {Kovetz}}, \bibinfo {author}
  {\bibfnamefont {A.}~\bibnamefont {Raccanelli}}, \ and\ \bibinfo {author}
  {\bibfnamefont {A.~G.}\ \bibnamefont {Riess}},\ }\href {\doibase
  10.1103/PhysRevLett.116.201301} {\bibfield  {journal} {\bibinfo  {journal}
  {Phys. Rev. Lett.}\ }\textbf {\bibinfo {volume} {116}},\ \bibinfo {pages}
  {201301} (\bibinfo {year} {2016})},\ \Eprint
  {http://arxiv.org/abs/1603.00464} {arXiv:1603.00464 [astro-ph.CO]}
  \BibitemShut {NoStop}%
%%CITATION = ARXIV:1603.00464;%%
\bibitem [{\citenamefont {Clesse}\ and\ \citenamefont
  {García-Bellido}(2017)}]{Clesse:2016vqa}%
  \BibitemOpen
  \bibfield  {author} {\bibinfo {author} {\bibfnamefont {S.}~\bibnamefont
  {Clesse}}\ and\ \bibinfo {author} {\bibfnamefont {J.}~\bibnamefont
  {García-Bellido}},\ }\href {\doibase 10.1016/j.dark.2016.10.002} {\bibfield
  {journal} {\bibinfo  {journal} {Phys. Dark Univ.}\ }\textbf {\bibinfo
  {volume} {15}},\ \bibinfo {pages} {142} (\bibinfo {year} {2017})},\ \Eprint
  {http://arxiv.org/abs/1603.05234} {arXiv:1603.05234 [astro-ph.CO]}
  \BibitemShut {NoStop}%
%%CITATION = ARXIV:1603.05234;%%
\bibitem [{\citenamefont {{Blinnikov}}\ \emph {et~al.}(2016)\citenamefont
  {{Blinnikov}}, \citenamefont {{Dolgov}}, \citenamefont {{Porayko}},\ and\
  \citenamefont {{Postnov}}}]{2016JCAP...11..036B}%
  \BibitemOpen
  \bibfield  {author} {\bibinfo {author} {\bibfnamefont {S.}~\bibnamefont
  {{Blinnikov}}}, \bibinfo {author} {\bibfnamefont {A.}~\bibnamefont
  {{Dolgov}}}, \bibinfo {author} {\bibfnamefont {N.~K.}\ \bibnamefont
  {{Porayko}}}, \ and\ \bibinfo {author} {\bibfnamefont {K.}~\bibnamefont
  {{Postnov}}},\ }\href {\doibase 10.1088/1475-7516/2016/11/036} {\bibfield
  {journal} {\bibinfo  {journal} {\jcap}\ }\textbf {\bibinfo {volume} {2016}},\
  \bibinfo {eid} {036} (\bibinfo {year} {2016})},\ \Eprint
  {http://arxiv.org/abs/1611.00541} {arXiv:1611.00541 [astro-ph.HE]}
  \BibitemShut {NoStop}%
\bibitem [{\citenamefont {Bringmann}\ \emph {et~al.}(2019)\citenamefont
  {Bringmann}, \citenamefont {Depta}, \citenamefont {Domcke},\ and\
  \citenamefont {Schmidt-Hoberg}}]{Bringmann:2018mxj}%
  \BibitemOpen
  \bibfield  {author} {\bibinfo {author} {\bibfnamefont {T.}~\bibnamefont
  {Bringmann}}, \bibinfo {author} {\bibfnamefont {P.~F.}\ \bibnamefont
  {Depta}}, \bibinfo {author} {\bibfnamefont {V.}~\bibnamefont {Domcke}}, \
  and\ \bibinfo {author} {\bibfnamefont {K.}~\bibnamefont {Schmidt-Hoberg}},\
  }\href {\doibase 10.1103/PhysRevD.99.063532} {\bibfield  {journal} {\bibinfo
  {journal} {Phys. Rev.}\ }\textbf {\bibinfo {volume} {D99}},\ \bibinfo {pages}
  {063532} (\bibinfo {year} {2019})},\ \Eprint
  {http://arxiv.org/abs/1808.05910} {arXiv:1808.05910 [astro-ph.CO]}
  \BibitemShut {NoStop}%
%%CITATION = ARXIV:1808.05910;%%
\bibitem [{\citenamefont {Raidal}\ \emph {et~al.}(2019)\citenamefont {Raidal},
  \citenamefont {Spethmann}, \citenamefont {Vaskonen},\ and\ \citenamefont
  {Veermäe}}]{Raidal:2018bbj}%
  \BibitemOpen
  \bibfield  {author} {\bibinfo {author} {\bibfnamefont {M.}~\bibnamefont
  {Raidal}}, \bibinfo {author} {\bibfnamefont {C.}~\bibnamefont {Spethmann}},
  \bibinfo {author} {\bibfnamefont {V.}~\bibnamefont {Vaskonen}}, \ and\
  \bibinfo {author} {\bibfnamefont {H.}~\bibnamefont {Veermäe}},\ }\href
  {\doibase 10.1088/1475-7516/2019/02/018} {\bibfield  {journal} {\bibinfo
  {journal} {JCAP}\ }\textbf {\bibinfo {volume} {1902}},\ \bibinfo {pages}
  {018} (\bibinfo {year} {2019})},\ \Eprint {http://arxiv.org/abs/1812.01930}
  {arXiv:1812.01930 [astro-ph.CO]} \BibitemShut {NoStop}%
%%CITATION = ARXIV:1812.01930;%%
\bibitem [{\citenamefont {Ballesteros}\ \emph {et~al.}(2018)\citenamefont
  {Ballesteros}, \citenamefont {Serpico},\ and\ \citenamefont
  {Taoso}}]{Ballesteros:2018swv}%
  \BibitemOpen
  \bibfield  {author} {\bibinfo {author} {\bibfnamefont {G.}~\bibnamefont
  {Ballesteros}}, \bibinfo {author} {\bibfnamefont {P.~D.}\ \bibnamefont
  {Serpico}}, \ and\ \bibinfo {author} {\bibfnamefont {M.}~\bibnamefont
  {Taoso}},\ }\href {\doibase 10.1088/1475-7516/2018/10/043} {\bibfield
  {journal} {\bibinfo  {journal} {JCAP}\ }\textbf {\bibinfo {volume} {1810}},\
  \bibinfo {pages} {043} (\bibinfo {year} {2018})},\ \Eprint
  {http://arxiv.org/abs/1807.02084} {arXiv:1807.02084 [astro-ph.CO]}
  \BibitemShut {NoStop}%
%%CITATION = ARXIV:1807.02084;%%
\bibitem [{\citenamefont {Raidal}\ \emph {et~al.}(2017)\citenamefont {Raidal},
  \citenamefont {Vaskonen},\ and\ \citenamefont {Veermäe}}]{Raidal:2017mfl}%
  \BibitemOpen
  \bibfield  {author} {\bibinfo {author} {\bibfnamefont {M.}~\bibnamefont
  {Raidal}}, \bibinfo {author} {\bibfnamefont {V.}~\bibnamefont {Vaskonen}}, \
  and\ \bibinfo {author} {\bibfnamefont {H.}~\bibnamefont {Veermäe}},\ }\href
  {\doibase 10.1088/1475-7516/2017/09/037} {\bibfield  {journal} {\bibinfo
  {journal} {JCAP}\ }\textbf {\bibinfo {volume} {1709}},\ \bibinfo {pages}
  {037} (\bibinfo {year} {2017})},\ \Eprint {http://arxiv.org/abs/1707.01480}
  {arXiv:1707.01480 [astro-ph.CO]} \BibitemShut {NoStop}%
%%CITATION = ARXIV:1707.01480;%%
\bibitem [{\citenamefont {Chen}\ and\ \citenamefont
  {Huang}(2018)}]{Chen:2018czv}%
  \BibitemOpen
  \bibfield  {author} {\bibinfo {author} {\bibfnamefont {Z.-C.}\ \bibnamefont
  {Chen}}\ and\ \bibinfo {author} {\bibfnamefont {Q.-G.}\ \bibnamefont
  {Huang}},\ }\href {\doibase 10.3847/1538-4357/aad6e2} {\bibfield  {journal}
  {\bibinfo  {journal} {Astrophys. J.}\ }\textbf {\bibinfo {volume} {864}},\
  \bibinfo {pages} {61} (\bibinfo {year} {2018})},\ \Eprint
  {http://arxiv.org/abs/1801.10327} {arXiv:1801.10327 [astro-ph.CO]}
  \BibitemShut {NoStop}%
%%CITATION = ARXIV:1801.10327;%%
\bibitem [{\citenamefont {Ali-Haïmoud}\ \emph {et~al.}(2017)\citenamefont
  {Ali-Haïmoud}, \citenamefont {Kovetz},\ and\ \citenamefont
  {Kamionkowski}}]{Ali-Haimoud:2017rtz}%
  \BibitemOpen
  \bibfield  {author} {\bibinfo {author} {\bibfnamefont {Y.}~\bibnamefont
  {Ali-Haïmoud}}, \bibinfo {author} {\bibfnamefont {E.~D.}\ \bibnamefont
  {Kovetz}}, \ and\ \bibinfo {author} {\bibfnamefont {M.}~\bibnamefont
  {Kamionkowski}},\ }\href {\doibase 10.1103/PhysRevD.96.123523} {\bibfield
  {journal} {\bibinfo  {journal} {Phys. Rev.}\ }\textbf {\bibinfo {volume}
  {D96}},\ \bibinfo {pages} {123523} (\bibinfo {year} {2017})},\ \Eprint
  {http://arxiv.org/abs/1709.06576} {arXiv:1709.06576 [astro-ph.CO]}
  \BibitemShut {NoStop}%
%%CITATION = ARXIV:1709.06576;%%
\bibitem [{\citenamefont {Sasaki}\ \emph {et~al.}(2016)\citenamefont {Sasaki},
  \citenamefont {Suyama}, \citenamefont {Tanaka},\ and\ \citenamefont
  {Yokoyama}}]{Sasaki:2016jop}%
  \BibitemOpen
  \bibfield  {author} {\bibinfo {author} {\bibfnamefont {M.}~\bibnamefont
  {Sasaki}}, \bibinfo {author} {\bibfnamefont {T.}~\bibnamefont {Suyama}},
  \bibinfo {author} {\bibfnamefont {T.}~\bibnamefont {Tanaka}}, \ and\ \bibinfo
  {author} {\bibfnamefont {S.}~\bibnamefont {Yokoyama}},\ }\href {\doibase
  10.1103/PhysRevLett.121.059901, 10.1103/PhysRevLett.117.061101} {\bibfield
  {journal} {\bibinfo  {journal} {Phys. Rev. Lett.}\ }\textbf {\bibinfo
  {volume} {117}},\ \bibinfo {pages} {061101} (\bibinfo {year} {2016})},\
  \bibinfo {note} {[erratum: Phys. Rev. Lett.121,no.5,059901(2018)]},\ \Eprint
  {http://arxiv.org/abs/1603.08338} {arXiv:1603.08338 [astro-ph.CO]}
  \BibitemShut {NoStop}%
%%CITATION = ARXIV:1603.08338;%%
\bibitem [{\citenamefont {De~Luca}\ \emph
  {et~al.}(2020{\natexlab{b}})\citenamefont {De~Luca}, \citenamefont
  {Franciolini}, \citenamefont {Pani},\ and\ \citenamefont
  {Riotto}}]{DeLuca:2020qqa}%
  \BibitemOpen
  \bibfield  {author} {\bibinfo {author} {\bibfnamefont {V.}~\bibnamefont
  {De~Luca}}, \bibinfo {author} {\bibfnamefont {G.}~\bibnamefont
  {Franciolini}}, \bibinfo {author} {\bibfnamefont {P.}~\bibnamefont {Pani}}, \
  and\ \bibinfo {author} {\bibfnamefont {A.}~\bibnamefont {Riotto}},\ }\href
  {\doibase 10.1088/1475-7516/2020/06/044} {\bibfield  {journal} {\bibinfo
  {journal} {JCAP}\ }\textbf {\bibinfo {volume} {06}},\ \bibinfo {pages} {044}
  (\bibinfo {year} {2020}{\natexlab{b}})},\ \Eprint
  {http://arxiv.org/abs/2005.05641} {arXiv:2005.05641 [astro-ph.CO]}
  \BibitemShut {NoStop}%
\bibitem [{\citenamefont {Vaskonen}\ and\ \citenamefont
  {Veermäe}(2020)}]{Vaskonen:2019jpv}%
  \BibitemOpen
  \bibfield  {author} {\bibinfo {author} {\bibfnamefont {V.}~\bibnamefont
  {Vaskonen}}\ and\ \bibinfo {author} {\bibfnamefont {H.}~\bibnamefont
  {Veermäe}},\ }\href {\doibase 10.1103/PhysRevD.101.043015} {\bibfield
  {journal} {\bibinfo  {journal} {Phys. Rev. D}\ }\textbf {\bibinfo {volume}
  {101}},\ \bibinfo {pages} {043015} (\bibinfo {year} {2020})},\ \Eprint
  {http://arxiv.org/abs/1908.09752} {arXiv:1908.09752 [astro-ph.CO]}
  \BibitemShut {NoStop}%
\bibitem [{\citenamefont {{Heggie}}(1975)}]{1975MNRAS.173..729H}%
  \BibitemOpen
  \bibfield  {author} {\bibinfo {author} {\bibfnamefont {D.~C.}\ \bibnamefont
  {{Heggie}}},\ }\href {\doibase 10.1093/mnras/173.3.729} {\bibfield  {journal}
  {\bibinfo  {journal} {\mnras}\ }\textbf {\bibinfo {volume} {173}},\ \bibinfo
  {pages} {729} (\bibinfo {year} {1975})}\BibitemShut {NoStop}%
\bibitem [{\citenamefont {{Heggie}}\ and\ \citenamefont
  {{Rasio}}(1996)}]{1996MNRAS.282.1064H}%
  \BibitemOpen
  \bibfield  {author} {\bibinfo {author} {\bibfnamefont {D.~C.}\ \bibnamefont
  {{Heggie}}}\ and\ \bibinfo {author} {\bibfnamefont {F.~A.}\ \bibnamefont
  {{Rasio}}},\ }\href {\doibase 10.1093/mnras/282.3.1064} {\bibfield  {journal}
  {\bibinfo  {journal} {\mnras}\ }\textbf {\bibinfo {volume} {282}},\ \bibinfo
  {pages} {1064} (\bibinfo {year} {1996})},\ \Eprint
  {http://arxiv.org/abs/astro-ph/9506082} {astro-ph/9506082} \BibitemShut
  {NoStop}%
\bibitem [{\citenamefont {{Spurzem}}\ \emph {et~al.}(2009)\citenamefont
  {{Spurzem}}, \citenamefont {{Giersz}}, \citenamefont {{Heggie}},\ and\
  \citenamefont {{Lin}}}]{2009ApJ...697..458S}%
  \BibitemOpen
  \bibfield  {author} {\bibinfo {author} {\bibfnamefont {R.}~\bibnamefont
  {{Spurzem}}}, \bibinfo {author} {\bibfnamefont {M.}~\bibnamefont {{Giersz}}},
  \bibinfo {author} {\bibfnamefont {D.~C.}\ \bibnamefont {{Heggie}}}, \ and\
  \bibinfo {author} {\bibfnamefont {D.~N.~C.}\ \bibnamefont {{Lin}}},\ }\href
  {\doibase 10.1088/0004-637X/697/1/458} {\bibfield  {journal} {\bibinfo
  {journal} {\apj}\ }\textbf {\bibinfo {volume} {697}},\ \bibinfo {pages} {458}
  (\bibinfo {year} {2009})},\ \Eprint {http://arxiv.org/abs/astro-ph/0612757}
  {astro-ph/0612757} \BibitemShut {NoStop}%
\bibitem [{\citenamefont {{Hamers}}(2018)}]{2018MNRAS.476.4139H}%
  \BibitemOpen
  \bibfield  {author} {\bibinfo {author} {\bibfnamefont {A.~S.}\ \bibnamefont
  {{Hamers}}},\ }\href {\doibase 10.1093/mnras/sty428} {\bibfield  {journal}
  {\bibinfo  {journal} {\mnras}\ }\textbf {\bibinfo {volume} {476}},\ \bibinfo
  {pages} {4139} (\bibinfo {year} {2018})},\ \Eprint
  {http://arxiv.org/abs/1802.05716} {arXiv:1802.05716 [astro-ph.SR]}
  \BibitemShut {NoStop}%
\bibitem [{\citenamefont {{Hamers}}\ and\ \citenamefont
  {{Samsing}}(2019{\natexlab{b}})}]{2019MNRAS.488.5192H}%
  \BibitemOpen
  \bibfield  {author} {\bibinfo {author} {\bibfnamefont {A.~S.}\ \bibnamefont
  {{Hamers}}}\ and\ \bibinfo {author} {\bibfnamefont {J.}~\bibnamefont
  {{Samsing}}},\ }\href {\doibase 10.1093/mnras/stz2029} {\bibfield  {journal}
  {\bibinfo  {journal} {\mnras}\ }\textbf {\bibinfo {volume} {488}},\ \bibinfo
  {pages} {5192} (\bibinfo {year} {2019}{\natexlab{b}})},\ \Eprint
  {http://arxiv.org/abs/1906.08666} {arXiv:1906.08666 [astro-ph.HE]}
  \BibitemShut {NoStop}%
\bibitem [{\citenamefont {Bullock}\ and\ \citenamefont
  {Primack}(1997)}]{Bullock:1996at}%
  \BibitemOpen
  \bibfield  {author} {\bibinfo {author} {\bibfnamefont {J.~S.}\ \bibnamefont
  {Bullock}}\ and\ \bibinfo {author} {\bibfnamefont {J.~R.}\ \bibnamefont
  {Primack}},\ }\href {\doibase 10.1103/PhysRevD.55.7423} {\bibfield  {journal}
  {\bibinfo  {journal} {Phys. Rev.}\ }\textbf {\bibinfo {volume} {D55}},\
  \bibinfo {pages} {7423} (\bibinfo {year} {1997})},\ \Eprint
  {http://arxiv.org/abs/astro-ph/9611106} {arXiv:astro-ph/9611106 [astro-ph]}
  \BibitemShut {NoStop}%
%%CITATION = ASTRO-PH/9611106;%%
\bibitem [{\citenamefont {Ivanov}(1998)}]{Ivanov:1997ia}%
  \BibitemOpen
  \bibfield  {author} {\bibinfo {author} {\bibfnamefont {P.}~\bibnamefont
  {Ivanov}},\ }\href {\doibase 10.1103/PhysRevD.57.7145} {\bibfield  {journal}
  {\bibinfo  {journal} {Phys. Rev.}\ }\textbf {\bibinfo {volume} {D57}},\
  \bibinfo {pages} {7145} (\bibinfo {year} {1998})},\ \Eprint
  {http://arxiv.org/abs/astro-ph/9708224} {arXiv:astro-ph/9708224 [astro-ph]}
  \BibitemShut {NoStop}%
%%CITATION = ASTRO-PH/9708224;%%
\bibitem [{\citenamefont {Byrnes}\ \emph {et~al.}(2012)\citenamefont {Byrnes},
  \citenamefont {Copeland},\ and\ \citenamefont {Green}}]{Byrnes:2012yx}%
  \BibitemOpen
  \bibfield  {author} {\bibinfo {author} {\bibfnamefont {C.~T.}\ \bibnamefont
  {Byrnes}}, \bibinfo {author} {\bibfnamefont {E.~J.}\ \bibnamefont
  {Copeland}}, \ and\ \bibinfo {author} {\bibfnamefont {A.~M.}\ \bibnamefont
  {Green}},\ }\href {\doibase 10.1103/PhysRevD.86.043512} {\bibfield  {journal}
  {\bibinfo  {journal} {Phys. Rev.}\ }\textbf {\bibinfo {volume} {D86}},\
  \bibinfo {pages} {043512} (\bibinfo {year} {2012})},\ \Eprint
  {http://arxiv.org/abs/1206.4188} {arXiv:1206.4188 [astro-ph.CO]} \BibitemShut
  {NoStop}%
%%CITATION = ARXIV:1206.4188;%%
\bibitem [{\citenamefont {Shandera}\ \emph {et~al.}(2013)\citenamefont
  {Shandera}, \citenamefont {Erickcek}, \citenamefont {Scott},\ and\
  \citenamefont {Galarza}}]{Shandera:2012ke}%
  \BibitemOpen
  \bibfield  {author} {\bibinfo {author} {\bibfnamefont {S.}~\bibnamefont
  {Shandera}}, \bibinfo {author} {\bibfnamefont {A.~L.}\ \bibnamefont
  {Erickcek}}, \bibinfo {author} {\bibfnamefont {P.}~\bibnamefont {Scott}}, \
  and\ \bibinfo {author} {\bibfnamefont {J.~Y.}\ \bibnamefont {Galarza}},\
  }\href {\doibase 10.1103/PhysRevD.88.103506} {\bibfield  {journal} {\bibinfo
  {journal} {Phys. Rev.}\ }\textbf {\bibinfo {volume} {D88}},\ \bibinfo {pages}
  {103506} (\bibinfo {year} {2013})},\ \Eprint {http://arxiv.org/abs/'}
  {arXiv:' [astro-ph.CO]} \BibitemShut {NoStop}%
%%CITATION = ARXIV:1211.7361;%%
\bibitem [{\citenamefont {Young}\ and\ \citenamefont
  {Byrnes}(2013)}]{Young:2013oia}%
  \BibitemOpen
  \bibfield  {author} {\bibinfo {author} {\bibfnamefont {S.}~\bibnamefont
  {Young}}\ and\ \bibinfo {author} {\bibfnamefont {C.~T.}\ \bibnamefont
  {Byrnes}},\ }\href {\doibase 10.1088/1475-7516/2013/08/052} {\bibfield
  {journal} {\bibinfo  {journal} {JCAP}\ }\textbf {\bibinfo {volume} {1308}},\
  \bibinfo {pages} {052} (\bibinfo {year} {2013})},\ \Eprint
  {http://arxiv.org/abs/1307.4995} {arXiv:1307.4995 [astro-ph.CO]} \BibitemShut
  {NoStop}%
%%CITATION = ARXIV:1307.4995;%%
\bibitem [{\citenamefont {Young}\ and\ \citenamefont
  {Byrnes}(2015{\natexlab{a}})}]{Young:2014oea}%
  \BibitemOpen
  \bibfield  {author} {\bibinfo {author} {\bibfnamefont {S.}~\bibnamefont
  {Young}}\ and\ \bibinfo {author} {\bibfnamefont {C.~T.}\ \bibnamefont
  {Byrnes}},\ }\href {\doibase 10.1103/PhysRevD.91.083521} {\bibfield
  {journal} {\bibinfo  {journal} {Phys. Rev. D}\ }\textbf {\bibinfo {volume}
  {91}},\ \bibinfo {pages} {083521} (\bibinfo {year} {2015}{\natexlab{a}})},\
  \Eprint {http://arxiv.org/abs/1411.4620} {arXiv:1411.4620 [astro-ph.CO]}
  \BibitemShut {NoStop}%
\bibitem [{\citenamefont {Young}\ \emph {et~al.}(2016)\citenamefont {Young},
  \citenamefont {Regan},\ and\ \citenamefont {Byrnes}}]{Young:2015cyn}%
  \BibitemOpen
  \bibfield  {author} {\bibinfo {author} {\bibfnamefont {S.}~\bibnamefont
  {Young}}, \bibinfo {author} {\bibfnamefont {D.}~\bibnamefont {Regan}}, \ and\
  \bibinfo {author} {\bibfnamefont {C.~T.}\ \bibnamefont {Byrnes}},\ }\href
  {\doibase 10.1088/1475-7516/2016/02/029} {\bibfield  {journal} {\bibinfo
  {journal} {JCAP}\ }\textbf {\bibinfo {volume} {1602}},\ \bibinfo {pages}
  {029} (\bibinfo {year} {2016})},\ \Eprint {http://arxiv.org/abs/1512.07224}
  {arXiv:1512.07224 [astro-ph.CO]} \BibitemShut {NoStop}%
%%CITATION = ARXIV:1512.07224;%%
\bibitem [{\citenamefont {Franciolini}\ \emph {et~al.}(2018)\citenamefont
  {Franciolini}, \citenamefont {Kehagias}, \citenamefont {Matarrese},\ and\
  \citenamefont {Riotto}}]{Franciolini:2018vbk}%
  \BibitemOpen
  \bibfield  {author} {\bibinfo {author} {\bibfnamefont {G.}~\bibnamefont
  {Franciolini}}, \bibinfo {author} {\bibfnamefont {A.}~\bibnamefont
  {Kehagias}}, \bibinfo {author} {\bibfnamefont {S.}~\bibnamefont {Matarrese}},
  \ and\ \bibinfo {author} {\bibfnamefont {A.}~\bibnamefont {Riotto}},\ }\href
  {\doibase 10.1088/1475-7516/2018/03/016} {\bibfield  {journal} {\bibinfo
  {journal} {JCAP}\ }\textbf {\bibinfo {volume} {1803}},\ \bibinfo {pages}
  {016} (\bibinfo {year} {2018})},\ \Eprint {http://arxiv.org/abs/1801.09415}
  {arXiv:1801.09415 [astro-ph.CO]} \BibitemShut {NoStop}%
%%CITATION = ARXIV:1801.09415;%%
\bibitem [{\citenamefont {Yoo}\ \emph {et~al.}(2019)\citenamefont {Yoo},
  \citenamefont {Gong},\ and\ \citenamefont {Yokoyama}}]{Yoo:2019pma}%
  \BibitemOpen
  \bibfield  {author} {\bibinfo {author} {\bibfnamefont {C.-M.}\ \bibnamefont
  {Yoo}}, \bibinfo {author} {\bibfnamefont {J.-O.}\ \bibnamefont {Gong}}, \
  and\ \bibinfo {author} {\bibfnamefont {S.}~\bibnamefont {Yokoyama}},\ }\href
  {\doibase 10.1088/1475-7516/2019/09/033} {\bibfield  {journal} {\bibinfo
  {journal} {JCAP}\ }\textbf {\bibinfo {volume} {09}},\ \bibinfo {pages} {033}
  (\bibinfo {year} {2019})},\ \Eprint {http://arxiv.org/abs/1906.06790}
  {arXiv:1906.06790 [astro-ph.CO]} \BibitemShut {NoStop}%
\bibitem [{\citenamefont {Atal}\ \emph {et~al.}(2019)\citenamefont {Atal},
  \citenamefont {Garriga},\ and\ \citenamefont
  {Marcos-Caballero}}]{Atal:2019cdz}%
  \BibitemOpen
  \bibfield  {author} {\bibinfo {author} {\bibfnamefont {V.}~\bibnamefont
  {Atal}}, \bibinfo {author} {\bibfnamefont {J.}~\bibnamefont {Garriga}}, \
  and\ \bibinfo {author} {\bibfnamefont {A.}~\bibnamefont {Marcos-Caballero}},\
  }\href {\doibase 10.1088/1475-7516/2019/09/073} {\bibfield  {journal}
  {\bibinfo  {journal} {JCAP}\ }\textbf {\bibinfo {volume} {09}},\ \bibinfo
  {pages} {073} (\bibinfo {year} {2019})},\ \Eprint
  {http://arxiv.org/abs/1905.13202} {arXiv:1905.13202 [astro-ph.CO]}
  \BibitemShut {NoStop}%
\bibitem [{\citenamefont {Atal}\ and\ \citenamefont
  {Germani}(2019)}]{Atal:2018neu}%
  \BibitemOpen
  \bibfield  {author} {\bibinfo {author} {\bibfnamefont {V.}~\bibnamefont
  {Atal}}\ and\ \bibinfo {author} {\bibfnamefont {C.}~\bibnamefont {Germani}},\
  }\href {\doibase 10.1016/j.dark.2019.100275} {\bibfield  {journal} {\bibinfo
  {journal} {Phys. Dark Univ.}\ }\textbf {\bibinfo {volume} {24}},\ \bibinfo
  {pages} {100275} (\bibinfo {year} {2019})},\ \Eprint
  {http://arxiv.org/abs/1811.07857} {arXiv:1811.07857 [astro-ph.CO]}
  \BibitemShut {NoStop}%
\bibitem [{\citenamefont {Tada}\ and\ \citenamefont
  {Yokoyama}(2015)}]{Tada:2015noa}%
  \BibitemOpen
  \bibfield  {author} {\bibinfo {author} {\bibfnamefont {Y.}~\bibnamefont
  {Tada}}\ and\ \bibinfo {author} {\bibfnamefont {S.}~\bibnamefont
  {Yokoyama}},\ }\href {\doibase 10.1103/PhysRevD.91.123534} {\bibfield
  {journal} {\bibinfo  {journal} {Phys. Rev. D}\ }\textbf {\bibinfo {volume}
  {91}},\ \bibinfo {pages} {123534} (\bibinfo {year} {2015})},\ \Eprint
  {http://arxiv.org/abs/1502.01124} {arXiv:1502.01124 [astro-ph.CO]}
  \BibitemShut {NoStop}%
\bibitem [{\citenamefont {Young}\ and\ \citenamefont
  {Byrnes}(2015{\natexlab{b}})}]{Young:2015kda}%
  \BibitemOpen
  \bibfield  {author} {\bibinfo {author} {\bibfnamefont {S.}~\bibnamefont
  {Young}}\ and\ \bibinfo {author} {\bibfnamefont {C.~T.}\ \bibnamefont
  {Byrnes}},\ }\href {\doibase 10.1088/1475-7516/2015/04/034} {\bibfield
  {journal} {\bibinfo  {journal} {JCAP}\ }\textbf {\bibinfo {volume} {04}},\
  \bibinfo {pages} {034} (\bibinfo {year} {2015}{\natexlab{b}})},\ \Eprint
  {http://arxiv.org/abs/1503.01505} {arXiv:1503.01505 [astro-ph.CO]}
  \BibitemShut {NoStop}%
\bibitem [{\citenamefont {Suyama}\ and\ \citenamefont
  {Yokoyama}(2019)}]{Suyama:2019cst}%
  \BibitemOpen
  \bibfield  {author} {\bibinfo {author} {\bibfnamefont {T.}~\bibnamefont
  {Suyama}}\ and\ \bibinfo {author} {\bibfnamefont {S.}~\bibnamefont
  {Yokoyama}},\ }\href {\doibase 10.1093/ptep/ptz105} {\bibfield  {journal}
  {\bibinfo  {journal} {PTEP}\ }\textbf {\bibinfo {volume} {2019}},\ \bibinfo
  {pages} {103E02} (\bibinfo {year} {2019})},\ \Eprint
  {http://arxiv.org/abs/1906.04958} {arXiv:1906.04958 [astro-ph.CO]}
  \BibitemShut {NoStop}%
\bibitem [{\citenamefont {Young}\ and\ \citenamefont
  {Byrnes}(2020)}]{Young:2019gfc}%
  \BibitemOpen
  \bibfield  {author} {\bibinfo {author} {\bibfnamefont {S.}~\bibnamefont
  {Young}}\ and\ \bibinfo {author} {\bibfnamefont {C.~T.}\ \bibnamefont
  {Byrnes}},\ }\href {\doibase 10.1088/1475-7516/2020/03/004} {\bibfield
  {journal} {\bibinfo  {journal} {JCAP}\ }\textbf {\bibinfo {volume} {03}},\
  \bibinfo {pages} {004} (\bibinfo {year} {2020})},\ \Eprint
  {http://arxiv.org/abs/1910.06077} {arXiv:1910.06077 [astro-ph.CO]}
  \BibitemShut {NoStop}%
\bibitem [{\citenamefont {Young}\ \emph {et~al.}(2019)\citenamefont {Young},
  \citenamefont {Musco},\ and\ \citenamefont {Byrnes}}]{Young:2019yug}%
  \BibitemOpen
  \bibfield  {author} {\bibinfo {author} {\bibfnamefont {S.}~\bibnamefont
  {Young}}, \bibinfo {author} {\bibfnamefont {I.}~\bibnamefont {Musco}}, \ and\
  \bibinfo {author} {\bibfnamefont {C.~T.}\ \bibnamefont {Byrnes}},\ }\href
  {\doibase 10.1088/1475-7516/2019/11/012} {\bibfield  {journal} {\bibinfo
  {journal} {JCAP}\ }\textbf {\bibinfo {volume} {11}},\ \bibinfo {pages} {012}
  (\bibinfo {year} {2019})},\ \Eprint {http://arxiv.org/abs/1904.00984}
  {arXiv:1904.00984 [astro-ph.CO]} \BibitemShut {NoStop}%
\bibitem [{\citenamefont {Yoo}\ \emph {et~al.}(2018)\citenamefont {Yoo},
  \citenamefont {Harada}, \citenamefont {Garriga},\ and\ \citenamefont
  {Kohri}}]{Yoo:2018kvb}%
  \BibitemOpen
  \bibfield  {author} {\bibinfo {author} {\bibfnamefont {C.-M.}\ \bibnamefont
  {Yoo}}, \bibinfo {author} {\bibfnamefont {T.}~\bibnamefont {Harada}},
  \bibinfo {author} {\bibfnamefont {J.}~\bibnamefont {Garriga}}, \ and\
  \bibinfo {author} {\bibfnamefont {K.}~\bibnamefont {Kohri}},\ }\href
  {\doibase 10.1093/ptep/pty120} {\bibfield  {journal} {\bibinfo  {journal}
  {PTEP}\ }\textbf {\bibinfo {volume} {2018}},\ \bibinfo {pages} {123E01}
  (\bibinfo {year} {2018})},\ \Eprint {http://arxiv.org/abs/1805.03946}
  {arXiv:1805.03946 [astro-ph.CO]} \BibitemShut {NoStop}%
\bibitem [{\citenamefont {Kawasaki}\ and\ \citenamefont
  {Nakatsuka}(2019)}]{Kawasaki:2019mbl}%
  \BibitemOpen
  \bibfield  {author} {\bibinfo {author} {\bibfnamefont {M.}~\bibnamefont
  {Kawasaki}}\ and\ \bibinfo {author} {\bibfnamefont {H.}~\bibnamefont
  {Nakatsuka}},\ }\href {\doibase 10.1103/PhysRevD.99.123501} {\bibfield
  {journal} {\bibinfo  {journal} {Phys. Rev. D}\ }\textbf {\bibinfo {volume}
  {99}},\ \bibinfo {pages} {123501} (\bibinfo {year} {2019})},\ \Eprint
  {http://arxiv.org/abs/1903.02994} {arXiv:1903.02994 [astro-ph.CO]}
  \BibitemShut {NoStop}%
\bibitem [{\citenamefont {Kalaja}\ \emph {et~al.}(2019)\citenamefont {Kalaja},
  \citenamefont {Bellomo}, \citenamefont {Bartolo}, \citenamefont {Bertacca},
  \citenamefont {Matarrese}, \citenamefont {Musco}, \citenamefont
  {Raccanelli},\ and\ \citenamefont {Verde}}]{Kalaja:2019uju}%
  \BibitemOpen
  \bibfield  {author} {\bibinfo {author} {\bibfnamefont {A.}~\bibnamefont
  {Kalaja}}, \bibinfo {author} {\bibfnamefont {N.}~\bibnamefont {Bellomo}},
  \bibinfo {author} {\bibfnamefont {N.}~\bibnamefont {Bartolo}}, \bibinfo
  {author} {\bibfnamefont {D.}~\bibnamefont {Bertacca}}, \bibinfo {author}
  {\bibfnamefont {S.}~\bibnamefont {Matarrese}}, \bibinfo {author}
  {\bibfnamefont {I.}~\bibnamefont {Musco}}, \bibinfo {author} {\bibfnamefont
  {A.}~\bibnamefont {Raccanelli}}, \ and\ \bibinfo {author} {\bibfnamefont
  {L.}~\bibnamefont {Verde}},\ }\href {\doibase 10.1088/1475-7516/2019/10/031}
  {\bibfield  {journal} {\bibinfo  {journal} {JCAP}\ }\textbf {\bibinfo
  {volume} {10}},\ \bibinfo {pages} {031} (\bibinfo {year} {2019})},\ \Eprint
  {http://arxiv.org/abs/1908.03596} {arXiv:1908.03596 [astro-ph.CO]}
  \BibitemShut {NoStop}%
\bibitem [{\citenamefont {De~Luca}\ \emph
  {et~al.}(2019{\natexlab{b}})\citenamefont {De~Luca}, \citenamefont
  {Franciolini}, \citenamefont {Kehagias}, \citenamefont {Peloso},
  \citenamefont {Riotto},\ and\ \citenamefont {Ünal}}]{DeLuca:2019qsy}%
  \BibitemOpen
  \bibfield  {author} {\bibinfo {author} {\bibfnamefont {V.}~\bibnamefont
  {De~Luca}}, \bibinfo {author} {\bibfnamefont {G.}~\bibnamefont
  {Franciolini}}, \bibinfo {author} {\bibfnamefont {A.}~\bibnamefont
  {Kehagias}}, \bibinfo {author} {\bibfnamefont {M.}~\bibnamefont {Peloso}},
  \bibinfo {author} {\bibfnamefont {A.}~\bibnamefont {Riotto}}, \ and\ \bibinfo
  {author} {\bibfnamefont {C.}~\bibnamefont {Ünal}},\ }\href {\doibase
  10.1088/1475-7516/2019/07/048} {\bibfield  {journal} {\bibinfo  {journal}
  {JCAP}\ }\textbf {\bibinfo {volume} {07}},\ \bibinfo {pages} {048} (\bibinfo
  {year} {2019}{\natexlab{b}})},\ \Eprint {http://arxiv.org/abs/1904.00970}
  {arXiv:1904.00970 [astro-ph.CO]} \BibitemShut {NoStop}%
\bibitem [{\citenamefont {Gow}\ \emph {et~al.}(2020)\citenamefont {Gow},
  \citenamefont {Byrnes}, \citenamefont {Cole},\ and\ \citenamefont
  {Young}}]{Gow:2020bzo}%
  \BibitemOpen
  \bibfield  {author} {\bibinfo {author} {\bibfnamefont {A.~D.}\ \bibnamefont
  {Gow}}, \bibinfo {author} {\bibfnamefont {C.~T.}\ \bibnamefont {Byrnes}},
  \bibinfo {author} {\bibfnamefont {P.~S.}\ \bibnamefont {Cole}}, \ and\
  \bibinfo {author} {\bibfnamefont {S.}~\bibnamefont {Young}},\ }\href@noop {}
  {\  (\bibinfo {year} {2020})},\ \Eprint {http://arxiv.org/abs/2008.03289}
  {arXiv:2008.03289 [astro-ph.CO]} \BibitemShut {NoStop}%
\bibitem [{\citenamefont {{Hut}}\ and\ \citenamefont
  {{Bahcall}}(1983)}]{1983ApJ...268..319H}%
  \BibitemOpen
  \bibfield  {author} {\bibinfo {author} {\bibfnamefont {P.}~\bibnamefont
  {{Hut}}}\ and\ \bibinfo {author} {\bibfnamefont {J.~N.}\ \bibnamefont
  {{Bahcall}}},\ }\href {\doibase 10.1086/160956} {\bibfield  {journal}
  {\bibinfo  {journal} {\apj}\ }\textbf {\bibinfo {volume} {268}},\ \bibinfo
  {pages} {319} (\bibinfo {year} {1983})}\BibitemShut {NoStop}%
\bibitem [{\citenamefont {{Hut}}(1983)}]{1983ApJ...268..342H}%
  \BibitemOpen
  \bibfield  {author} {\bibinfo {author} {\bibfnamefont {P.}~\bibnamefont
  {{Hut}}},\ }\href {\doibase 10.1086/160957} {\bibfield  {journal} {\bibinfo
  {journal} {\apj}\ }\textbf {\bibinfo {volume} {268}},\ \bibinfo {pages} {342}
  (\bibinfo {year} {1983})}\BibitemShut {NoStop}%
\bibitem [{\citenamefont {{Heggie}}\ and\ \citenamefont
  {{Sweatman}}(1991)}]{1991MNRAS.250..555H}%
  \BibitemOpen
  \bibfield  {author} {\bibinfo {author} {\bibfnamefont {D.~C.}\ \bibnamefont
  {{Heggie}}}\ and\ \bibinfo {author} {\bibfnamefont {W.~L.}\ \bibnamefont
  {{Sweatman}}},\ }\href {\doibase 10.1093/mnras/250.3.555} {\bibfield
  {journal} {\bibinfo  {journal} {\mnras}\ }\textbf {\bibinfo {volume} {250}},\
  \bibinfo {pages} {555} (\bibinfo {year} {1991})}\BibitemShut {NoStop}%
\bibitem [{\citenamefont {{Heggie}}\ and\ \citenamefont
  {{Hut}}(1993)}]{1993ApJS...85..347H}%
  \BibitemOpen
  \bibfield  {author} {\bibinfo {author} {\bibfnamefont {D.~C.}\ \bibnamefont
  {{Heggie}}}\ and\ \bibinfo {author} {\bibfnamefont {P.}~\bibnamefont
  {{Hut}}},\ }\href {\doibase 10.1086/191768} {\bibfield  {journal} {\bibinfo
  {journal} {\apjs}\ }\textbf {\bibinfo {volume} {85}},\ \bibinfo {pages} {347}
  (\bibinfo {year} {1993})}\BibitemShut {NoStop}%
\bibitem [{\citenamefont {{Hut}}(1993)}]{1993ApJ...403..256H}%
  \BibitemOpen
  \bibfield  {author} {\bibinfo {author} {\bibfnamefont {P.}~\bibnamefont
  {{Hut}}},\ }\href {\doibase 10.1086/172199} {\bibfield  {journal} {\bibinfo
  {journal} {\apj}\ }\textbf {\bibinfo {volume} {403}},\ \bibinfo {pages} {256}
  (\bibinfo {year} {1993})}\BibitemShut {NoStop}%
\bibitem [{\citenamefont {{Goodman}}\ and\ \citenamefont
  {{Hut}}(1993)}]{1993ApJ...403..271G}%
  \BibitemOpen
  \bibfield  {author} {\bibinfo {author} {\bibfnamefont {J.}~\bibnamefont
  {{Goodman}}}\ and\ \bibinfo {author} {\bibfnamefont {P.}~\bibnamefont
  {{Hut}}},\ }\href {\doibase 10.1086/172200} {\bibfield  {journal} {\bibinfo
  {journal} {\apj}\ }\textbf {\bibinfo {volume} {403}},\ \bibinfo {pages} {271}
  (\bibinfo {year} {1993})}\BibitemShut {NoStop}%
\bibitem [{\citenamefont {{Sigurdsson}}\ and\ \citenamefont
  {{Phinney}}(1993)}]{1993ApJ...415..631S}%
  \BibitemOpen
  \bibfield  {author} {\bibinfo {author} {\bibfnamefont {S.}~\bibnamefont
  {{Sigurdsson}}}\ and\ \bibinfo {author} {\bibfnamefont {E.~S.}\ \bibnamefont
  {{Phinney}}},\ }\href {\doibase 10.1086/173190} {\bibfield  {journal}
  {\bibinfo  {journal} {\apj}\ }\textbf {\bibinfo {volume} {415}},\ \bibinfo
  {pages} {631} (\bibinfo {year} {1993})}\BibitemShut {NoStop}%
\bibitem [{\citenamefont {{Davies}}\ \emph {et~al.}(1993)\citenamefont
  {{Davies}}, \citenamefont {{Benz}},\ and\ \citenamefont
  {{Hills}}}]{1993ApJ...411..285D}%
  \BibitemOpen
  \bibfield  {author} {\bibinfo {author} {\bibfnamefont {M.~B.}\ \bibnamefont
  {{Davies}}}, \bibinfo {author} {\bibfnamefont {W.}~\bibnamefont {{Benz}}}, \
  and\ \bibinfo {author} {\bibfnamefont {J.~G.}\ \bibnamefont {{Hills}}},\
  }\href {\doibase 10.1086/172828} {\bibfield  {journal} {\bibinfo  {journal}
  {\apj}\ }\textbf {\bibinfo {volume} {411}},\ \bibinfo {pages} {285} (\bibinfo
  {year} {1993})}\BibitemShut {NoStop}%
\bibitem [{\citenamefont {{McMillan}}\ and\ \citenamefont
  {{Hut}}(1996)}]{1996ApJ...467..348M}%
  \BibitemOpen
  \bibfield  {author} {\bibinfo {author} {\bibfnamefont {S.~L.~W.}\
  \bibnamefont {{McMillan}}}\ and\ \bibinfo {author} {\bibfnamefont
  {P.}~\bibnamefont {{Hut}}},\ }\href {\doibase 10.1086/177610} {\bibfield
  {journal} {\bibinfo  {journal} {\apj}\ }\textbf {\bibinfo {volume} {467}},\
  \bibinfo {pages} {348} (\bibinfo {year} {1996})},\ \Eprint
  {http://arxiv.org/abs/astro-ph/9604015} {arXiv:astro-ph/9604015 [astro-ph]}
  \BibitemShut {NoStop}%
\bibitem [{\citenamefont {{Heggie}}\ \emph {et~al.}(1996)\citenamefont
  {{Heggie}}, \citenamefont {{Hut}},\ and\ \citenamefont
  {{McMillan}}}]{1996ApJ...467..359H}%
  \BibitemOpen
  \bibfield  {author} {\bibinfo {author} {\bibfnamefont {D.~C.}\ \bibnamefont
  {{Heggie}}}, \bibinfo {author} {\bibfnamefont {P.}~\bibnamefont {{Hut}}}, \
  and\ \bibinfo {author} {\bibfnamefont {S.~L.~W.}\ \bibnamefont
  {{McMillan}}},\ }\href {\doibase 10.1086/177611} {\bibfield  {journal}
  {\bibinfo  {journal} {\apj}\ }\textbf {\bibinfo {volume} {467}},\ \bibinfo
  {pages} {359} (\bibinfo {year} {1996})}\BibitemShut {NoStop}%
\bibitem [{\citenamefont {{Kocsis}}\ and\ \citenamefont
  {{Levin}}(2012)}]{2012PhRvD..85l3005K}%
  \BibitemOpen
  \bibfield  {author} {\bibinfo {author} {\bibfnamefont {B.}~\bibnamefont
  {{Kocsis}}}\ and\ \bibinfo {author} {\bibfnamefont {J.}~\bibnamefont
  {{Levin}}},\ }\href {\doibase 10.1103/PhysRevD.85.123005} {\bibfield
  {journal} {\bibinfo  {journal} {\prd}\ }\textbf {\bibinfo {volume} {85}},\
  \bibinfo {eid} {123005} (\bibinfo {year} {2012})},\ \Eprint
  {http://arxiv.org/abs/1109.4170} {arXiv:1109.4170 [astro-ph.CO]} \BibitemShut
  {NoStop}%
\bibitem [{\citenamefont {{Samsing}}\ \emph
  {et~al.}(2018{\natexlab{c}})\citenamefont {{Samsing}}, \citenamefont
  {{D'Orazio}}, \citenamefont {{Askar}},\ and\ \citenamefont
  {{Giersz}}}]{2018arXiv180208654S}%
  \BibitemOpen
  \bibfield  {author} {\bibinfo {author} {\bibfnamefont {J.}~\bibnamefont
  {{Samsing}}}, \bibinfo {author} {\bibfnamefont {D.~J.}\ \bibnamefont
  {{D'Orazio}}}, \bibinfo {author} {\bibfnamefont {A.}~\bibnamefont {{Askar}}},
  \ and\ \bibinfo {author} {\bibfnamefont {M.}~\bibnamefont {{Giersz}}},\
  }\href@noop {} {\bibfield  {journal} {\bibinfo  {journal} {arXiv e-prints}\
  ,\ \bibinfo {eid} {arXiv:1802.08654}} (\bibinfo {year}
  {2018}{\natexlab{c}})},\ \Eprint {http://arxiv.org/abs/1802.08654}
  {arXiv:1802.08654 [astro-ph.HE]} \BibitemShut {NoStop}%
\bibitem [{\citenamefont {{Peters}}(1964)}]{1964PhRv..136.1224P}%
  \BibitemOpen
  \bibfield  {author} {\bibinfo {author} {\bibfnamefont {P.~C.}\ \bibnamefont
  {{Peters}}},\ }\href {\doibase 10.1103/PhysRev.136.B1224} {\bibfield
  {journal} {\bibinfo  {journal} {Physical Review}\ }\textbf {\bibinfo {volume}
  {136}},\ \bibinfo {pages} {1224} (\bibinfo {year} {1964})}\BibitemShut
  {NoStop}%
\bibitem [{\citenamefont {{Cohen}}\ \emph {et~al.}(1996)\citenamefont
  {{Cohen}}, \citenamefont {{Hindmarsh}},\ and\ \citenamefont
  {{Dubois}}}]{1996ComPh..10..138C}%
  \BibitemOpen
  \bibfield  {author} {\bibinfo {author} {\bibfnamefont {S.~D.}\ \bibnamefont
  {{Cohen}}}, \bibinfo {author} {\bibfnamefont {A.~C.}\ \bibnamefont
  {{Hindmarsh}}}, \ and\ \bibinfo {author} {\bibfnamefont {P.~F.}\ \bibnamefont
  {{Dubois}}},\ }\href@noop {} {\bibfield  {journal} {\bibinfo  {journal}
  {Computers in Physics}\ }\textbf {\bibinfo {volume} {10}},\ \bibinfo {pages}
  {138} (\bibinfo {year} {1996})}\BibitemShut {NoStop}%
\bibitem [{\citenamefont {Inman}\ and\ \citenamefont
  {Ali-Haïmoud}(2019)}]{Inman:2019wvr}%
  \BibitemOpen
  \bibfield  {author} {\bibinfo {author} {\bibfnamefont {D.}~\bibnamefont
  {Inman}}\ and\ \bibinfo {author} {\bibfnamefont {Y.}~\bibnamefont
  {Ali-Haïmoud}},\ }\href {\doibase 10.1103/PhysRevD.100.083528} {\bibfield
  {journal} {\bibinfo  {journal} {Phys. Rev. D}\ }\textbf {\bibinfo {volume}
  {100}},\ \bibinfo {pages} {083528} (\bibinfo {year} {2019})},\ \Eprint
  {http://arxiv.org/abs/1907.08129} {arXiv:1907.08129 [astro-ph.CO]}
  \BibitemShut {NoStop}%
\bibitem [{\citenamefont {Hoeft}\ \emph {et~al.}(2004)\citenamefont {Hoeft},
  \citenamefont {Mucket},\ and\ \citenamefont {Gottlober}}]{Hoeft:2003ea}%
  \BibitemOpen
  \bibfield  {author} {\bibinfo {author} {\bibfnamefont {M.}~\bibnamefont
  {Hoeft}}, \bibinfo {author} {\bibfnamefont {J.}~\bibnamefont {Mucket}}, \
  and\ \bibinfo {author} {\bibfnamefont {S.}~\bibnamefont {Gottlober}},\ }\href
  {\doibase 10.1086/380990} {\bibfield  {journal} {\bibinfo  {journal}
  {Astrophys. J.}\ }\textbf {\bibinfo {volume} {602}},\ \bibinfo {pages} {162}
  (\bibinfo {year} {2004})},\ \Eprint {http://arxiv.org/abs/astro-ph/0311083}
  {arXiv:astro-ph/0311083} \BibitemShut {NoStop}%
\bibitem [{\citenamefont {{Hamers}}\ and\ \citenamefont
  {{Samsing}}(2020)}]{2020MNRAS.494..850H}%
  \BibitemOpen
  \bibfield  {author} {\bibinfo {author} {\bibfnamefont {A.~S.}\ \bibnamefont
  {{Hamers}}}\ and\ \bibinfo {author} {\bibfnamefont {J.}~\bibnamefont
  {{Samsing}}},\ }\href {\doibase 10.1093/mnras/staa691} {\bibfield  {journal}
  {\bibinfo  {journal} {\mnras}\ }\textbf {\bibinfo {volume} {494}},\ \bibinfo
  {pages} {850} (\bibinfo {year} {2020})},\ \Eprint
  {http://arxiv.org/abs/2002.04950} {arXiv:2002.04950 [astro-ph.HE]}
  \BibitemShut {NoStop}%
\bibitem [{\citenamefont {Boucenna}\ \emph {et~al.}(2018)\citenamefont
  {Boucenna}, \citenamefont {Kuhnel}, \citenamefont {Ohlsson},\ and\
  \citenamefont {Visinelli}}]{Boucenna:2017ghj}%
  \BibitemOpen
  \bibfield  {author} {\bibinfo {author} {\bibfnamefont {S.~M.}\ \bibnamefont
  {Boucenna}}, \bibinfo {author} {\bibfnamefont {F.}~\bibnamefont {Kuhnel}},
  \bibinfo {author} {\bibfnamefont {T.}~\bibnamefont {Ohlsson}}, \ and\
  \bibinfo {author} {\bibfnamefont {L.}~\bibnamefont {Visinelli}},\ }\href
  {\doibase 10.1088/1475-7516/2018/07/003} {\bibfield  {journal} {\bibinfo
  {journal} {JCAP}\ }\textbf {\bibinfo {volume} {07}},\ \bibinfo {pages} {003}
  (\bibinfo {year} {2018})},\ \Eprint {http://arxiv.org/abs/1712.06383}
  {arXiv:1712.06383 [hep-ph]} \BibitemShut {NoStop}%
\bibitem [{\citenamefont {Adamek}\ \emph {et~al.}(2019)\citenamefont {Adamek},
  \citenamefont {Byrnes}, \citenamefont {Gosenca},\ and\ \citenamefont
  {Hotchkiss}}]{Adamek:2019gns}%
  \BibitemOpen
  \bibfield  {author} {\bibinfo {author} {\bibfnamefont {J.}~\bibnamefont
  {Adamek}}, \bibinfo {author} {\bibfnamefont {C.~T.}\ \bibnamefont {Byrnes}},
  \bibinfo {author} {\bibfnamefont {M.}~\bibnamefont {Gosenca}}, \ and\
  \bibinfo {author} {\bibfnamefont {S.}~\bibnamefont {Hotchkiss}},\ }\href
  {\doibase 10.1103/PhysRevD.100.023506} {\bibfield  {journal} {\bibinfo
  {journal} {Phys. Rev. D}\ }\textbf {\bibinfo {volume} {100}},\ \bibinfo
  {pages} {023506} (\bibinfo {year} {2019})},\ \Eprint
  {http://arxiv.org/abs/1901.08528} {arXiv:1901.08528 [astro-ph.CO]}
  \BibitemShut {NoStop}%
\bibitem [{\citenamefont {Kavanagh}\ \emph {et~al.}(2020)\citenamefont
  {Kavanagh}, \citenamefont {Nichols}, \citenamefont {Bertone},\ and\
  \citenamefont {Gaggero}}]{Kavanagh:2020cfn}%
  \BibitemOpen
  \bibfield  {author} {\bibinfo {author} {\bibfnamefont {B.~J.}\ \bibnamefont
  {Kavanagh}}, \bibinfo {author} {\bibfnamefont {D.~A.}\ \bibnamefont
  {Nichols}}, \bibinfo {author} {\bibfnamefont {G.}~\bibnamefont {Bertone}}, \
  and\ \bibinfo {author} {\bibfnamefont {D.}~\bibnamefont {Gaggero}},\
  }\href@noop {} {\  (\bibinfo {year} {2020})},\ \Eprint
  {http://arxiv.org/abs/2002.12811} {arXiv:2002.12811 [gr-qc]} \BibitemShut
  {NoStop}%
\bibitem [{\citenamefont {Jedamzik}(2020)}]{Jedamzik:2020ypm}%
  \BibitemOpen
  \bibfield  {author} {\bibinfo {author} {\bibfnamefont {K.}~\bibnamefont
  {Jedamzik}},\ }\href@noop {} {\  (\bibinfo {year} {2020})},\ \Eprint
  {http://arxiv.org/abs/2006.11172} {arXiv:2006.11172 [astro-ph.CO]}
  \BibitemShut {NoStop}%
\end{thebibliography}%

\end{document}